\documentclass[sigconf]{acmart}

\usepackage[ruled,vlined]{algorithm2e}
\usepackage{amsmath}
\usepackage{multirow}
\usepackage{enumitem}
\usepackage{bm}
\usepackage{tabularx}
\usepackage{balance}
\usepackage{acmart-taps}



\newcommand{\gz}{{\color{lightgray}0}}

\AtBeginDocument{%
  \providecommand\BibTeX{{%
    \normalfont B\kern-0.5em{\scshape i\kern-0.25em b}\kern-0.8em\TeX}}}

\setcopyright{acmlicensed}
\copyrightyear{2026}
\acmYear{2026}
\setcopyright{cc}
\setcctype{by}
\acmConference[CHI '26]{Proceedings of the 2026 CHI Conference on Human Factors in Computing Systems}{April 13--17, 2026}{Barcelona, Spain}
\acmBooktitle{Proceedings of the 2026 CHI Conference on Human Factors in Computing Systems (CHI '26), April 13--17, 2026, Barcelona, Spain}
\acmPrice{}
\acmDOI{10.1145/3772318.3791375}
\acmISBN{979-8-4007-2278-3/2026/04}

\acmSubmissionID{8619}

\begin{document}

% \title{Editable XAI: Toward Bidirectional Human-AI Alignment with Co-Editable Explanations}
\title{Editable XAI: Toward Bidirectional Human-AI Alignment with\\Co-Editable Explanations of Interpretable Attributes}


\author{Haoyang Chen}
\orcid{0000-0001-7342-0549}
\affiliation{
\department{Department of Computer Science}\institution{National University of Singapore}
\city{Singapore}
\country{Singapore}}
\email{chen_haoyang@u.nus.edu}

\author{Jingwen Bai}
\orcid{0000-0001-5118-993X}
\affiliation{
\department{Department of Computer Science}\institution{National University of Singapore}
\city{Singapore}
\country{Singapore}}
\email{jingwenbai@u.nus.edu}

\author{Fang Tian}
\orcid{0009-0004-9549-0694}
\affiliation{
\department{College of Design and Engineering}\institution{National University of Singapore}
\city{Singapore}
\country{Singapore}}
\email{tfang@u.nus.edu}

\author{Brian Y Lim}
\authornote{Corresponding author}
\orcid{0000-0002-0543-2414}
\affiliation{
\department{Department of Computer Science}\institution{National University of Singapore}
\city{Singapore}
\country{Singapore}}
\email{brianlim@nus.edu.sg}

\renewcommand{\shortauthors}{Chen, Bai, Fang, and Lim}

\begin{abstract}

  While Explainable AI (XAI) helps users understand AI decisions, misalignment in domain knowledge can lead to disagreement. This inconsistency hinders understanding, and because explanations are often read-only, users lack the control to improve alignment. We propose making XAI editable, allowing users to write rules to improve control and gain deeper understanding through the generation effect of active learning. We developed CoExplain, leveraging a neural network for universal representation and symbolic rules for intuitive reasoning on interpretable attributes. CoExplain explains the neural network with a faithful proxy decision tree, parses user-written rules as an equivalent neural network graph, and collaboratively optimizes the decision tree. In a user study (N=43), CoExplain and manually editable XAI improved user understanding and model alignment compared to read-only XAI. CoExplain was easier to use with fewer edits and less time. This work contributes Editable XAI for bidirectional AI alignment, improving understanding and control.

\end{abstract}





\maketitle
\section{Introduction}

Explainable AI (XAI) helps users understand the reasoning behind AI decisions. However, when the AI is incorrect or incongruent with a user’s domain knowledge or beliefs, this results in human–AI misalignment, which can cause distrust, errors, and extra work for users to verify or override AI decisions~\cite{ji2023ai, zhuang2020consequences}. XAI can also exacerbate the problem: by presenting explanations that appear reasonable, it can reinforce automation bias, where users place undue trust in AI outputs and overlook potential errors~\cite{cabitza2024explanations}. When explanations are static and read-only, misalignment persists, leaving users unable to correct errors or incorporate their own knowledge.

We propose \textbf{Editable XAI} and hypothesize that allowing users to rewrite the rules of AI reasoning can improve alignment. By allowing users and AI to collaboratively inspect and revise a shared medium explanation, users can better understand the AI behavior and identify the misalignment. Beyond improving alignment, the act of editing explanations engages users in actively producing content rather than passively consuming it, leveraging the \textit{generation effect} from active learning~\cite{slamecka1978generation,wittrock1992generative}. In this way, editable explanations promote bi-directional alignment~\cite{shen2024towards,shen2025bidirectional}: 
I) users can adapt AI reasoning to their domain knowledge,
II) while simultaneously deepening their own understanding of how the AI makes decisions.

To explore this idea, we conducted an elicitation study to understand how users may want to edit or override AI reasoning. From this study, we identified several design needs for \textbf{Editable XAI}. 
First, explanations should be contextualized in a \textit{writable format}, such as rules, so that users can directly adjust rather than only read them. 
Second, systems should support \textit{user-written} rules that capture domain knowledge with user familiar terms, rather than relying solely on data-driven patterns. 
Third, AI assistance should provide \textit{automated enhancements} to user rules, such as adjusting thresholds or reorganizing rule topology, to help close knowledge gaps. 
Finally, automated edits should be \textit{constrained} to remain consistent with user intent and avoid introducing unnecessary complexity. These findings directly informed the design of our system.

\begin{figure}[t]
    \centering
    \includegraphics[width=6.0cm]{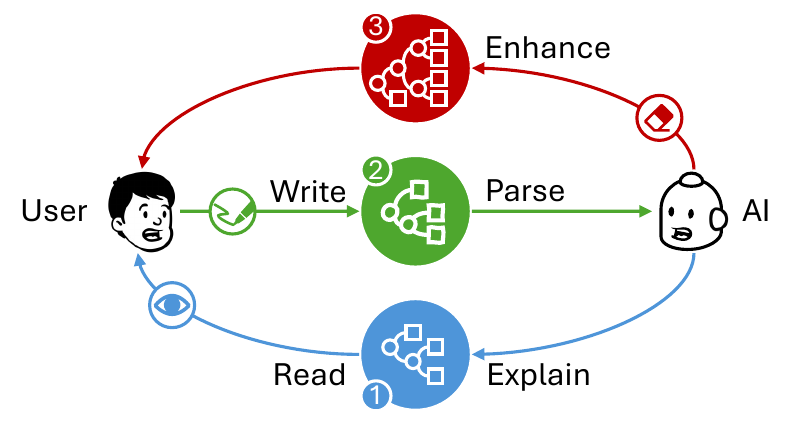}
    % \vspace{-0.4cm}
    \caption{The three interaction modes for Editable XAI.
    (1) Read: users inspect explanations generated by the AI, (2) Write: users modify explanations to guide the AI, and (3) Enhance: users and the AI collaboratively refine explanations. Creating a bidirectional alignment between the user and AI.}
    \label{fig:conceptual-overview}
    \Description[Circular workflow of Editable XAI's three interaction modes]{Circular diagram with three colored nodes (blue, green, red) and arrows forming a cycle between a user and an AI. Blue node (1): "Explain" leads to user "Read" via an eye icon. Green node (2): User "Write" (pencil icon) goes to "Parse" for AI. Red node (3): "Enhance" cycles back to user. Shows bidirectional user-AI alignment.}
    % \vspace{-0.4cm}
\end{figure}

Based on these requirements, we introduce \textbf{CoExplain}, a framework for \textbf{Editable XAI} that supports three modes of interaction: \textbf{Read}, \textbf{Write}, and \textbf{Enhance}. We use decision tree rules as the interaction medium as they are human-interpretable and human-editable while compatible with machine learning models. 
We use neural networks as the underlying AI model to exploit its expressivity and flexible training with gradient descent.
For \textbf{Read}, CoExplain distills a neural network into a decision tree to provide faithful and accessible explanations of model decisions. 
For \textbf{Write}, user-authored rules are parsed into an equivalent neural network, enabling the system to incorporate domain knowledge while maintaining trainability. 
For \textbf{Enhance}, CoExplain augments user-written rules in two ways: conservatively refining threshold values to adjust decision boundaries, and more aggressively reorganizing the tree’s topology to improve predictive performance. 
These actions are linked through neurosymbolic approaches that maintain equivalence between decision trees and neural networks via distillation, parsing, backpropagation, and regularization, thereby enabling bi-directional collaboration between humans and AI.

We evaluated CoExplain in a user study with 43 participants, compared it with Read-only XAI and manually Editable XAI without AI enhancements. 
Both editable techniques improved understanding and alignment compared to read-only explanations. 
Qualitative findings showed that writing rules helped users align AI reasoning with their own knowledge and deepened their understanding on model behavior, while AI enhancements were valued for refining thresholds and restructuring rules under user control. 
Quantitative results showed that both Editable and CoExplain improved understanding, with CoExplain striking a balance between maintaining alignment with users’ initial knowledge and achieving near-optimal model performance. 
CoExplain also reduced editing effort by providing on-demand enhancements, saving users time on refining thresholds and topology. 
Participants generally perceived it as easier to use and appreciated its collaborative nature. 

In summary, this paper makes two \textbf{contributions}:
\aptLtoX{\begin{enumerate}
    \item[1)] Insights on how users edit explanatory rules, both independently and collaboratively with AI, highlighting needs for \textbf{Editable XAI}, AI-guided enhancements, and constrained automation.
    \item[1)] \textbf{CoExplain}, a neurosymbolic framework and interactive XAI tool that supports read, write, and enhance rule-based explanation, enabling bi-directional human-AI alignment.
\end{enumerate}}{\begin{enumerate}[leftmargin=*, noitemsep, label=\arabic*)]
    \item Insights on how users edit explanatory rules, both independently and collaboratively with AI, highlighting needs for \textbf{Editable XAI}, AI-guided enhancements, and constrained automation.
    \item \textbf{CoExplain}, a neurosymbolic framework and interactive XAI tool that supports read, write, and enhance rule-based explanation, enabling bi-directional human-AI alignment.
\end{enumerate}}

Our evaluations showed that editable explanations, CoExplain in particular, foster more effective and efficient human–AI alignment than static, read-only explanations.
We discuss the scope, limitations, and generalization of Editable XAI, and outline future directions toward effective human–AI collaboration and alignment.

\section{Related Work}
We discuss related work on 
how explainable AI is mostly used for reading only, 
interactive machine learning methods for human-AI alignment, 
neurosymbolic methods to interface semantic rules with expressive AI models
and end-user development to increase access and control of smart systems. Prior works in these areas are mostly one directional in solving the misalignment problem, while Editable XAI approach it in a bi-directional way by deepening both the user's understanding of AI and the AI's alignment with user. 

\subsection{Usage of Explainable Artificial Intelligence}
Explainable AI (XAI) has been a key research area to increase transparency in machine learning systems. Most existing approaches focus on generating post-hoc justifications for model outputs, such as saliency maps~\cite{selvaraju2017grad,adebayo2018sanity,alqaraawi2020evaluating,hong2015online} or gradient-based attributions~\cite{sundararajan2017axiomatic, ancona2019gradient, ancona2017towards}, primarily targeting technical audiences. These methods often leave non-expert users underserved, particularly in high-stake domains where understanding both the structure and reasoning of AI systems is crucial~\cite{laato2022explain, zhang2022towards}. Such post-hoc explanations are \textbf{one-way communications} that show users what the model did, rather than inviting them to question or modify. Even seemingly user-friendly formats like natural language rationalization~\cite{ehsan2018rationalization} or example-based explanations~\cite{cheng2019explaining, keane2020good, bien2011prototype, kim2014bayesian} limit interactivity and rarely expose the underlying logical structure of the model. 

Recent work emphasizes the importance of interactive explanations, where users are not just recipients of reasoning but active participants in shaping it~\cite{sokol2020one, lee2021human}. By allowing users to modify explanations, systems can support deeper understanding~\cite{lakkaraju2019faithful, guo2022building}, however, these works treat user input as optional or post-hoc, without letting user input affect the AI's reasoning process. On the other hand, \textbf{Editable XAI} includes users as \textbf{active collaborators} with the AI system, rather than as passive receivers of the static explanations, treating user input as a core part of model reasoning and refinement. By allowing users to \textbf{directly inspect and modify} the AI's reasoning with decision tree explanations~\cite{chou2019explainable, blanco2019machine}, \textbf{CoExplain} creates a \textbf{bi-directional alignment} between the user and the AI.

\subsection{Interactive Machine Learning for Human–AI Alignment}
While XAI improves transparency, aligning AI systems with user expectations often requires \textbf{interactive interventions} rather than passive explanations. Interactive Machine Learning (IML) places users in the loop to adjust models, leveraging domain knowledge and improving alignment under scarce or noisy data~\cite{fails2003interactive, mosqueira2023human}.

Most prior IML systems focus on \textbf{data-level feedback}, such as labeling instances~\cite{bernard2017comparing, chegini2019interactive}, modifying features~\cite{wu2019local, brooks2015featureinsight}, or curating datasets~\cite{piorkowski2023aimee, berg2019ilastik}. ~\cite{sinitsin2020editable} explored using gradient to correct specific mistakes, but cannot inspect and correct the model's reasoning as a whole. Such instance-based corrections are \textbf{local}, requiring users to inspect examples individually and making global misalignment patterns difficult to detect.

In contrast, \textbf{model-level interactions}, where users directly shape a model’s structure or reasoning logic, remain underexplored~\cite{mosqueira2023human}. Recent work using rules or decision trees offers interpretable interfaces~\cite{guo2022building, daly2021user, kulesza2015principles, kulesza2013too, piorkowski2023aimee}, but these are typically post-hoc aids rather than manipulable model components. 

\textbf{Editable XAI} advances this space by treating explanations as \textbf{both interpretable and editable substrates}. Users can inspect and revise rules that govern \textbf{model-wide behavior}, creating a collaborative, continuous feedback loop. This enables \textbf{global alignment}, correcting reasoning patterns rather than isolated instances, making explanations not only clarify the model’s logic but also serve as a manipulable substrate for alignment.

\subsection{Integrating Human Knowledge in Neural Networks with Neuro-Symbolic Learning}
\textbf{Neural-symbolic learning} bridges human reasoning and statistical learning by integrating symbolic rules into neural models~\cite{d2009neural, garcez2015neural}. These hybrid approaches enhance interpretability and allow networks to incorporate logical constraints, aligning model behavior more closely with human expectations.

Prior work typically embeds rules through \textbf{formal logic encodings}~\cite{zhang2021neural, crouse2021deep, riegel2020logical, badreddine2022logic} or \textbf{constraint-based objectives}, such as differentiable logic or regularization~\cite{hu2016harnessing, li2019augmenting}. While effective, these methods demand technical expertise to specify rules and integrate them into training pipelines, limiting access for domain experts or end-users whose knowledge is critical for meaningful alignment.

This gap highlights the need for \textbf{user-accessible} neuro-symbolic methods that allow \textbf{intuitive knowledge input} and \textbf{iterative refinement}. By enabling symbolic rules to be directly editable and interpretable, \textbf{Editable XAI} supports collaborative human–AI reasoning, allowing users to guide model behavior without requiring technical expertise. This shifts Neural-symbolic methods from technically encoded rules to \textbf{human-centered, editable knowledge}.

\subsection{End-User Development for Editable Systems}
% Many real-world applications involve \textbf{non-technical users} who possess domain knowledge but lack tools to integrate it into AI systems. \textbf{End-User Development (EUD)} extends interactive approaches to these users, enabling them to modify system behavior without programming expertise~\cite{lieberman2006end,fischer2012end}. Rule-based interfaces are common in EUD as they offer a \textbf{transparent representation of user intent}~\cite{dietvorst2018overcoming, barricelli2019end}. Prior works demonstrates EUD in personalizing smart environments, IoT behaviors, and adaptive systems~\cite{houben2016physikit, ghiani2017personalization, guo2011toward}.
Many real-world applications involve \textbf{non-technical users} who possess domain knowledge but lack effective means to integrate it into AI systems. \textbf{End-User Development (EUD)} addresses this gap by enabling users to modify system behavior without programming expertise~\cite{lieberman2006end,fischer2012end}. Rule-based interfaces are common in EUD as they provide a \textbf{transparent representation of user intent}~\cite{dietvorst2018overcoming,barricelli2019end}. Prior work demonstrates EUD applications in personalizing smart environments, IoT behaviors, and adaptive systems~\cite{houben2016physikit,ghiani2017personalization,guo2011toward}.

However, these efforts generally focus on \textbf{simple if–then rules}, which limits their applicability to complex decision structures or integration with AI reasoning. Extending EUD to support \textbf{richer rule authoring and editing} would allow end-users to contribute domain knowledge in a way that directly influences model behavior, while maintaining accessibility and interpretability. This perspective positions EUD not just as a personalization tool, but as a potential bridge for \textbf{human-guided, editable AI systems}.

\section{Elicitation User Study}
\label{sec:elicitation}
Despite advancements in tools for rule editing, end-user programming, and interactive machine learning~~\cite{daly2021user,ghiani2017personalization,guo2022building}, a significant gap exists in the support for editable AI explanations. 
To address this, we developed a basic rule editor tool informed by Tree Edit Distance (TED, Zhang and Shasha ~\cite{zhang1989simple}) to include tree editing operations for inserting, deleting, and updating rules. 
We then used this tool as a \textit{design probe} in a user study to elicit user needs for automatic and AI-supported editing of XAI explanations.

\subsection{Method}

\subsubsection{Probe apparatus and user task} 
This study was conducted using a basic editing tool we developed with an attribute list where users can select the attributes to use and a canvas where they can edit their rule with drag and drop as well as text input. The participants were asked to use this tool to create and edit a rule that they can understand and agree with, while trying to increase the rule's accuracy.

\subsubsection{Participants} 
% shared with Manual Edit condition in Evaluation Section.
We recruited 14 participants from a local university for the elicitation study. It was conducted over Zoom for 60 minutes. Participants were compensated \$16 USD in local currency.
They were genders 10 females and 4 males, ages 20 to 29 years old (M = 22.3, SD = 2.3) from 10 different discipline majors (3 in Computer Science and 3 in Business, and 1 for each of Chemistry, Medicine, Political Science).

\subsubsection{Study procedure} 
The overall study consisted of two parts: i) editing stage where users were asked to create and edit a rule on their own to understand their interaction behavior using rules, ii) forward simulation stage to estimate their understanding of an AI model trained using their rules. In this section we focus only on the editing stage while the results from the forward simulation were collected and analysed in Section~\ref{sec:evaluation}. The participants were shown 3 scenarios in random order: Income Prediction, House Price Prediction and Heart Disease Prediction. They built the rule and did the forward simulation for each of the scenarios.

During the editing stage, for each of the scenarios, they were given a guideline rule as background knowledge of the scenario, we used the same dataset and preparation procedure as described in Section~\ref{sec:evaluation}. An accuracy indicator was available to them while they perform the editing. The participants were asked to think aloud to articulate their rationale behind their editing and to describe any difficulties they have encountered as well as their need of assistance.

\subsection{Findings}
All 14 participants successfully created and edited their rules for all 3 scenarios, we conducted a thematic analysis on participant behaviors and report key findings.

% \subsubsection{Human-understandable rules are preferred over purely accuracy-driven rules}
\subsubsection{Preference for human-written rules over AI learned rules}
Although the guideline rules achieved higher accuracy, most participants preferred to create and edit rules that reflected their own logic and domain knowledge. They valued explanations that “made sense” to them over rules that merely optimized prediction accuracy.

For instance, E5 reported that she \textit{“actually just followed the guideline rule because I think it makes sense,”} but clarified that this was only because the rule’s logic aligned with her reasoning: \textit{“the size of the living area bigger than [a threshold] means higher value, and these also correspond with the grade. If it's older but in good condition, then it's high [price].”} In contrast, many others deliberately diverged from the guideline rules. E2 emphasized that logical consistency mattered more than accuracy, drawing on her own real-life knowledge: \textit{“I started off with a high grade. And then I thought, what people want is a bigger living room… even if the living room is smaller, and the grade is smaller, more number of bedrooms can give high price.”} She further explained, \textit{“the more I focused on accuracy, like, the less I used logic. If I only focused on that, I might, like, come up with illogical choices.”} Similarly, E3 rejected the guideline rules outright, noting, \textit{“I'm following my thoughts. Because I think looking at the guideline rules, to me, it didn't really… make sense.”}

Together, these accounts highlight that participants wanted the freedom to use rules that reflected their reasoning, even if these rules were less accurate. The ability to create and edit rules empowered them to maintain ownership of the decision logic, rather than feeling forced to adopt explanations they disagreed with.

% \paragraph{b) Participants iterated on edits but struggled to achieve high accuracy.}
\subsubsection{Capped performance despite iterative edits}
During the forward simulation stage, participants were allowed to revise their rules after seeing real instances and labels, and we logged their interactions for analysis. Most participants engaged in multiple rounds of editing, gradually revising their rules in search of improvement. For example, E2 described a step-by-step process of modifying thresholds and adding features until reaching a final model: \textit{“I changed these 3 to… this from 4 to 3, and this from 3 to 2. And then I added this. The age, because I felt like high and low weren't doing, like, enough.”} Similarly, E9 explained that he \textit{“modified [the framework] based on the provided real labels because I noticed factors like house age and living area were not initially considered. I gradually revised the branches to improve it.”}
Despite such efforts, participants often struggled to achieve satisfactory accuracy. Iterative editing was described as tedious, and some resorted to drastic changes after several rounds of refinement. For instance, E8 abandoned her original rules after four rounds, stating that she \textit{“wants to try out a different set of rule[s].”} Likewise, E7 and E14 restructured their rules late in the process when incremental edits failed to improve outcomes. These behaviors suggest that while iterative refinement is a natural strategy, it is insufficient without system support. Users require not only editing flexibility, but also suggestions and feedback tailored to their rules to sustain productive refinement.

\subsubsection{Difficulty in determining threshold values and need for external advice.}
One of the most frequently requested forms of assistance was help with selecting appropriate threshold values for attributes. Participants often experimented with different cutoffs around an initial guess to see if accuracy improved. They sought guidance on ranges that were logical or grounded in domain knowledge, while still wanting to preserve the topology of their own rules. For instance, E1 remarked, \textit{“I need to see [from the guideline], what range of values are logical,”} emphasizing the difficulty of choosing cutoffs without reference points. E7 similarly questioned what particular thresholds to be used, noting, \textit{“I heard that you don't need to have a degree in order to get a high-paying job, but I don't know what’s the cutoff for the age also.”} Likewise, E11 expressed a desire for external assistance, suggesting:\textit{“some help can tell you, for example, in the heart case, what is the normal resting [blood pressure] rate.”}

These reflections underscore a recurring challenge: participants valued ownership of their rules but lacked confidence in setting precise numerical thresholds. This points to a clear design implication: the systems should provide threshold suggestions around the users' initial guess, while being contextualized by domain knowledge or data distributions, preserving the topology defined by users.

\subsubsection{Need for reorganization suggestions while preserving some priors.}
Participants also recognized that their rules could benefit from topological improvements. Rather than discarding their work, they expressed interest in revisions: refining specific branches or adding detail while preserving the core of their original creation. For example, E3 reflected on his rule as \textit{“I think it’s incomplete.”}, and would like to see where and how can improvements be made. Similarly, E5 struggled with how to make certain improvements, admitting, \textit{“I’m not so sure about this [heart disease rule], I don’t know exactly where to improve,”} and suggested that \textit{“maybe the second resting blood pressure [needs improvement], but I don’t know how.”} E10 desired greater granularity on his rule, noting, \textit{“I think my criteria could be more detailed, like maybe having more branches.”}

These comments suggest that while participants valued the interpretability of their own topology, they lacked confidence in knowing where or how to revise them. This highlights the need for system support that can propose structural refinements, while maintaining alignment with users’ original rule.

\subsection{Design Guidelines}
%  focused on rules
% reading (default assumption), editing, enhancement (thresholds, reorganize), constraints
Based on these findings, we propose design guidelines for \textbf{Editable XAI} to facilitate human-AI alignment:
\aptLtoX{\begin{enumerate}
    \item[1)] \textbf{Contextualize explanations in a writable format} such as rules so that users can contribute to adjusting the explanations rather than just read them.
    \item[2)] \textbf{Support user-written rules} by explaining in terms of rules established in the domain, or defined by users as they need, instead of purely learned from data. 
    \item[3)] \textbf{Provide automated enhancements of user-written rules} to alleviate knowledge deficits of users, and augment their reasoning.
    The system can recommend adjustments to rule thresholds (parameters), or even reorganize the rule topology (decision tree structure) to significantly improve the explanation accuracy.
    \item[4)] \textbf{Constrain enhancements} to preserve alignment with user-driven rules and limit explanation complexity.
\end{enumerate}}{\begin{enumerate}[leftmargin=*, noitemsep, label=\arabic*)]
    \item \textbf{Contextualize explanations in a writable format} such as rules so that users can contribute to adjusting the explanations rather than just read them.
    \item \textbf{Support user-written rules} by explaining in terms of rules established in the domain, or defined by users as they need, instead of purely learned from data. 
    \item \textbf{Provide automated enhancements of user-written rules} to alleviate knowledge deficits of users, and augment their reasoning.
    The system can recommend adjustments to rule thresholds (parameters), or even reorganize the rule topology (decision tree structure) to significantly improve the explanation accuracy.
    \item \textbf{Constrain enhancements} to preserve alignment with user-driven rules and limit explanation complexity.
\end{enumerate}}

\section{Technical Approach}
% To achieve bi-directional collaboration between humans and AI, we draw insights from cognitive science and neural-symbolic learning by using decision trees as an interaction medium. However, unlike prior work that enabled users to customize straightforward glass-box models, our challenge is to connect the easy-to-understand and editable decision tree model with the hard-to-interpret and hard-to-modify neural network. 
% This section covers the 3 interaction modes offered by \textbf{CoExplain} (Fig. \ref{fig:technical-overview}) as \textbf{Read}, \textbf{Write} and \textbf{Enhance}. We first introduce how we extract decision tree explanations from neural networks through distillation to support user \textbf{Read}, transform user-created decision tree rules into neural network through parsing for user \textbf{Write}. Then we break down \textbf{Enhance} into two features as threshold refinement and structural reorganization following the design guidelines from Section~\ref{sec:elicitation}, and introduce their details respectively.

\begin{figure}[t]
    \centering
    \includegraphics[width=8.0cm]{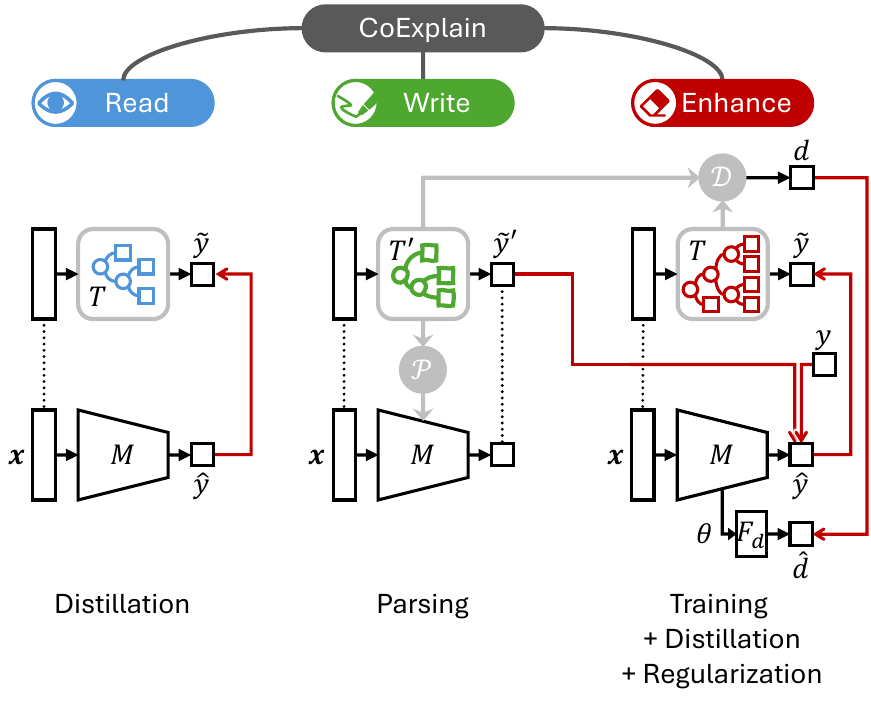}
    % \vspace{-0.3cm}
    \caption{Overview of CoExplain’s three interaction modes and their underlying mechanisms. \textbf{Read}: explanations are distilled from the neural network $M$ into a decision tree $T$, using the input $x$ and network prediction $\hat{y}$ to make sure tree prediction $\tilde{y}$ aligns with $M$. \textbf{Write}: user-authored rules $T'$ are transformed into a neural network $M$ through a parser $\mathcal{P}$. \textbf{Enhance}: user collaborates with AI edits on the thresholds and topology. AI refines the rules through training with regularization, aligning predictions $\hat{y}$ with both user-defined rules' prediction $\tilde{y}'$ and data-driven adjustments with $y$, the topology of the explanation $T$ is aligned with $T'$ using a proxy model $F_d$ mapping the network parameters $\theta$ to their Tree Edit Distance $d$ calculated by $\mathcal{D}$.}
    \label{fig:technical-overview}
    \Description[Technical overview of CoExplain’s three interaction modes]{Technical overview with three sections: Read (Distillation), Write (Parsing), and Enhance (Training + Distillation + Regularization). In Read, input x flows to neural network M (producing prediction y hat) and decision tree T (producing prediction y tilde), with a red arrow aligning y tilde and y hat. In Write, input x flows to M and a user-authored tree T prime, with a gray arrow from T prime through parser P to M. In Enhance, input x flows to M (y hat) and tree T (y tilde). Red arrows connect y hat to ground truth y, y tilde, and d hat from function F sub d (using M's parameters theta). Component D calculates tree edit distance d between T and T prime, with an arrow from d to d hat. All sections show data flows to align artificial intelligence models with user created rules across the three interaction modes.}
    % \vspace{-0.5cm}
\end{figure}

To satisfy the design guidelines, we introduce \textbf{CoExplain} for collaborative, editable explainable AI (XAI).
We model predictions using neural networks to leverage its representational power as universal function approximators~\cite{scarselli1998universal}, and gradient descent methods for model training.
We explain the model using decision trees to leverage its intuitive representation of semantic relationships~\cite{luvstrek2016makes}.
As illustrated in Fig. \ref{fig:technical-overview}, 
CoExplain enables users to 
\textit{read} explanations through \textbf{distillation} (Section~\ref{subSec:distillation}), 
\textit{write} explanations through \textbf{parsing} (Section~\ref{subSec:parsing}), and
the AI can \textit{enhance} explanations through \textbf{training} with backpropagation and regularization (Section~\ref{subSec:training}).

% \subsection{Decision Tree Rules Explanation Distillation}
\subsection{Distillation: Neural Network Model to Decision Tree Rule Explanation}
\label{subSec:distillation}
% \bad{Neural networks are widely applied across domains, but they remain difficult to interpret. The central challenge is to transform a neural network into an interpretable decision tree without losing faithfulness to the original model. Motivated by prior work on distilling trained neural networks into decision trees for human understanding ~\cite{craven1995extracting, nguyen2020towards}, we adopt a similar approach, with the additional step of checking candidate trees against the user’s original tree to maintain structural similarity.}

% - model agnostic explanations popular to explain black box AI
% - distillation to condense models
% - but can also be used to estimate proxy models (e.g., \todo{LIME~\cite{ribeiro}} does this locally)

Model agnostic explanations, such as LIME~\cite{ribeiro2016should}, are commonly used to explain black box AI models like neural networks.
These can be done for local explanations with instances near a target, or global explanations for the overall model decisions.
Generally, this approach is \textit{model distillation}, most commonly used to condense large models (e.g., student-teacher models~\cite{wang2021knowledge, gou2021knowledge}), but can also be used for explanation since distilled models are simpler.
For this work, we focus on explaining neural networks with a global decision tree using distillation~\cite{craven1995extracting, nguyen2020towards, schmitz1999ann, krishnan1999extracting, boz2002extracting}.

Consider the AI model (neural network) to be explained $M$, and explainer XAI model (decision tree) $T$. 
Predictor model $M$ is trained on instances $\bm{x}$ with ground truth label $y$, and predicts label $\hat{y}$, i.e., $\hat{y} = M(\bm{x})$.
Instead of training explainer model $T$ on ground truth labels, we train it on the prediction label $\hat{y}$ to predict explanation label $\tilde{y}$, i.e.,
$\tilde{y} = T(\bm{x})$.
We performed hyperparameter tuning of the tree depth to obtain the most faithful explainer model $T$ to predictor model $M$.
%
% Since the models are not the same type, they do not have the same training algorithms.
The neural network $M$ is trained via backpropagation, while the decision tree $T$ is trained via CART~\cite{loh2011classification}.

\subsection{Parsing: Decision Tree Rule Explanation to Neural Network Model}
\label{subSec:parsing}

% Decision trees provide an interpretable, structured representation of decision logic. Their flowchart-like structure makes human knowledge easy to inspect, modify, and validate. However, their fixed and discrete nature limits their expressive capacity, especially for modeling complex, non-linear decision boundaries. Neural networks, by contrast, serve as universal function approximators with much greater representational power ~\cite{scarselli1998universal}, but they often suffer from opacity. 

% Our goal is to bridge these paradigms. We map a user-authored decision tree into a neural network such that the network, at initialization, \textit{aligns with the tree’s predictive behavior and structure} while retaining the ability to learn and generalize. This approach builds on prior work in tree-to-network structural initialization ~\cite{humbird2018deep} and aligning neural networks with symbolic logic ~\cite{li2019augmenting}, with a focus on preserving both the \textbf{structural hierarchy} and \textbf{decision behavior} of the original tree.

% Decision trees provide an interpretable, structured representation of decision logic. Their flowchart-like design makes it easy for humans to inspect, modify, and validate decisions. However, their discrete nature limits expressiveness, particularly for modeling complex, non-linear relationships. Neural networks, in contrast, offer greater representational power as universal function approximators~\cite{scarselli1998universal}, but at the cost of transparency.

While the distillation focuses on explaining a pre-existing neural network, parsing starts with a pre-existing decision tree written by a user.
Moreover, while distillation is model agnostic and approximate, parsing is heuristic and can \textit{exactly convert} a tree into a neural network, exploiting the property that both are graph structures.

% Our goal is to bridge these paradigms by mapping a user-authored decision tree into a neural network. The network is initialized to \textit{preserve the tree’s predictive behavior and hierarchical structure} while remaining trainable for generalization. This approach builds on prior work in tree-to-network initialization~\cite{humbird2018deep} and neural-symbolic alignment~\cite{li2019augmenting}, emphasizing the preservation of both the \textbf{structural hierarchy} and \textbf{decision behavior} of the original tree.

We build on prior work in tree-to-network initialization~\cite{humbird2018deep} and neural-symbolic alignment~\cite{li2019augmenting} to parse a user-defined decision tree into a topologically-equivalent neural network.
Note that while both trees and neural networks have graph structures, their aspects do not represent the same concepts. 
Table~\ref{tab:comparison_dt_nn} compares key aspects between these models.
Next, we describe how to parse aspects of the decision tree (decision nodes, trace conjoined branches, disjoint traces), and how to ensure extensibility of the tree-based neural network.
See Fig.~\ref{fig:parsing} for an illustrated example of parsing.

\aptLtoX{\begin{table}
    \caption{Comparison of graph structures between a decision tree and its topologically equivalent neural network.}\label{tab:comparison_dt_nn}
    \begin{tabular}{lll}
        \toprule
        Aspect & Decision Tree (DT) & Neural Network (NN) \\
        \midrule
        Node & 
        \textbf{Decision Node} representing a test on a single attribute against a threshold. \textbf{Leaf Node:} Represents the predicted label from the preceding branch outcome (e.g., $x_1 > 5$). & 
        \textbf{Neuron} representing a unit that receives weighted inputs and activates an output. \\
        Edge & 
        \textbf{Branch} representing the binary prediction outcome, connecting to the next decision test. & 
        \textbf{Connection} representing the weight of the prior input (neuron). \\
        Path & 
        \textbf{Trace} through multiple decision nodes represents the conjunction (AND) of tests. 
        If multiple traces have the same predicted label in their leaf nodes, then the traces can be combined as a disjunction (OR). & 
        \textbf{Propagation} of multiple computed operations from input to output. \\
        Level & 
        \textbf{Depth} of a node corresponds to the number of attribute tests required to reach it. & 
        \textbf{Layer:} For the parsed NN, the first layer represents straight (linear) decision boundaries, and subsequent layers are weighted, piecewise combinations of these boundaries. General NNs learn nonlinear decision boundaries or more separable feature spaces. \\
        Computation & \textbf{Symbolic:} Each node performs a symbolic test (e.g., $x_2 > 8$). & \textbf{Continuous:} Each neuron performs a nonlinear operation (weighted sum and activation function). \\
        \bottomrule
    \end{tabular}
\end{table}}{\begin{table*}[t!]
    \centering
    \caption{Comparison of graph structures between a decision tree and its topologically equivalent neural network.}
    \label{tab:comparison_dt_nn}
    \begin{tabularx}{\textwidth}{l >{\raggedright\arraybackslash}X >{\raggedright\arraybackslash}X}
        \toprule
        Aspect & Decision Tree (DT) & Neural Network (NN) \\
        \midrule
        Node & 
        \textbf{Decision Node} representing a test on a single attribute against a threshold. \newline \textbf{Leaf Node:} Represents the predicted label from the preceding branch outcome (e.g., $x_1 > 5$). & 
        \textbf{Neuron} representing a unit that receives weighted inputs and activates an output. \\
        \addlinespace
        Edge & 
        \textbf{Branch} representing the binary prediction outcome, connecting to the next decision test. & 
        \textbf{Connection} representing the weight of the prior input (neuron). \\
        \addlinespace
        Path & 
        \textbf{Trace} through multiple decision nodes represents the conjunction (AND) of tests. 
        If multiple traces have the same predicted label in their leaf nodes, then the traces can be combined as a disjunction (OR). & 
        \textbf{Propagation} of multiple computed operations from input to output. \\
        \addlinespace
        Level & 
        \textbf{Depth} of a node corresponds to the number of attribute tests required to reach it. & 
        \textbf{Layer:} For the parsed NN, the first layer represents straight (linear) decision boundaries, and subsequent layers are weighted, piecewise combinations of these boundaries. \newline General NNs learn nonlinear decision boundaries or more separable feature spaces. \\
        \addlinespace
        Computation & \textbf{Symbolic:} Each node performs a symbolic test (e.g., $x_2 > 8$). & \textbf{Continuous:} Each neuron performs a nonlinear operation (weighted sum and activation function). \\
        \bottomrule
    \end{tabularx}
\end{table*}}

\subsubsection{Decision Nodes and Consequent Branches}
For a binary decision tree, each decision node results in a True (T) or False (F) outcome. We model these in the first layer of the neural network, where
each outcome is associated with a neuron, and each decision node has a pair of neurons. So, $k$ decision nodes results in a first NN layer with $2k$ neurons.

From Fig. \ref{fig:parsing}, consider the blue decision node with test $x_1 \ge \tau_1$. The test is True when $x_1 - \tau_1 \ge 0$ and False when $-(x_1 - \tau_1) > 0$.
These results can be represented in neurons as $a_{11} = \mathbb{I}(x_1 - \tau_1) = \mathbb{I}((-1, +1) \cdot (\tau_1, x_1))$ and $a_{12} = \mathbb{I}(-(x_1 - \tau_1)) = \mathbb{I}((+1, -1) \cdot (\tau_1, x_1))$, respectively; where $\mathbb{I}(P)$ is the indicator function which is 1 when $P$ is true and 0 otherwise.
Similarly, the green decision node $x_2 \ge \tau_2$ is represented by neurons $a_{13} = \mathbb{I}((-1, +1) \cdot (\tau_1, x_1))$ and $a_{14} = \mathbb{I}((+1, -1) \cdot (\tau_1, x_1))$.
Represented in parallel, in NN layer 1 and subsituting $\mathbb{I}$ with a differentiable smooth sigmoid activation function ($\sigma$), both decision nodes are:
\begin{equation}
    \begin{split}
    \bm{a}_1 &= \sigma \left( \bm{W}_1^\top \bm{x} + \bm{b}_1 \right),\\
    \begin{pmatrix} a_{11}\\a_{12}\\a_{13}\\a_{14} \end{pmatrix} &= \sigma \left( \begin{pmatrix} +1 & \gz \\ -1 & \gz \\ \gz & +1\\ \gz & -1 \end{pmatrix} \begin{pmatrix} x_{1}\\x_{2} \end{pmatrix} + \begin{pmatrix} -\tau_1\\+\tau_1\\-\tau_2\\+\tau_2 \end{pmatrix} \right),
    \end{split}
    \label{eq:parse_decision_node}
\end{equation}
with input $\bm{x}$, weights $\bm{W}_1$, and biases $\bm{b}_1$.
% where input $\bm{x} = (x_1, x_2)^\top$, weights $\bm{W}_1 = (\begin{smallmatrix} +1 & -1 & 0 & 0 \\ 0 & 0 & +1 & -1 \end{smallmatrix})
% $, and biases $\bm{b}_1 = (-\tau_1, +\tau_1, -\tau_2, +\tau_2)^\top$.

\subsubsection{Conjunction along a Trace}
\begin{figure}[t]
    \centering
    \includegraphics[width=8.5cm]{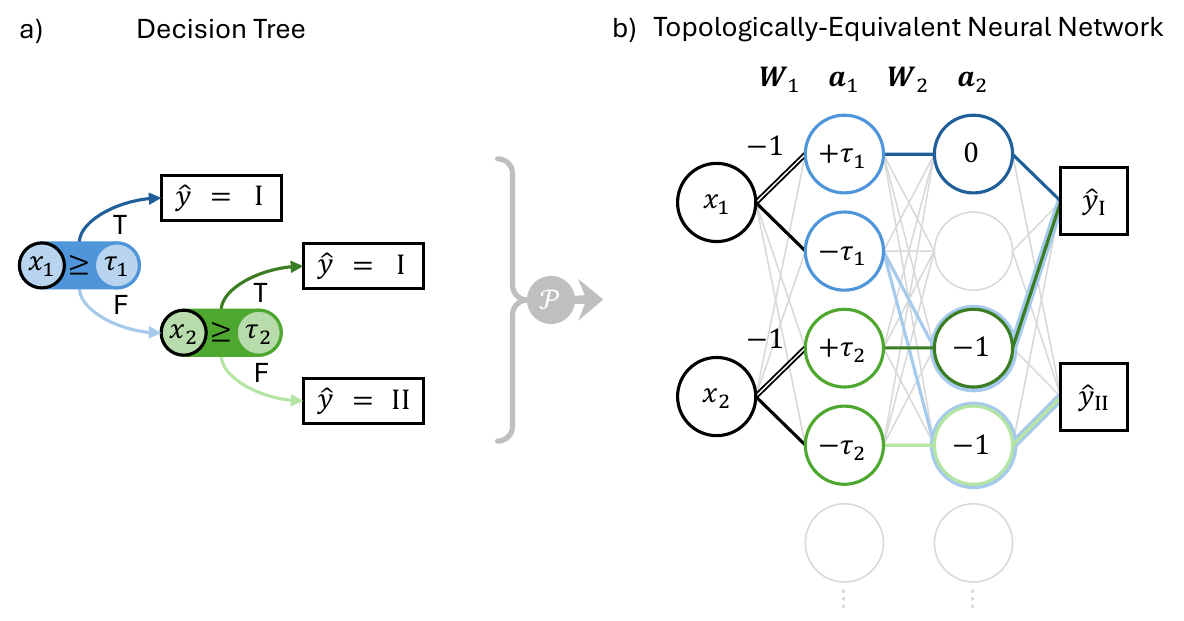}
    % \vspace{-0.7cm}
    \caption{Transforming a decision tree into a topologically equivalent neural network via a parser $\mathcal{P}$.
a) A decision tree with two internal nodes testing feature thresholds.
b) The corresponding neural network, where each tree node is encoded by a pair of first-layer neurons with biases $\pm \tau_i$ representing the threshold test. Subsequent layers mirror the tree’s decision paths: connections are preserved with unit weights, and neuron biases encode the logical relationships. Colored connections trace decision paths, the output layer aggregates signals to reproduce the tree’s leaf predictions.}
    \label{fig:parsing}
    \Description[Transformation of a decision tree into a neural network via parser P]{Diagram with two parts: (a) Decision Tree: Two internal nodes test feature thresholds. The first node checks if feature x one is greater than or equal to threshold tau one. If true, the prediction is y hat equals one. If false, a second node checks if feature x two is greater than or equal to threshold tau two; if true, prediction is y hat equals one, and if false, prediction is y hat equals two. (b) Topologically-Equivalent Neural Network: Inputs x one and x two connect to neurons in the first hidden layer. These neurons have biases plus tau one (blue) and minus tau one, plus tau two (green) and minus tau two. Later layers match the decision paths of the tree, with connections and biases set to match logical relationships. Colored connections show decision paths, and the output layer (y hat one and y hat two) reproduces the tree’s leaf predictions. A gray arrow labeled P shows the transformation from the tree to the network.}
    % \vspace{-0.5cm}
\end{figure}

Although represented together in $\bm{a}_1$ (Eq. \ref{eq:parse_decision_node}), each branch of each decision node is treated separately.
To represent traversing down the tree through multiple decision nodes, we need to encode logical \textit{conjunction} (AND $\land$).
We do so with {\L}ukasiewicz conjunction to handle continuous values:
\begin{equation}
    P \land Q = \max(0, p + q - 1),
\end{equation}
where $p = 1$ when $P$ is True and 0 otherwise, and same for $q$ with $Q$.
We encode this operation in the second NN layer by representing $p + q - 1$ in a computational graph with ReLU activation function since $\text{ReLU}(x) = \max(0,x)$.
As an example, the conjunction $(x_1 \ge \tau_1) \land (x_2 \ge \tau_2)$ in the second depth of the decision tree in Fig.~\ref{fig:parsing} is represented by neuron $a_{23} = \text{ReLU}(a_{12} + a_{13} - 1)$.
All decision branches are represented in the NN layer 2 as:
\begin{equation}
    \begin{split}
    \bm{a}_2 &= \text{ReLU} \left( \bm{W}_2^\top \bm{a}_1 + \bm{b}_2 \right),\\
    \begin{pmatrix} a_{21}\\a_{23}\\a_{24} \end{pmatrix} &= \text{ReLU} \left( \begin{pmatrix} 1 & \gz & \gz & \gz \\ \gz & 1 & 1 & \gz \\ \gz & 1 & \gz & 1 \end{pmatrix} \begin{pmatrix} a_{11}\\a_{12}\\a_{13}\\a_{14} \end{pmatrix} + \begin{pmatrix} \gz\\-1\\-1 \end{pmatrix} \right),
    \end{split}
    \label{eq:parse_decision_node}
\end{equation}
with input $\bm{a}_1$, weights $\bm{W}_2$, and biases $\bm{b}_2$.

For DT traces that terminate at the first or earlier tree depths, the NN propagations \textit{pass through} the layer without modification (i.e., weights = 1 and biases = 0).
For example, decision branch $x_1 \ge \tau_1 \models \hat{y}$ has the NN propagation $\hat{y}_{\text{I}} = 1 \cdot a_{11} + 0$.

\subsubsection{Disjunction of Traces with Same Leaf Node Predictions}
We represent multiclass classification using one-hot encoding, where each neuron in the NN output layer is a separate class.
When multiple traces in the decision tree share the same predictions in their leaf nodes, this indicates alternative antecedents for the same consequent, which is a logical \textit{disjunction} (OR $\lor$).
We encode this numerically with {\L}ukasiewicz disjunction:
\begin{equation}
    P_1 \lor P_2 \lor \dots = \min(1, \sum_{t \in T_c} p_{t}),
\end{equation}
where $t \in T_c$ is the $t$th trace that predicts class $c$, and $p_t = 1$ when $P$ is True and 0 otherwise.
We encode this in the output layer by summing relevant activations from the penultimate layer with Clipped ReLU (cReLU) activation function: $\text{cReLU}(x) = \min(1,x)$ for clip at 1 and $0 < x \le 1$.
All decision traces are represented in the output layer as:
\begin{equation}
    \begin{split}
    \hat{\bm{y}} &= \text{cReLU} \left( \bm{W}_y^\top \bm{a}_2 + \bm{b}_y \right),\\
    \begin{pmatrix} \hat{y}_\text{I}\\\hat{y}_\text{II} \end{pmatrix} &= \text{cReLU} \left( \begin{pmatrix} 1 & 1 & \gz \\ \gz & \gz & 1 \end{pmatrix} \begin{pmatrix} a_{21}\\a_{23}\\a_{24} \end{pmatrix} + \begin{pmatrix} \gz\\\gz \end{pmatrix} \right),
    \end{split}
    \label{eq:parse_decision_node}
\end{equation}
with input $\bm{a}_2$, weights $\bm{W}_y$, and biases $\bm{b}_y$.

\subsubsection{Model Capacity and Zero Padding}
The aforementioned methods describe how to set parameter values (weights and biases) in the neural network, but do not specify the model capacity needed to represent the parsed tree or future extensions.
At minimum, 
i) the number of neurons in each NN layer is twice the number of DT decision nodes, and
ii) the number of NN layers is determined by the depth of the DT.
The number of neurons in NN output layer is the number of prediction classes.

However, to support extensibility whereby the model can be retrained to be more expressive and accurate, we need to support more relationships and decision boundaries.
We do so by \textit{zero padding} with more neurons per layer (weights = 0, bias = 0), and more layers with \textit{pass through}.
Section~\ref{subSec:training} describes how training to update these additional parameters can optimize the model further.

\subsection{Training: Optimizing Model for Performance and Alignment}
\label{subSec:training}

User-created decision trees provide an interpretable starting point, but their predictive performance is often limited by the greedy\footnote{Due to CART-based learning with information gain.}, local optimization used during construction. 
To manage user sense of control and moderate edits, CoExplain supports conservative and disruptive update methods: Threshold Enhancement and Topology Enhancement.
These change the underlying neural network model slightly or by a lot and explain with a new decision tree as updated.
Threshold Enhancement will retain the same decision tree structure (topology), but may change the threshold value (e.g., $x_1 \ge 5$ may become $x_1 \ge 5.5$).
Topology Enhancement will learn complex, nonlinear, multivariate relationships, resulting in a change in topology, such as adding and rearranging decision nodes, and using additional attributes.
To limit the changes, constraints can be applied.
% With the bidirectional mapping between trees and neural networks, CoExplain enables these trees to be further refined through neural training. We distinguish two types of enhancement (Fig.~\ref{fig:enhance}) as a) Threshold Enhancement, and b) Topological Enhancement:  
Next, we describe how CoExplain enhance through model training, distills explanations for validation, and aligns to constraints via regularization (see Fig. \ref{fig:technical-overview} Enhancement for architecture).

\begin{figure}[t]
    \centering
    \includegraphics[width=8.0cm]{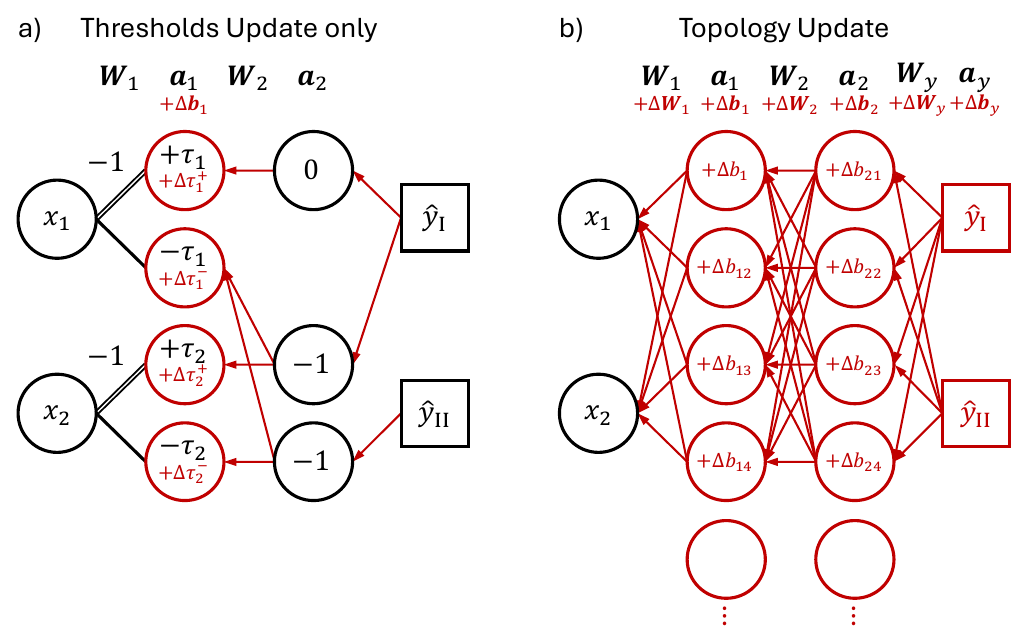}
    % \vspace{-0.3cm}
    \caption{Two types of enhancement from CoExplain, a) Threshold Update, only the threshold is updated, while keeping the topology still. b) Topology Update, both the thresholds and connections are trained, we utilize additional neurons and connections to support decision tree extension. Backpropagation and parameter update are marked as red.}
    \label{fig:enhance}
    \Description[Two enhancement types in CoExplain]{Diagram with two parts. (a) Threshold Update only: A neural network with inputs x one and x two. The first hidden layer has neurons with biases plus tau one, minus tau one (with plus delta tau one positive, plus delta tau one negative) and plus tau two, minus tau two (with plus delta tau two positive, plus delta tau two negative). Connections and later layers (with biases zero, negative one) stay black; red highlights updates to thresholds. Outputs are y hat one and y hat II. (b) Topology Update: A neural network with inputs x one and x two, multiple hidden layers (with biases plus delta b one, plus delta b twelve, plus delta b thirteen, plus delta b fourteen, plus delta b twenty one, plus delta b twenty two, plus delta b twenty three, plus delta b twenty four), and an output layer with y hat one and y hat II. All connections, neurons, and additional grayed-out neurons are red, showing updates to both thresholds and connections. Red indicates backpropagation and parameter updates.}
    % \vspace{-0.4cm}
\end{figure}

\subsubsection{Training}
We retrain the neural network in the standard approach using supervised learning and gradient descent via backpropagation on a training dataset using cross-entropy loss between the model's prediction $\hat{y}$ and ground truth $y$, i.e., $\mathcal{L}_\text{data}(\hat{y}, y)$.
Enhancements for thresholds and topology are handled differently.

\textbf{Threshold Enhancement:} 
To enhance only the DT thresholds, we freeze all parameters except for the bias terms in the NN layer 1 that contain the DT thresholds (see Fig.~\ref{fig:enhance}a).
For example, $b_{11} = +\tau_1$ can be updated with $\Delta \tau_1$.
Moreover, a user may wish to lock in an edited threshold. 
To preserve this user-defined threshold, and ensure local consistency, we freeze the bias terms of its two corresponding neurons in the first layer.
For example, locking the node $x_1 \geq \tau_1$ in Fig.~\ref{fig:parsing}a will freeze the bias terms in the top two neurons of layer $\bm{a}_1$ in Fig.\ref{fig:enhance}a.
% \toreview{To preserve certain user defined thresholds, we selectively freeze biases corresponding to the node on user specification, so that the threshold remain unchanged during training.}
% \toreview{For specific user modified thresholds that prefer the AI not to change, the corresponding bias will also be frozen to prevent changing during training.}

\textbf{Topology Enhancement:} 
To further enhance the DT toplogy, we unfreeze all layers and parameters and allow full weight and bias updates (see Fig. \ref{fig:enhance}b).
This includes updating the zero-padded neurons. 
Moreover, instead of letting CoExplain fully alter the full decision tree, the user may want only some rules to be enhanced and others restricted.
To preserve the restricted rules, we selectively freeze those rules as we freeze threshold neuron biases. However, this may be overly restrictive, leading to low accuracy.
Instead, to discourage the AI from changing restricted rules, 
we increase their weight in the Tree Edit Distance regularization, so that changes to restricted rules have a higher penalty than changes to other rules.
% we apply a training loss regularization penalty on changing their respective bias terms.

% \toreview{To preserve certain user defined subtrees, we increase the changing cost with a higher editing distance for editing operations changing the user specified subtrees, thus making the enhancement more conservative with certain changes.} 
% \todo{R1, How are threshold modifications constrained so that they affect only the intended rule, without cascading effects on other related criteria? }
% \todo{R1: Can authors do control test or some experiment to validate?}
% \todo{1AC: Also validate that threshold changes remain locally consistent or fairness of the resulting decision boundary.}
% \todo{subtree...}
% \toreview{For certain parts of the structure that the user specified and do not wish to be changed, the edit distance will be set higher than other parts when the AI proposes to change it, thus preserving user preferences.}

\subsubsection{Distillation}
To generate the decision tree explanation $T$ of the retrained model $M$, we use the same distillation method described in Section~\ref{subSec:distillation}.
However, we do this iteratively at each training batch as an intermediate check on the performance and topology of the explainer model.
% If the tree has a suboptimal outcome, the neural network will be retrained to mitigate that.

\subsubsection{Regularization}
Although standard supervised learning focuses on model performance, CoExplain also supports alignment to user-defined rules.
This is accomplished by referring to the user's written decision tree $T'$, and using self-supervised learning to ensure that the distilled tree explanation $T$ has similar behavior ($\tilde{y} \approx \tilde{y}'$) and topology design.
Fig. \ref{fig:technical-overview}
We regularize behavior with cross-entropy loss between the DT explainer's prediction $\tilde{y}$ and user DT prediction $\tilde{y}$, i.e., $\mathcal{L}_\text{behavior}(\tilde{y}, \tilde{y}')$.

To determine the difference in topology between $T$ and $T'$, we calculate the Tree Edit Distance (TED) $d$ which counts the number of edit operations (insert, delete, update) needed to change from one tree to the other. Specifically, we used the ZSS algorithm~\cite{zhang1989simple} implemented in Python.
Since this distillation to obtain $T$ from $M$ is external to the neural network, we cannot use $T$ directly for training with gradient descent.
Instead, we trained a proxy model $F_d$ that uses the parameters $\theta$ in $M$ as input to predict the TED distance $\hat{d}$.
Hence, $M$ is a multitask model to predict $(\hat{y}, \hat{d})$ and it updates its parameters to optimize for both tasks.
We regularize topology with MSE loss $\mathcal{L}_\text{topology}(\hat{d}, d)$.

In summary, to align the model to training data and user-defined rules we have multiple loss terms:
\begin{equation}
    \mathcal{L} = \mathcal{L}_\text{data}(\hat{y}, y) + \lambda_\text{b} \mathcal{L}_\text{behavior}(\tilde{y}, \tilde{y}') + \lambda_\text{t} \mathcal{L}_\text{topology}(\hat{d}, d)
    \label{eq:loss}
\end{equation}
where $\lambda_\text{b}$ and $\lambda_\text{t}$ are hyperparameters that we let users control as similarity settings.

\begin{figure*}[t]
    \centering
    \includegraphics[width=16.0cm]{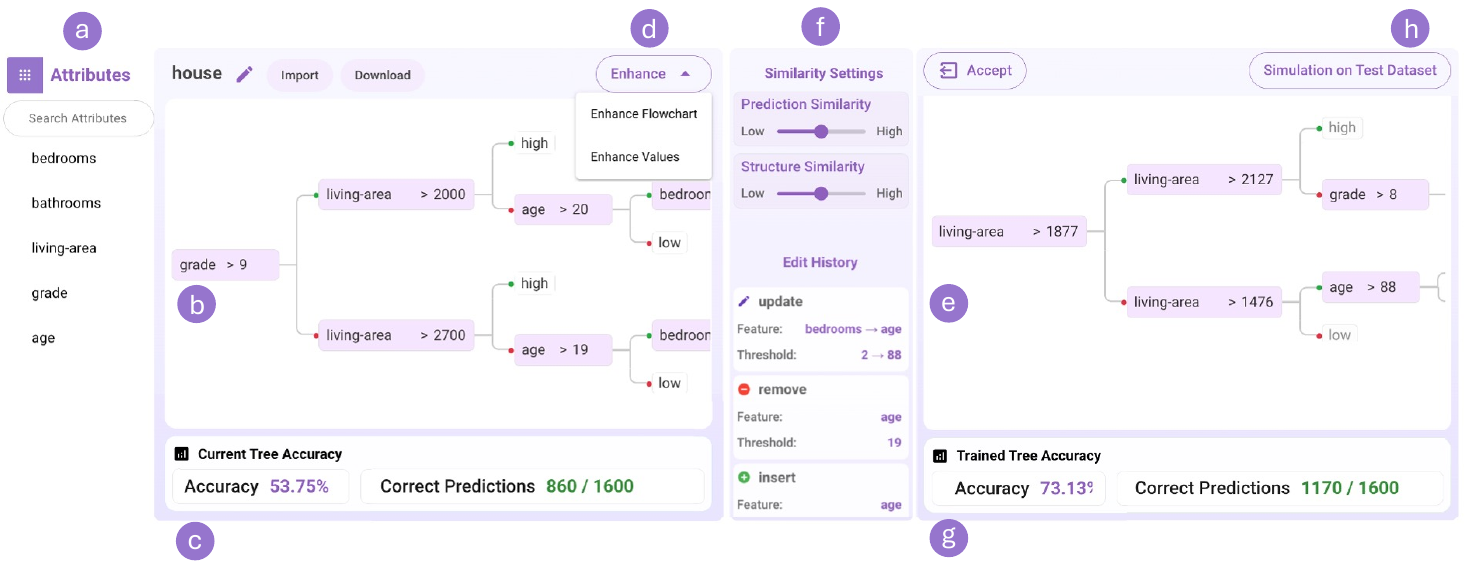}
    \caption{Interface of CoExplain. a) Data attributes. b) User’s explanation rules canvas. 
    c) User rule performance metrics. d) Enhancement actions. 
    e) AI-enhanced explanation rules canvas. f) Enhancement constraints and edit history. 
    g) Enhanced rule performance metrics. h) Simulation on test dataset.}
    \label{fig:interface}
    \Description[Layout of CoExplain’s user interface]{Interface organized into eight labeled sections across a single screen. (a) Left panel: "Data attributes" with a search bar and a list of data attributes (e.g., bedrooms, bathrooms, living area). (b) Middle-left canvas: User’s current explanation rules displayed as a decision tree with branching nodes (testing features like grade, living area) and leaf predictions (e.g., "high"/"low"). (c) Bottom-left: "Model performance metrics of the current tree" showing metrics for the user’s tree. (d) Top-middle: "Enhancement actions menu" as a dropdown with options (e.g., Enhance Flowchart, Enhance Values). (e) Middle-right canvas: AI-enhanced explanation rules displayed as a decision tree, with branching nodes (testing features like living area, grade) and leaf predictions mirroring the user’s tree structure. (f) Top-right panel: "Enhancement constraints and edit history" containing sliders (for Prediction Similarity, Structure Similarity) and a list of edit actions (e.g., update, remove, insert for features or thresholds). (g) Bottom-right: "Trained model performance metrics" showing metrics for the AI-enhanced model. (h) Top-right corner: "Simulation on test dataset" button. The layout balances user-generated and AI-enhanced content with supporting metrics and controls.}
    % \vspace{-0.2cm}
\end{figure*}

\subsection{Implementation Details}

We implemented the decision tree components using \texttt{scikit-learn}, using CART~\cite{breiman2017classification} as tree learner.  
The neural network components were implemented in \texttt{PyTorch} and optimized with Adam (learning rate $1\times10^{-2}$). All experiments were conducted on a workstation equipped with an NVIDIA H100 PCIe GPU (81,559\,MiB memory) and an AMD EPYC 9654 96-core processor. 

% sklearn for decision tree training
% pytorch for NN
% adam optimizer lr= 1e-2
% GPU NVIDIA H100 PCIe 81559MiB
% CPU AMD EPYC 9654 96-Core Processor 
% CPU max MHz:          2400.0000
% CPU min MHz:          1500.0000
% BogoMIPS:             4792.54
% tree pruned with sklearn API of ccp alpha in np.arange(0.02, 0.004, -0.001)

\section{Interface Design}
\label{sec:interface}

Fig.~\ref{fig:interface} shows the user interface for the CoExplain rule explanation editor.
The interface has a key component for reading and editing rules, complemented by panels that display dataset attributes, model performance, and AI enhancement.

% \subsubsection{Data Attributes} 
% This lists the attributes in the dataset that the user or AI can use to construct the rule explanation.

\subsection{Data Attributes}
The attribute list (Fig.~\ref{fig:interface}a) lists available dataset features. Both user and the AI use attributes from this list when editing.

% \subsection{Explanation Rules Canvas}
% - user read the AI explanation as rules formatted in a decision tree.
% - presented as a graphical flow chart, nodes represent decision antecedents, and edges are true (green top) or false (red bottom) consequents.
% - we considered an alternative format programming syntax (see Appendix...) that is commonly used in end-user development, but early pilot users preferred the flow charts.
% - we do not make the neural network explicit and do not mention this, since that is extraneous to understanding the model decisions. In the subsequent user study, no participant demanded to see this; one technical participant who did ask were surprised to learn about the implementation was a neural network, but did not need to see further.

\subsection{Explanation Rules Canvas}
The core component is the rules canvas (Fig.~\ref{fig:interface}b), where users read and edit the AI explanation as rules represented in a decision tree.  
Each rule is visualized as a graphical flowchart: nodes denote decision antecedents, while edges represent true (green, top) or false (red, bottom) consequents.  
We considered presenting the rules in programming syntax, a format often used in end-user development. However, pilot users found the flowchart representation more intuitive.  
We abstract away the underlying neural network implementation. In the subsequent user study, participants did not request such details; the one technical participant who inquired was surprised to learn about the relationship between his rules and the neural network, but expressed no need for further inspection.

% \subsection{Data and Model Components}
% These components show meta-data about the dataset to train the model, and the model performance after training.

\subsection{Model Performance Metrics}
Performance metrics are displayed below each tree (Fig.~\ref{fig:interface}c and g). These include accuracy on the training set and the number of correctly predicted cases, allowing users to compare their current tree with the AI-enhanced alternative.

% \subsection{Enhancement Aids}
% - also has a rule canvas to show rules with enhancements suggested by the AI
% - user can Accept (button) if happy with all suggestions, or selectively write parts that they agree with.

\subsection{Enhancement Aids}
To support the user with CoExplain enhancements, the interface displays a parallel AI-edited rule canvas (Fig.~\ref{fig:interface}e). Users can choose to accept the entire AI editing or selectively incorporate parts that they agree with.

% \subsubsection{Enhancement Actions}
% - main button, with dropdowns to 2 sub-actions
% - parameter (threshold) enhancement
% - topology (flow chart structure) enhancement

\subsubsection{Enhancement Actions}
Enhancement actions are accessed via a main button with a dropdown (Fig.~\ref{fig:interface}d).  
Two sub-actions are available:  
1) Threshold Parameters (Value) Enhancement, and 
2) Topology (Flowchart) Enhancement.

% \subsubsection{Enhancement Constraints}
% - Tree Prediction Similarity slider controls how similar the AI-enhanced decision tree should predict as the current decision tree on the same test dataset.
% - Tree Structure Similarity slider controls how similar in topology (structure) the two decision trees should be.
% The two slider values determine the hyperparameter of the regularization terms in the loss function of Eq. \ref{}, i.e., ? and ?, respectively.

\subsubsection{Enhancement Constraints}
Users can regulate how the AI proposes changes via two similarity sliders (Fig.~\ref{fig:interface}f).  
\emph{Prediction Similarity} controls how closely the enhanced tree should match the current tree’s predictions on the test dataset.  
\emph{Structure Similarity} controls how similar the enhanced topology should remain to the user’s current tree.  
Together, these values determine the hyperparameters of the regularization terms in Eq.~\ref{eq:loss}.

% \subsubsection{Edit Operations History}

\subsubsection{Edit Operations History}
The interface records a complete edit history performed by the AI (Fig.~\ref{fig:interface}f), showing insert, update, and remove operations. This allows users to track how the enhanced tree diverges from the original one.

% \subsection{Simulation on Test Dataset}
% - on clicking the Simulate button...
% - purpose: for users to examine performance on an explicit test dataset
% - can inspect which instances shared the same decision outcomes and which deferred, to help with debugging
% - another Fig?

\subsection{Simulation on Test Dataset}
Finally, the simulation button (Fig.~\ref{fig:interface}h) lets users test their tree on an explicit test dataset.

\section{Evaluation}
\label{sec:evaluation}
We conducted a user study to evaluate the degree of alignment and interpretability of our proposed editable explanation compared with Read-only explanations and manual editing variations. Our method and its variations were tested on three scenarios with real-world datasets, qualitative and quantitative analysis were conducted on the collected data.

\subsection{Method}
% We begin by introducing our choice of datasets, the extraction of guideline rules, and the variations of our proposed method. We then describe our participants and the experimental procedure.
We describe the method of our experiment, including the experiment design, application tasks, background knowledge context, experiment procedure, and recruited participants.

\subsubsection{Experiment Design}

We compare 3 XAI Types---variants of the explainable AI interface with different explanation capabilities:
\begin{itemize}[leftmargin=*, noitemsep]
    \item \textbf{Read-only} shows a global explanation of the AI model as a decision tree, which the user can only \textbf{Read} and not modify, examples of Read-only explanation rules can be found in Appendix~\ref{Appendix:readonly}. 
    \item \textbf{Editable} represents a non-collaborative setting, where participants \textbf{Write} and edit their rules without AI assistance. They are provided with an accuracy indicator of their rules, consistent with the one used in Section~\ref{sec:elicitation}.
    \item \textbf{CoExplain} 
    % is fully collaborative, allowing users not only to express their own domain knowledge but also to observe and iteratively align the AI’s reasoning with their expectations.
    allows the participant to collaborate with the AI to co-edit the explanatory rules of the model. In addition to \textbf{Read} and \textbf{Write} facilities, the participant can ask CoExplain to \textbf{Enhance} the explained rules to increase performance on a training dataset.
\end{itemize}

The experiment design comprises 
\textbf{XAI Type} as the primary independent variable (IV), 
decision with or without XAI (\textbf{w/ XAI}) as a secondary IV, and
application \textbf{Dataset} as a random variable (RV).
We presented each XAI Type \textit{between-subjects} to avoid the learning effect from prior exposure to different explanation features, and misattributing opinions to wrong systems.
To measure participant understanding, we ask participants to estimate the AI decision (measures described in Section~\ref{sec:experiment-measures}, procedure described in Section~\ref{sec:experiment-procedure}) without then with XAI, \textit{sequentially} to avoid the learning effect of being shown more information first.
For generality, we also varied the application task for which the participant uses the XAI UI. This was presented \textit{within-subjects} in random order (counterbalanced for three applications) to allow repeated measures of different applications per participant to reduce individual variance.

% These variations allow us to isolate and test the effects of editing and collaboration on both the alignment between AI models and human reasoning, and on participants’ understanding of the AI model.
% CoExplain enables users to actively shape the AI’s reasoning through both \textbf{Write} (authoring rules) and \textbf{Enhance} (refining thresholds and topology with AI) interactions. This design allows users not only to express their own domain knowledge but also to observe and iteratively align the AI’s reasoning with their expectations.

% \toreview{We set the XAI type as a between-subjects factor, with participants randomly assigned to one of the three XAI variants (Read-only, Editable, or CoExplain). Each participant, however, worked with multiple datasets presented in a within-subjects randomized order, ensuring that exposure to different datasets was balanced and that any effects of dataset order were minimized. This design allows us to isolate the impact of explanation type on participants’ rule creation and understanding while controlling for potential order effects across datasets.} \todo{1AC, Report statistical controls, effect sizes, and any counterbalancing}

\subsubsection{Measures} \label{sec:experiment-measures}
We measured objective metrics of user understanding (U), AI performance (P), AI alignment (A), engagement (E), and subjective perceived ratings (Q):
\begin{enumerate}[leftmargin=0.6cm, noitemsep]
    \item[U1)] \textbf{User-AI Faithfulness} of whether the participants label was the same to the AI prediction label (1 if same, 0 if different).
    This indicates if they can correctly anticipate the AI's decision, as user understanding of the AI.
    
    \item[P2)] \textbf{AI Accuracy (in Pre-trained Distribution)} of how well the final AI model (if edited) performs on a test set from the Pre-trained distribution.
    This indicates if the AI performs well in its original training domain.
    
    \item[A3)] \textbf{AI Accuracy (in Guideline Distribution)} of how well the final AI model (if edited) performs on a test set the target Guideline distribution.
    This indicates if the AI is aligned to target domain as desired by the user.
    
    \item[A4)] \textbf{Distance between XAI and Guideline Rules} of how similar are the Guideline Rules and the distilled Decision Tree explanation of the final AI model. 
    We computed distance using the Tree Edit Distance (TED)~\cite{zhang1989simple}.
    This indicates if the XAI is aligned to target guidelines as desired by the user.
    
    \item[E5)] \textbf{\# Edit Operations} indicating how many changes the participant made to the XAI rules.
    
    \item[E6)] \textbf{\# Edit Iterations} indicating how many iterations the participant went through before finalizing the AI.
    
    \item[E7)] \textbf{Editing Time} indicating how long all editing took.
    We log transformed Editing Time due to the long tail of slower participants, to make the response better satisfy the normality assumption for statistical analysis.
    
    \item[R7)] \textbf{Perceived Ratings} along a 7-pt Likert scale ($-3$ Strongly Disagree to $+3$ Strongly Agree) on the participant perceptions regarding:
    \begin{itemize}
        \item Understanding of AI Rules (decision tree)
        \item Understanding of AI Decisions (predictions)
        \item Ease of Use to Edit XAI (Editable XAI and CoExplain)
        \item Ease of Use of CoExplain Enhancement
        \item Alignment of CoExplain Enhancement
    \end{itemize}
\end{enumerate}

\subsubsection{Application Tasks}
\label{subsub:task}
We selected three application tasks and their dataset for evaluation: \textit{Adult Income}~\cite{frank2010uci}, \textit{House Price Prediction} from the "House Sales in King County, USA" dataset~\cite{harlfoxem2016housesales}, and \textit{Heart Disease} from the UCI Machine Learning Repository~\cite{janosi1988heart}. The \textit{Adult Income} task has been widely used as a testbed for human-AI collaborative decision-making~\cite{ghai2021explainable, ma2023should, hase2020evaluating}, while the other two datasets follow prior work by Bo et al.~\cite{bo2024incremental}.
The datasets vary in domain knowledge requirements: the \textit{Adult Income} and \textit{House Price} tasks rely on general knowledge and are therefore more accessible to lay participants, while the \textit{Heart Disease} task requires professional training beyond commonsense. To balance complexity and usability for the participants~\cite{bo2024incremental, ghai2021explainable, ma2023should}, we selected a small set of features: 5 for \textit{Adult Income} (Age, Education, Marital status, Investment gain, Weekly hours), 5 for \textit{House Price} (Bedrooms, Bathrooms, Living area, Grade, Age), and 4 for \textit{Heart Disease} (Age, Resting blood pressure, Cholesterol, Max heart rate). We partition each application task’s data into two distributions: a guideline distribution used to provide background knowledge to participants, and a pre-trained distribution that participants haven’t been trained on while it’s used to train the AI. This creates a controlled misalignment between the human and AI, as discussed in Appendix \ref{Appendix:partition}.

\subsubsection{Background Knowledge: Guideline Rules}
\label{subsub:guideline}
\begin{figure}
    \centering
    \includegraphics[width=8.5cm]{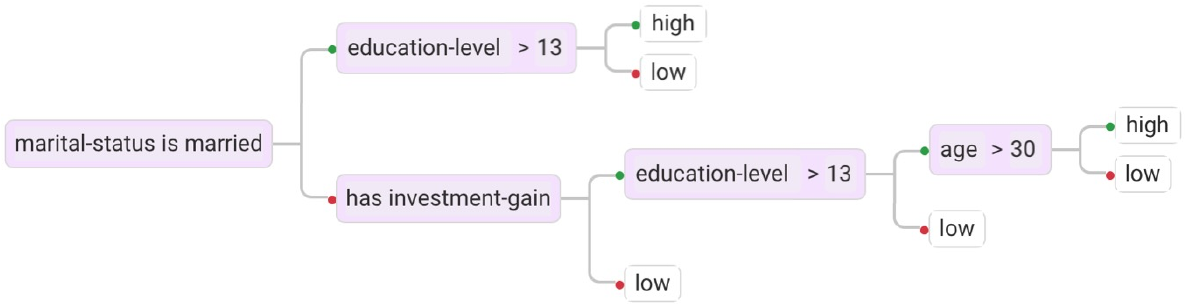}
    % \vspace{-0.5cm}
    \caption{Guideline rule for \textit{Adult Income}, green dot indicates true branch, red indicates false branch.}
    \label{fig:guideline}
    \Description[Decision tree rule guideline for Adult Income Prediction Task]{The tree starts with a root node testing if "marital-status is married" is true, the branch leads to a node testing "education-level greater than 13": if true, prediction is "high income"; if false, prediction is "low income". If "marital-status is married" is false, the branch leads to a node testing "has investment-gain": if true, the branch leads to a node testing "education-level greater than 13" if true, this leads to a node testing "age greater than 30" (true predicts "high income", false predicts "low income"); if "education-level greater than 13" is false, prediction is "low income". If "has investment-gain" is false, prediction is "low income". All false paths through nodes result in "low income" predictions.}
    % \vspace{-0.5cm}
\end{figure}

For each application task, we provide participants with a guideline rule as the starting point for constructing their rules. This guideline rule serves as a controlled background knowledge and is based on the guideline distribution. The rules are obtained by training a neural network on the guideline distribution for five iterations to ensure fairness with the Read-only, which adopt the same setting. Afterward, the distillation method described in Section~\ref{subSec:distillation} is applied to extract decision-tree rules from the trained network. Fig~\ref{fig:guideline} illustrates a guideline rule from the Adult Income Prediction Task. The guideline rules for the other tasks are provided in Appendix~\ref{Appendix:survey}.

% % \subsubsection{Variations}
% \subsubsection{Independent Variable: XAI Types}
% % We created 3 XAI types of our system: 
% We compare three variants of the explainable AI interface with different explanation capabilities:
% % \textbf{Read-only}, \textbf{Editable} and \textbf{CoExplain}. 
% \begin{itemize}
%     \item \textbf{Read-only} models one-directional explanations, where participants can view the AI’s decision rules but cannot provide feedback or modifications. The AI model and rules are trained following the procedure in Section~\ref{subsub:guideline}.
%     \item \textbf{Editable} represents a non-collaborative setting, where participants create and edit their own rules without AI assistance. They are provided only with an accuracy indicator of their rules, consistent with the one used in Section~\ref{sec:elicitation}.
%     \item \textbf{CoExplain} is fully collaborative, where participants iteratively edit their rules with user-controlled AI enhancement.
% \end{itemize}
% These variations allow us to isolate and test the effects of editing and collaboration on both the alignment between AI models and human reasoning, and on participants’ understanding of the AI model.

\subsubsection{Procedure} \label{sec:experiment-procedure}
Each participant went through the following:

\aptLtoX{\begin{enumerate}
    \item[1)] Introduction (5 min). Participants were briefed on the research background, motivation, and study protocol. Informed consent was obtained to record their interactions and screen-sharing for analysis, the University ethics review board approved human-subjects research and they approved this project.
    \item[2)] Tutorial and Screening (5 min). Participants received an introduction to decision rules and the AI system functions, followed by a short screening task to ensure comprehension.
    \item[3)] Main Session (50 min). Participants completed three scenarios in random order. For each scenario:
    \begin{enumerate}
        \item[a)] Introduction to the attribute definitions and guideline rules for the scenario.  
        \item[b)] Forward simulation across 20 case trials, consisting of:  
        \begin{enumerate}
            \item[i)] Rule editing on the explanation (Editable, CoExplain only).  
            \item[ii)] Estimation of the AI’s prediction without explanation.  
            \item[iii)] Estimation of the AI’s prediction with explanation.  
            \item[iv)] Observation of the AI’s prediction and ground-truth label.  
        \end{enumerate}
        \item[c)] Perceived ratings and a short interview.  
    \end{enumerate}
    \item[4)] Demographics, Debrief, and Compensation. Participants completed a demographics questionnaire, were debriefed of their performance, and received compensation.
\end{enumerate}}{\begin{enumerate} [leftmargin=*, noitemsep, label=\arabic*)]
    \item Introduction (5 min). Participants were briefed on the research background, motivation, and study protocol. Informed consent was obtained to record their interactions and screen-sharing for analysis, the University ethics review board approved human-subjects research and they approved this project.
    \item Tutorial and Screening (5 min). Participants received an introduction to decision rules and the AI system functions, followed by a short screening task to ensure comprehension.
    \item Main Session (50 min). Participants completed three scenarios in random order. For each scenario:
    \begin{enumerate}[label=\alph*)]
        \item Introduction to the attribute definitions and guideline rules for the scenario.  
        \item Forward simulation across 20 case trials, consisting of:  
        \begin{enumerate}[label=\roman*)]
            \item Rule editing on the explanation (Editable, CoExplain only).  
            \item Estimation of the AI’s prediction without explanation.  
            \item Estimation of the AI’s prediction with explanation.  
            \item Observation of the AI’s prediction and ground-truth label.  
        \end{enumerate}
        \item Perceived ratings and a short interview.  
    \end{enumerate}
    \item Demographics, Debrief, and Compensation. Participants completed a demographics questionnaire, were debriefed of their performance, and received compensation.
\end{enumerate}}

\subsubsection{Participants} 
Via a university mailing list, we recruited 43 participants from a local university (16 males, 27 females), aged between 19 and 30 years old (M = 22.3, SD = 2.5), with diverse disciplinary backgrounds from 18 different majors (e.g., Psychology, Business, Political Science, Computing, Medicine). Participants were randomly assigned to one of three variants. We recruited 15 participants per condition, due to no-shows the final distribution was 14, 14, and 15 participants for \textbf{Editable} (E1--E14), \textbf{CoExplain} (C15--C28) and \textbf{Read-only} (R29--R43). This minor imbalance does not substantially affect the comparability of conditions. The study was conducted through an online audio call, with participants consent, we recorded the meeting and their screen-sharing during the study. The study lasted for around 1 hour and participants were compensated \$16 USD in local currency.

\subsection{Qualitative Findings}
To investigate how users engage with \textbf{CoExplain} and benefit from editable explanations, we conducted a qualitative analysis on their interactions and post-trial interview. 
% Unlike conventional Read-only XAI, CoExplain enables users to actively shape the AI’s reasoning through both \textbf{Write} (authoring rules) and \textbf{Enhance} (refining thresholds and structure with AI) interactions. This design allows users not only to express their own domain knowledge but also to observe and iteratively align the AI’s reasoning with their expectations. 
% We organize our findings into four key themes: 
% 1) editing to align AI with user knowledge, 
% 2) bi-directional understanding between users and AI, 
% 3) threshold enhancement, and 
% 4) structural reorganization and control. 
% Together, these themes illustrate how editable interactions support users in co-constructing model behavior and understanding the AI’s reasoning processes.
We organize our findings in terms of the edit operations of writing to i) align the AI to the user's knowledge, ii) actively engage user understanding of the AI decisions, and 
AI-enhancement to update iii) basic thresholds and iv) complex topology.

% \todo{NEED TO KNOW FOR EACH PARTICIPANT WHICH XAI Type they are in, and for each quote which dataset/domain they are discussing.}

\subsubsection{Editing Aligns AI Reasoning with Users’ Knowledge}

Participants found the decision rules \textbf{clear and easy to follow}, which made them accessible. R38 described the rules as \textit{“easily understandable”} with \textit{“clear indicators of what orders to follow according to quantifiable metrics”}. C23 found rule creation approachable, stating, \textit{“If you satisfy the first [condition], then look at the next, and so on.”}

Participants considered the AI’s outputs as closely \textbf{mirroring their own reasoning}. E12 explained her reasoning process on house price prediction, \textit{“I followed the logic of house age first (older = lower price), then grade, then living area, so the AI’s behavior matches my thoughts.”} E14 similarly observed, \textit{“AI’s predictions mirrored my thought process.”} These comments reveal that editing allowed participants to directly reflect their reasoning in the AI’s behavior.

Editing also enabled participants to align the AI with their expectations, which \textbf{increased predictability} and \textbf{perceived reliability}. C25 noted, \textit{“AI's prediction and my prediction are in line. I can trust AI's prediction.”} E13 added, \textit{“When I adjusted the rules, the AI’s predictions changed to match. It’s like the AI ‘thinks’ how I do.”} These reflections show that editing not only align AI with user reasoning but also fostered user confidence in the AI's reasoning. 

% Referring to his knowledge in real estate, C19 highlighted that alignment enhanced explainability: \textit{“For houses, AI is explainable. A grade 4 house is terrible, so it sells low; a 98-year-old house is at fire risk, this aligns with what I’d expect.”} 

% C18 further emphasized trust arising from alignment: \textit{“the algorithm (rule) and this AI are based on my real-life experiences, so I could trust it.”}

Some participants went beyond the provided guidelines, integrating their own domain knowledge. For instance, C21, a Pharmacy student, noted during the heart disease prediction task, \textit{“I think the AI is generally following me in terms of my beliefs, as well as the values that I’m using. Like, it’s quite congruent with what I’m thinking.”} This demonstrates that rule editing can empower participants to project personal expertise into the AI. However, alignment is not always beneficial: when users lack domain knowledge or misinterpret rules, the AI can reinforce incorrect reasoning.

Overall, these findings indicate that editing supports active engagement, allowing participants to shape the AI’s reasoning in ways that improve understanding, predictability, and trust.

\subsubsection{Editing Improves User Understanding through Active Engagement}

With \textbf{Read-only} Explanation, some participants struggled to make sense of the AI’s reasoning because they could not adjust the rules. As R42 noted, \textit{“some portions don’t really make sense, like how having more bedrooms at the end made it low when logically it should be high.”} Without the ability to intervene, participants often resorted to guessing patterns, as R43 described, \textit{“There were some attributes that were more repeatedly affecting decisions than other so [I] started focusing on them.”}

Allowing users to directly modify rules with \textbf{Editable} explanation substantially improved their understanding. Many emphasized that comprehension came naturally because they wrote the rules themselves. E10 explained, \textit{“The judgment process was written by myself, so I know which indicators will lead to which results.”} As E13 put it, \textit{“I know the rule well because I created it.”} These reflections show how the act of editing turned explanation into \textbf{active engagement}, strengthening users’ understanding about the AI.

% Similarly, E11 said, \textit{“I made this chart by myself. So, naturally, I will know what I’m making and what I’m thinking.”} 

\textbf{CoExplain} went further by combining user edits with AI enhancements, enabling participants to see beyond their own assumptions. C21 described a shift: \textit{“At first, I didn’t understand why the AI didn’t prioritize age… now I see age isn’t a strong factor, so it makes sense.”} C15 echoed this process of adaptation, acquiescing to the AI’s logic during the house price prediction task: \textit{“The AI kept rejecting [the age attribute], so maybe I should just follow the AI and see.”} Through such iterations, participants came to understand not only their own rules but also the AI’s reasoning. Yet, enhancements sometimes created confusion. C25 recalled, \textit{“During the enhancement, the AI suggested that I swap the age and the cholesterol level, it was not my original order, I think I am a bit confused.”} 

Together, these accounts show that editing, whether with self-authored rules or collaborative enhancement, helped participants grasp the AI’s reasoning more effectively than passive observation.

\subsubsection{Appreciation for the Advice of Conservative Updates on XAI Rule Parameters}
% restrained, conservative, moderate
% - need for domain knowledge, plausibility, conservative changes
Participants valued CoExplain's enhancement to thresholds when those adjustments aligned with their knowledge. When working on the medical task, C21 welcomed help with unfamiliar medical ranges, \textit{“Because I’m not familiar with the numbers. For example, the max heart rate, so I’ll accept that one.”} 
Participants regarded these edits as “light touches” that calibrated rather than replaced their intentions, as C23 put it, \textit{"the values don't differ by a lot. I think the values are actually quite close to mine"}.

When threshold changes aligned with plausible expectations, participants found them credible and useful. C19 grounded this reaction in housing knowledge, \textit{“A grade 4 house is terrible, so it sells low; a 98-year-old house is a fire risk—this aligns with what I’d expect.”} In health scenarios, concrete anchors made threshold choices feel self-evident: C17 noted, \textit{“If you see someone with a 300 cholesterol level, there’s something wrong.”} 

Participants also appreciated when suggested thresholds added clarity to their distinctions, as C25 reflected after getting his threshold rounded for easier interpretability (e.g. 38 changed to 40 for age) on the Income Prediction task, \textit{"straightforward with clear cut-offs.”} 

Not all changes, however, were viewed positively. A few participants felt some adjustments were either too minor to matter or misaligned with their expectations, for example, C25 noted on heart disease prediction task, \textit{“the value change for resting blood pressure is very small... and for the cholesterol level the change is very significant, but I don’t think it is correct.”} Such reactions suggest that conservative updates were most effective when they align with participants’ domain understanding.

Across these cases, threshold enhancements were accepted when they fit participants’ knowledge, clarified their boundaries, and addressed uncertainty without undermining ownership.

\subsubsection{Topological Enhancements Enables Users to Refine Rules under their Control}
Participants frequently described topological enhancement as making rules clearer and easier to work with, while still retaining control over key elements. 
C25 praised the \textbf{legibility} and \textbf{conciseness} of the enhanced rules which makes it easier to understand, \textit{“more simplified and easy to understand at a glance.”} 
% \todo{BTW also helps understanding}

% C16 framed reorganization as a debugging aid, \textit{“\dots it gives you another flowchart for cross-reference, so maybe some branches are redundant, and you could improve upon it.”} Participants used these structural views to streamline logic and remove clutter.

C16 framed reorganization as providing a second opinion for verification, \textit{“it gives you another flowchart for cross-reference, maybe some branches are redundant, and you could improve.”} Participants used these structural views to streamline logic and remove clutter.

At the same time, participants actively \textbf{moderated} how much restructuring to allow. 
C19 reported intentionally widening leeway, \textit{“I lowered the [structure] similarity to the lowest because I don’t need the AI to stick to my structure.”} 
C15 explained using constraints to preserve a belief, \textit{“I wanted to make the AI still follow the attribute, both similarities need to be high enough.”} Participants also articulated trade-offs. C26 reflected on domain-contingent deference, \textit{“I trusted my own judgment in familiar areas (like house prices),”} while still appreciating support where knowledge was thinner.

Resistance to structural edits surfaced when participants felt important features were dropped. 
C17 objected to removals, \textit{“Sometimes it didn’t really enhance it because it dropped the bathroom and bedroom attributes.”} 
C24 preferred complexity and accepted wholesale restructuring as, \textit{“I didn’t insist on my own structure because I think the AI will just give me a better one. It’s very complex, and my own one is very simple.”} Together these accounts show participants selectively embracing simplification and reorganization while exercising granular control over what should remain intact.

% \subsubsection{Take-away}
% Overall, participants used editing to bring the system’s reasoning into line with their own and to understand how and why it reasoned differently when it did. 
% They accepted threshold refinements that matched lived knowledge and relied on structural reorganization to clarify logic—while managing how much change to permit. Editable XAI thus supported bi-directional alignment: users shaped the AI with rules and thresholds they endorsed, and, through the same interactions, learned how the AI made sense of their domains.
% \todo{missing enhancement?}

\subsubsection{Take-away}
Overall, participants used editing to align the system’s reasoning with their own and to understand how and why it reasoned differently when it did. They welcomed threshold and structural enhancements that reflected their knowledge, while actively moderating changes. Editing and enhancements together supported \textbf{bi-directional alignment}: users shaped the AI with their rules and learnt how it refined and extended their reasoning.

\begin{figure*}[t]
    \centering
    \includegraphics[width=16.0cm]{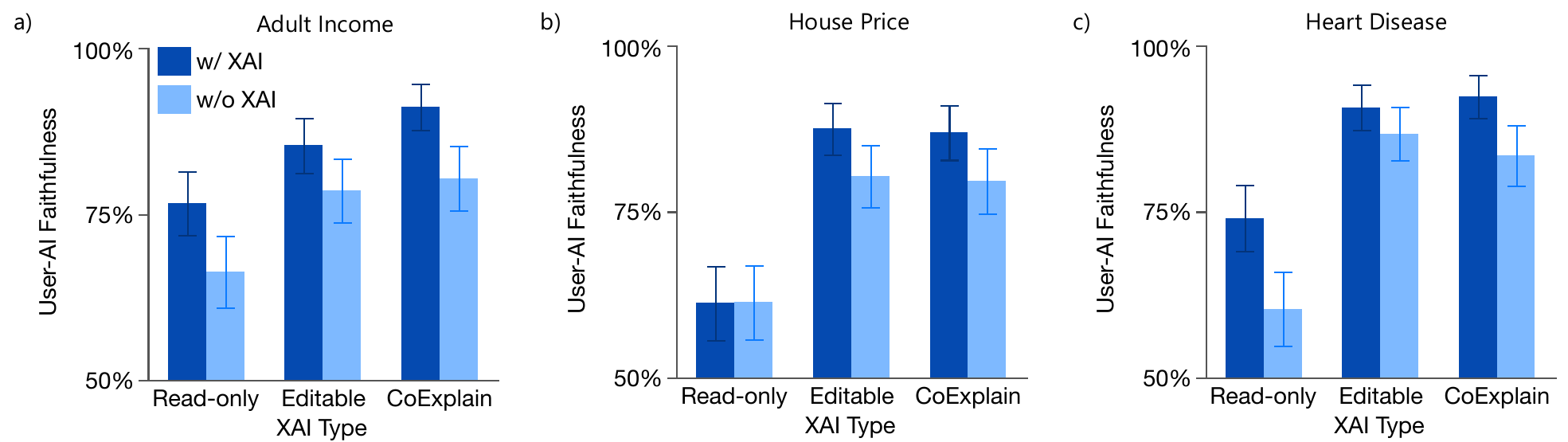}
    % \vspace{-0.2cm}
    \caption{Results from forward simulation trials for users to estimate the AI system's prediction with and without viewing explanations on a) Adult Income, b) House Price, and c) Heart Disease.}
    \label{fig:user-faithfulness}
    \Description[Three bar charts of User-AI Faithfulness across XAI Types]{Three line graphs labeled a) Adult Income, b) House Price, c) Heart Disease. Each graph plots "User-AI Faithfulness" (y-axis, 50\% to 100\%) against "XAI Type" (x-axis: Read-only, Editable, CoExplain). Two types of bar per graph: dark blue ("with XAI") and light blue ("without XAI"), with error bars. a) Adult Income: "with XAI" rises from around 77\% (Read-only) to around 85\% (Editable) to around 92\% (CoExplain). "without XAI" rises from around 65\% (Read-only) to around 80\% (Editable) and stays flat for CoExplain. b) House Price: "with XAI" jumps from around 65\% (Read-only) to around 90\% (Editable) then dips slightly for CoExplain. "without XAI" rises from around 65\% (Read-only) to around 80\% (Editable) then dips slightly for CoExplain. c) Heart Disease: "with XAI" rises from around 75\% (Read-only) to around 95\% (Editable) and stays flat for CoExplain. "without XAI" rises from around 60\% (Read-only) to around 90\% (Editable) then dips to around 85\% (CoExplain).}
    % \vspace{-0.3cm}
\end{figure*}

\subsection{Quantitative Results}
We evaluated how Editable and CoExplain explanations affect user understanding, alignment, and engagement compared with Read-only explanations. 
We consider results with $p \leq .002$ as statistically significant to account for up to 25 multiple comparisons up and avoid Type I error. 
Our analyses address four questions: i) do Editable and CoExplain improve user-AI faithfulness, ii) how do they influence alignment and performance, iii) how do they shape user engagement, and iv) how are these explanations perceived by users.

% \subsubsection{Statistical Analysis}
% For statistical analysis, binary measures P1, A2 and A3 were analyzed using a Generalized Linear Mixed Model (GLMM) with a binomial distribution and logit link. Fixed effects included \textit{XAI Type}, \textit{Dataset}, whether the trial was \textit{with or without XAI}, and their interactions, with Participant ID as a random effect. For all other measures, we fit Linear Mixed-Effects Models with \textit{XAI Type}, \textit{Dataset}, and their interaction as fixed effects, and Participant ID as a random effect, using restricted maximum likelihood (REML) estimation. 

% \subsubsection{Statistical Analysis}
% For statistical analysis, binary measures P1, A2, and A3 were analyzed using a Generalized Linear Mixed Model (GLMM) with a binomial distribution and logit link. Fixed effects included \textit{XAI Type}, \textit{Dataset}, whether the trial was \textit{with or without XAI}, and their interactions, with Participant ID as a random effect. For all other measures, we fit Linear Mixed-Effects Models with \textit{XAI Type}, \textit{Dataset}, and their interaction as fixed effects, and Participant ID as a random effect, using restricted maximum likelihood (REML) estimation. \toreview{In addition to model significance tests, we report pairwise effect sizes using Cohen's $d$ to quantify the magnitude of differences between XAI conditions.}

\subsubsection{Statistical Analysis}
For statistical analysis, binary measures P1, A2, and A3 were analyzed using a Generalized Linear Mixed Model (GLMM) with a binomial distribution and logit link. Fixed effects included \textit{XAI Type}, \textit{Dataset}, whether the trial was \textit{with or without XAI}, and their interactions, with Participant ID as a random effect. For all other measures, we fit Linear Mixed-Effects Models with \textit{XAI Type}, \textit{Dataset}, and their interaction as fixed effects, and Participant ID as a random effect, using restricted maximum likelihood (REML) estimation. 
We examined statistical controls with demographic and background variables (gender, major, age, domain familiarity, and prior AI experience) as covariates, but found no exhibited significant effects. We thus omitted them from our statsitical models.
% , nor did they change the significance or direction of the primary results. 
We report significant differences (p < .001), and their effect size using Cohen's $d$~\cite{cumming2013understanding} from the baseline Read-only condition.
% In addition to model significance tests, we report pairwise effect sizes using Cohen's $d$~\cite{cumming2013understanding} to quantify the magnitude of differences between XAI conditions}

\begin{figure*}[t]
    \centering
    \includegraphics[width=16.0cm]{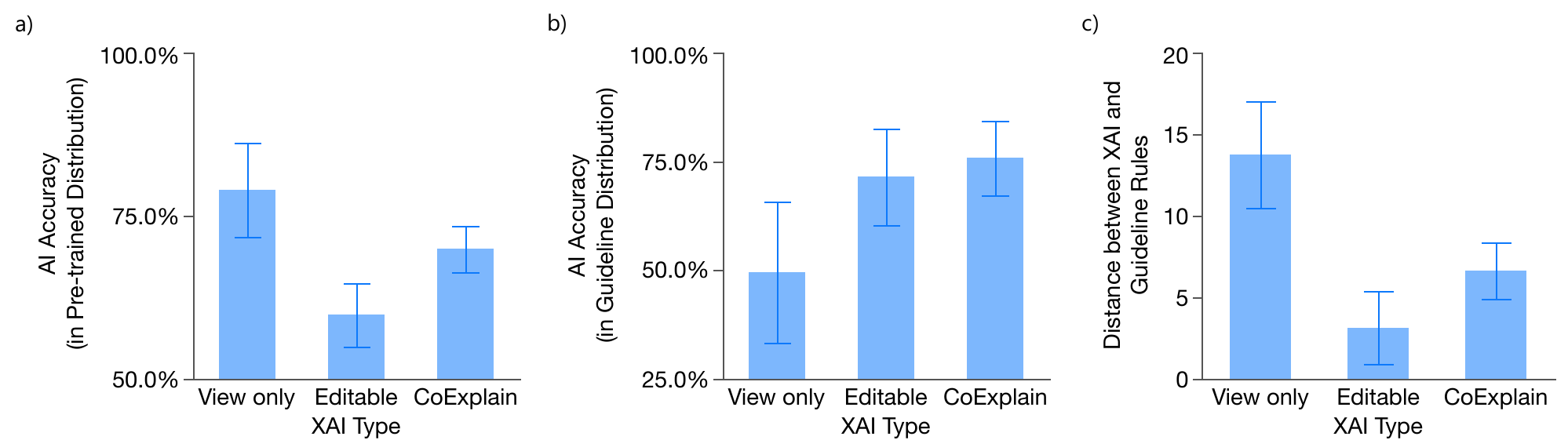}
    % \vspace{-0.3cm}
    \caption{Results measuring AI performance on pre-trained and guideline distribution and alignment with guideline rules across different XAI types: a) AI accuracy on the pre-trained distribution, b) AI accuracy in the guideline distribution, and c) distance between XAI explanations and guideline rules.}
    \label{fig:performance}
    \Description[AI performance across XAI types]{Three line graphs showing AI performance for Read-only, Editable, and CoExplain XAI types. a) AI accuracy on pre-trained distribution: Highest for Read-only (around 80\%), drops for Editable (around 60\%), then rises for CoExplain (around 70\%). b) AI accuracy in guideline distribution: Lowest for Read-only (around 50\%), rises for Editable (around 75\%), and increases slightly for CoExplain (around 78\%). c) Distance between XAI explanations and guideline rules: Highest for Read-only (around 14), lowest for Editable (around 3), then rises for CoExplain (around 7). Error bars show variability. Trends highlight tradeoffs: Editable improves alignment but reduces pre-trained accuracy; CoExplain balances across metrics.}
    % \vspace{-0.3cm}
\end{figure*}

\subsubsection{Editable Explanations Improve User-AI Faithfulness}
We found significant main effects of \textit{XAI Type} ($p<.0001$), \textit{w/ XAI} ($p<.0001$), and \textit{Dataset} ($p=0.0003$). Post-hoc contrast tests showed that both Editable ($p<.0001$) and CoExplain ($p<.0001$) significantly improved user-AI faithfulness compared to Read-only, while not differing significantly from each other.

\textit{Adult Income.}
As shown in Fig.~\ref{fig:user-faithfulness}a, Read-only participants performed the worst (M = 66.6\% without XAI; 77.1\% with XAI). Editable performed better overall (M = 80.6\% and 87.5\%), showing large advantages over Read-only in both settings ($d$ = 0.94 and $d$ = 0.87). CoExplain achieved similarly strong results (M = 81.4\% and 91.9\%), again with large differences relative to Read-only ($d$ = 0.98 and $d$ = 1.32). Both Editable and CoExplain thus fostered substantially better user understanding than Read-only.

\textit{House Price.}
Editable participants achieved high faithfulness (M = 82.5\% and 89.6\%), with large effects relative to Read-only ($d$ = 1.37 and $d$ = 2.01). CoExplain showed similarly strong performance (M = 80.6\% and 87.6\%), again yielding large differences from Read-only ($d$ = 1.21 and $d$ = 1.80). By contrast, Read-only participants showed minor improvement (M = 61.6\% and 61.9\%), as shown in Fig.~\ref{fig:user-faithfulness}b.

\textit{Heart Disease.}
Editable explanations were particularly beneficial in this task (Fig.~\ref{fig:user-faithfulness}c). Editable participants showed strong performance (M = 89.2\% and 92.8\%), with large effects relative to Read-only ($d$ = 1.99 and 1.62). CoExplain achieved similarly high faithfulness (M = 84.4\% and 93.9\%), also with large differences from Read-only ($d$ = 1.56 and 1.62). In contrast, Read-only participants performed much worse overall (M = 60.5\% and 74.4\%) and benefited far less from explanations that were not aligned through user editing.

% \textit{Heart Disease.}
% Editable explanations were particularly beneficial in this task (Fig.~\ref{fig:faithfulness}c). Editable (M = 92.8\%) and CoExplain (M = 93.9\%) participants began with strong baselines (M = 89.2\% and M = 84.4\%) and improved further with explanations, whereas Read-only participants started much lower (M = 60.5\%) and improved only to M = 74.4\%. \rev{Cohen's $d$ confirmed these patterns: Editable and CoExplain were nearly identical with XAI ($d = 0.06$) and both far outperformed Read-only (both $d = 1.62$). Without XAI, they differed modestly ($d = 0.43$) but each still exceeded Read-only by a large effect (Editable $d = 1.99$, CoExplain $d = 1.56$).}

Across all datasets, Editable and CoExplain supported higher initial levels of user-AI faithfulness and larger gains once explanations were available, demonstrating improved understanding. 
These results support our hypothesis that editable explanations leverage the \textit{generation effect} from active learning: by constructing and refining rules, users internalized the AI’s behavior more effectively, amplifying the benefits of explanation access.

\subsubsection{CoExplain Improve AI Accuracy.}
As shown in Fig.~\ref{fig:performance}a, the model revealed significant main effects of \textit{XAI Type} and \textit{Dataset} ($p<.0001$), with no significant interaction or effect of Edit Iterations. Read-only achieved the highest pre-trained accuracy (M = 78.9\%), Editable the lowest (M = 59.8\%), and CoExplain (M = 69.9\%) increased the accuracy from Editable with AI enhancement ($d$ = 1.25) and minimized the gap with data optimized Read-only ($d$ = 0.81), improving editing accuracy without sacrificing alignment.

\subsubsection{Editable Explanations Improve Prediction Alignment.}
There are strong effects of \textit{Dataset} and its interaction with \textit{XAI Type} ($p<.0001$). Although the main effect of \textit{XAI Type} was not significant, least-square means showed clear improvements for \textit{Editable} ($d$ = 0.80, M = 71.4\%) and \textit{CoExplain} ($d$ = 1.03, M = 75.7\%) over \textit{Read-only} (M = 49.5\%). Editing effort also mattered, with Iterations showing a robust effect ($p<.0001$).

\subsubsection{Editable Explanations improve Topological Alignment.}
Fig.~\ref{fig:performance}c shows the rule-level distance measure. The model revealed a strong main effect of \textit{XAI Type} ($p<.0001$) and an interaction with \textit{Dataset} ($p=.0016$). Without user editing, Read-only showed the farthest topological difference from the guideline rules (M =13.74 edits), comparing with it, Editable ($d$ = 1.92, M = 3.13) and CoExplain ($d$ = 1.39, M = 6.64) achieved a much shorter distance, showing a closer topological alignment with the guideline rules

% Editable rules were closest to guidelines (M = 3.13 edits), Read-only the farthest (M = 13.74), and CoExplain intermediate (M = 6.64). \toreview{Pairwise comparisons showed substantial differences: Editable vs.\ CoExplain ($d = 0.94$), Editable vs.\ Read-only ($d = 1.92$), and CoExplain vs.\ Read-only ($d = 1.39$), indicating stronger structural alignment for editable explanations.}

Together, guideline accuracy reflects alignment at the prediction level, while distance reflects alignment at the explanation level that users directly shape. Editable maximized alignment but sacrificed performance, Read-only maximized performance but ignored user knowledge, and CoExplain balanced both—achieving near-optimal accuracy with explanations that stayed closer to user rules.  

\begin{figure*}[t]
    \centering
    \includegraphics[width=13cm]{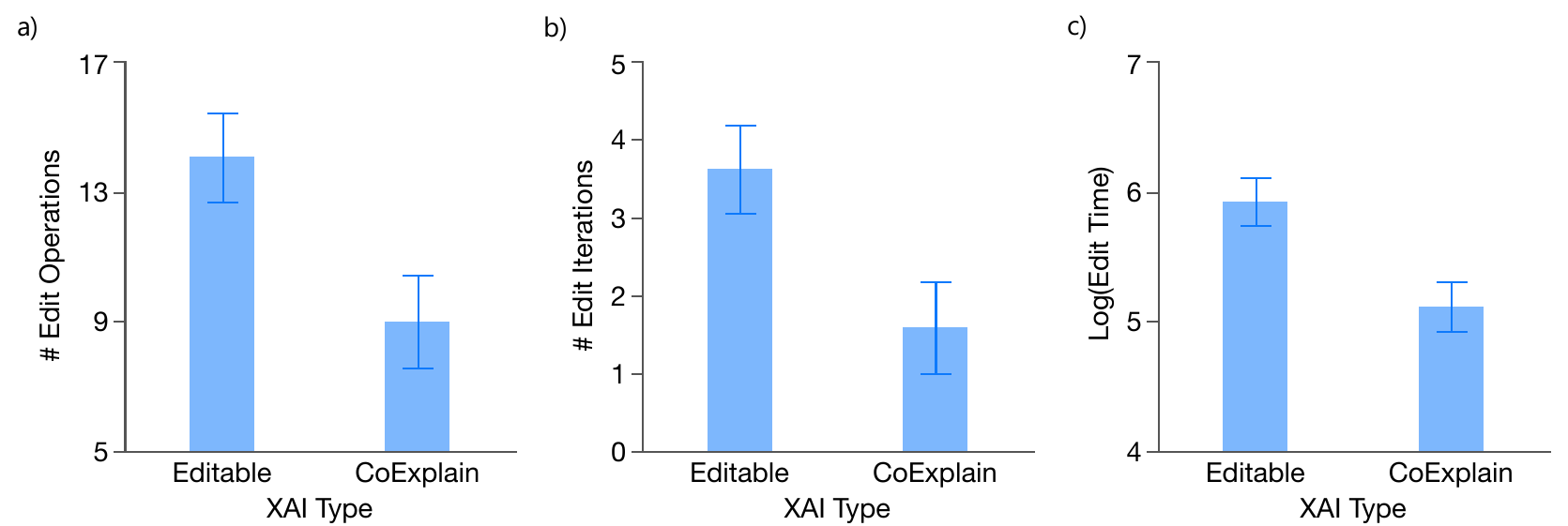}
    % \vspace{-0.4cm}
    \caption{Results measuring user engagement in terms of: a) the number of edit operations, b) the number of edit iterations, and c) log of the time spent on editing.}
    \label{fig:engagement}
    \Description[User editing effort for Editable vs CoExplain XAI types]{Three graphs show user editing effort for Editable (no AI enhancement) and CoExplain (with AI enhancement) XAI types: a) Number of edit operations: Editable requires the most operations (around 15), while CoExplain needs fewer (around 10). b) Number of edit iterations: Editable has the highest count (around 3.5), and CoExplain has fewer (around 1.5). c) Log of editing time: Editable takes the longest (around 6), with CoExplain taking less time (around 5). For all metrics, Editable shows greater editing effort compared to CoExplain, with error bars for each data point.}
    % \vspace{-0.4cm}
\end{figure*}

\subsubsection{User Engagement}
We analyze editing interaction logs to compare user engagement across XAI types in terms of a) the number of edit operations, b) the number of edit iterations, and c) editing time (Fig.~\ref{fig:engagement}).

\textit{\# Edit Operations.}
Fig.~\ref{fig:engagement}a showed a significant main effect of \textit{XAI Type} ($p<.0001$), with no effects of \textit{Dataset} or the interaction. Editable required the most operations (M = 14.05), while CoExplain reduced this load ($d$ = 1.94, M = 9.00), indicating more efficient editing with AI support.

\textit{\# Edit Iterations.}
As shown in Fig.~\ref{fig:engagement}b, \textit{XAI Type} had a strong effect ($p<.0001$), with no influence from \textit{Dataset} or the interaction. Editable users needed more revision rounds (M = 3.62) than CoExplain users ($d$ = 1.91, M = 1.59).

\textit{Editing Time.}
Fig.~\ref{fig:engagement}c similarly showed a main effect of \textit{XAI Type} ($p<.0001$). Editable users spent more time (M = 7.18 mins) than CoExplain users (M = 3.36 mins), a 53\% reduction with $d$ = 2.33, demonstrating substantial savings in editing effort.
 
Together, these results show that while Editable required the most edit operations, iterations, and time, CoExplain significantly reduced user effort while still allowing meaningful interaction. Combined with its performance and alignment results, CoExplain provided faithful and efficient AI enhanced editing.

\begin{figure}[t]
    \centering
    \includegraphics[width=8.5cm]{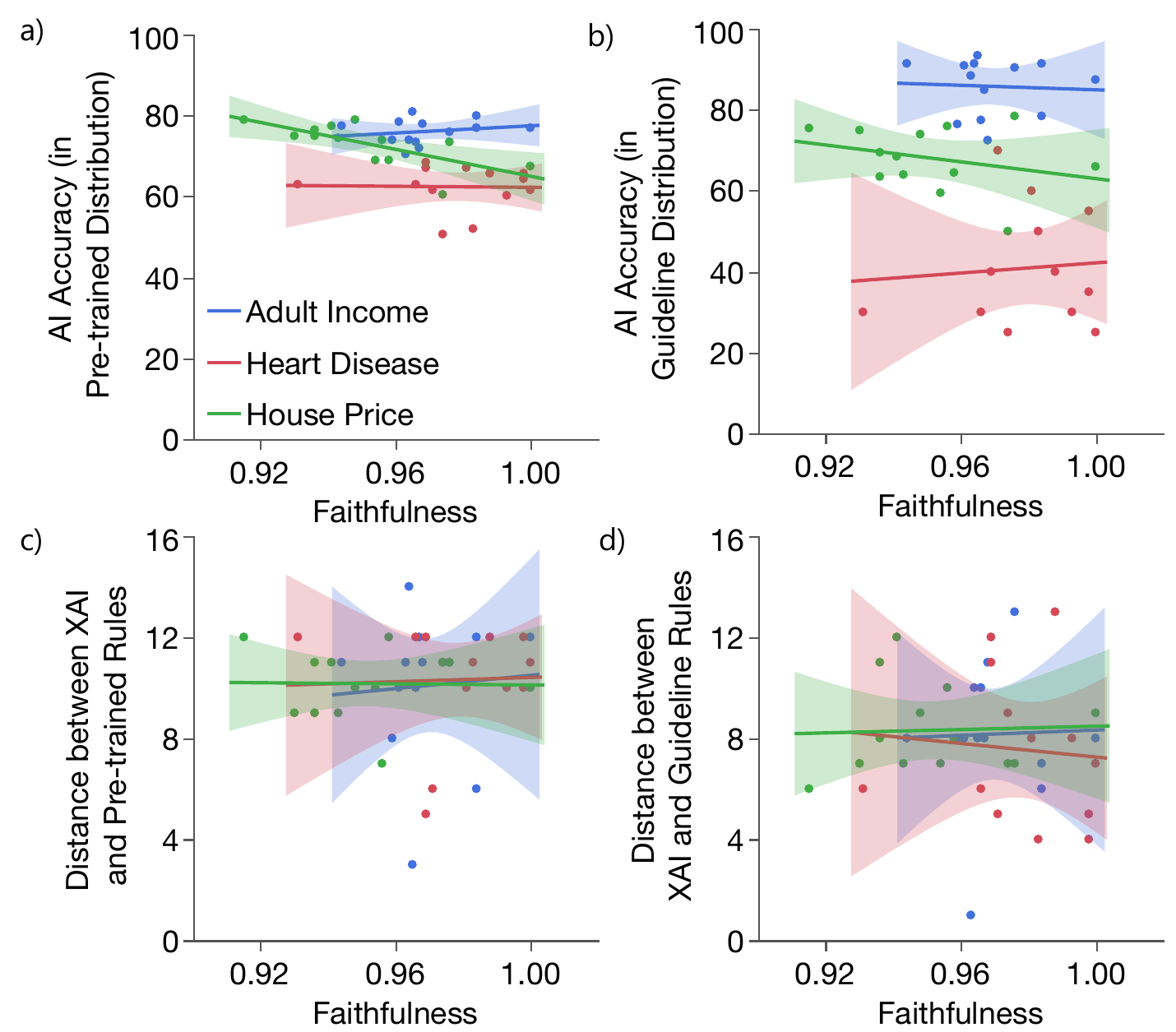}
    % \vspace{-0.7cm}
    \caption{Analysis of edit outcomes in relation to the faithfulness of the tree explanations to the underlying model. a) AI accuracy on the pre-trained distribution, b) AI accuracy on the guideline distribution, c) TED between XAI and pre-trained rules, and d) TED between XAI and guideline rules.}
    \label{fig:faithfulness}
    \Description[Analysis of edit outcomes in relation to the faithfulness of the tree explanations to the underlying model]{Scatter plots showing the relationship between faithfulness of tree explanations and four editing outcomes across three datasets: Adult Income, House Price, and Heart Disease. Faithfulness values are on the x-axis (ranging roughly 96\% to 98\%), and each plot shows one outcome on the y-axis: (a) AI accuracy on the pre-trained distribution, (b) AI accuracy on the guideline distribution, (c) tree edit distance between the XAI rules and pre-trained rules, and (d) tree edit distance between the XAI rules and guideline rules. Each dataset is represented with separate markers. Regression lines are shown for each dataset, but slopes are generally small and confidence intervals wide, indicating no significant effect of faithfulness on editing outcomes. Overall, the plots suggest that variation in faithfulness within the observed range did not systematically affect how well users edited the rules.}
    % \vspace{-0.7cm}
\end{figure}

\subsubsection{AI-Explanation Faithfulness and Edit Outcome Analysis}
Faithfulness quantifies how closely the distilled decision tree matches the neural network’s predictions. 
Since a distilled tree explanation may oversimplify the neural network model, the explanation may be unfaithful to the model. 
This unfaithfulness could mislead users to editing irrelevant attributes.
Hence, we investigate if faithfulness was low and whether it led to ineffective user edits.
Across 350 trials from the user study, the average faithfulness was high, $97.5\%$ for \textit{Adult Income}, $96.8\%$ for \textit{House Price}, and $96.53\%$ for \textit{Heart Disease}.
Figure~\ref{fig:faithfulness} plots faithfulness against four editing outcomes: 
a) AI accuracy on the pre-trained distribution,
b) AI accuracy on the guideline distribution, 
c) distance between the XAI rules and pre-trained rules, and
d) distance between the XAI rules and guideline rules. 
% Across all tasks and outcomes, the regression slopes are small and confidence intervals are wide
Almost all trends are not significantly different from 0, indicating no clear effect on edit outcomes due to faithfulness. 
Interestingly, for House Price, accuracy decreases as Faithfulness increases, suggesting that users may be less engaged in editing a larger tree.
In general, the results suggest that variation in faithfulness within the observed range did not systematically influence how well users edited the rules.
Future work could study this effect by aggressively pruning the tree explanation or training a model on a significantly more complex domain.

\begin{figure}[t]
    \centering
    \includegraphics[width=8.5cm]{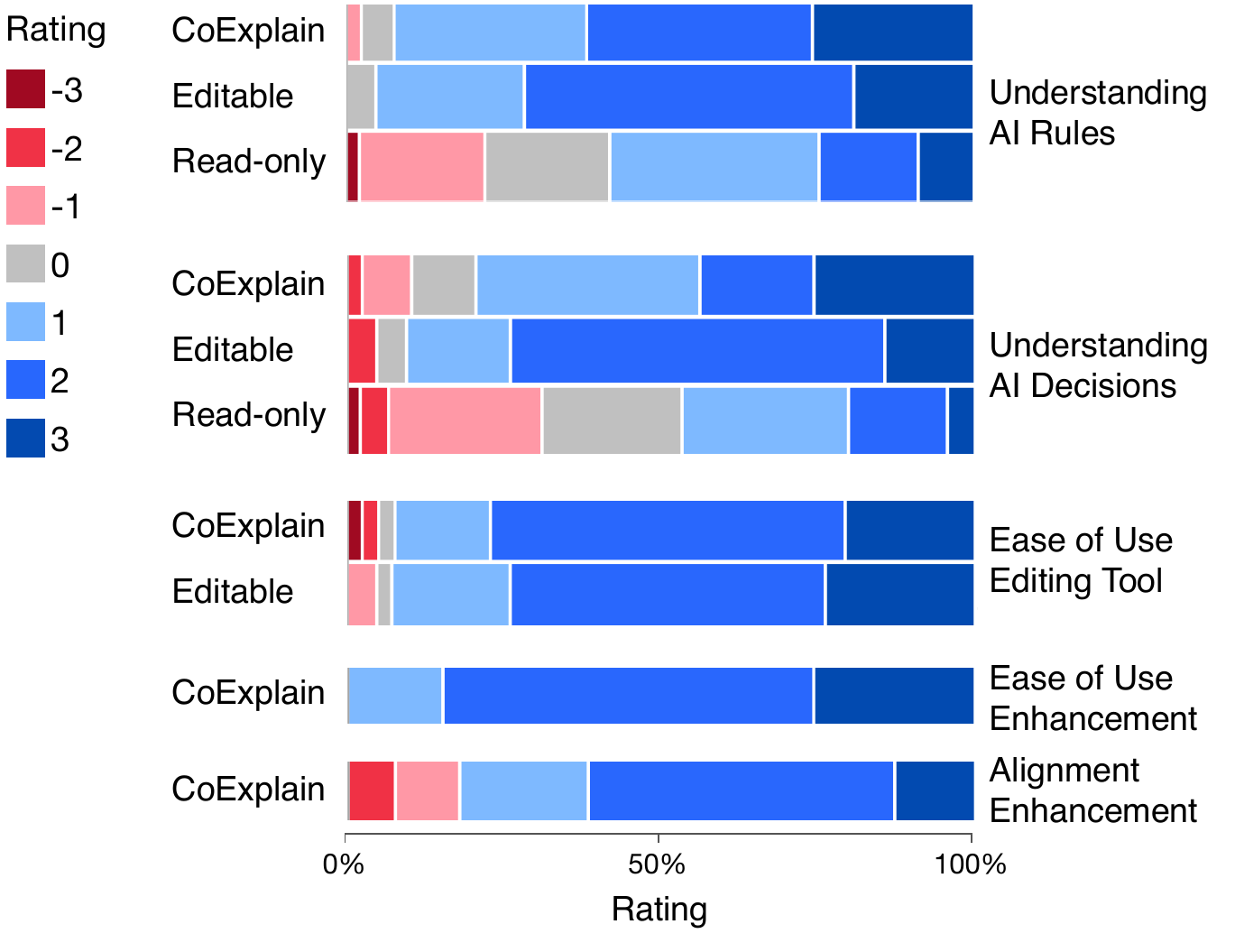}
    % \vspace{-0.9cm}
    \caption{Perceived ratings of understanding of AI rules and decisions, ease of use, and alignment of enhancements across XAI types. AI enhancement is only available in CoExplain, Read-only does not support editing.}
    \label{fig:perceived}
    \Description[Perceived ratings across XAI types]{Horizontal stacked bar chart showing perceived ratings for CoExplain, Editable, and Read-only XAI types across categories: Alignment AI Enhancement, Ease AI Enhancement, Understanding AI Behavior, and Understanding AI Rules. “Alignment AI Enhancement”, “Ease AI Enhancement” (exclusive to CoExplain) and "Ease Editing Tool" have high positive ratings. For “Understanding AI Behavior” and “Understanding AI Rules”, CoExplain and Editable have more positive ratings than Read-only.}
    % \vspace{-0.6cm}
\end{figure}

\subsubsection{Perceived Ratings}
We collected perceived ratings from participants about their understanding of the resulting AI model in terms of its rule explanations and decisions, the ease of use of the system, and the ease of use and alignment of CoExplain enhancements (Fig.~\ref{fig:perceived}).

For understanding the AI model, editable conditions were clearly preferred. A mixed-effects analysis revealed a significant effect of \textit{XAI Type} on perceived understanding of AI behavior ($p=0.0006$) and rules ($p=0.0008$). Post-hoc contrast tests showed that both Editable ($M=1.69$ ($\approx$ 2 Agree) for Understanding of AI Decisions; $M=1.86$ ($\approx$ 2 Agree) for Understanding of AI Rules) and CoExplain ($M=1.36$ ($\approx$ 1 Somewhat Agree); $M=1.77$ ($\approx$ 2 Agree)) were rated significantly higher than Read-only ($M=0.31$ ($\approx$ 0 Neither Agree nor Disagree); $M=0.64$ ($\approx$ 1 Somewhat Agree)).  

Participants generally rated the editing interface as easy to use. Most users found CoExplain’s enhancement suggestions controllable and well aligned with their rules with an acceptance rate of 89.74\% of enhancement edits. However, a few participants expressed dissatisfaction with certain AI suggestions, noting that they occasionally felt misaligned with their goals. C19, who was familiar with the real estate, complained about the AI on the house price prediction task as \textit{"I think it was way too simple."}

In summary, participants perceived interactive explanations as more supportive of understanding the AI model, easier to use, and better aligned with them. These perceptions reinforce the value of engaging users in a collaborative relationship with the system, rather than limiting them to static, read-only explanations.

\subsection{Summary of Results}

Our results show that Editable explanations improved user–AI faithfulness, understanding, and perceived ease of use compared to Read-only explanations. The summative study demonstrated that editable explanations led to higher alignment and better understanding of the AI logic, with CoExplain further reducing editing effort through co-editing enhancements. 
% Perceived ratings confirmed that participants found editing easier, more aligned with their intentions, and significantly better for understanding AI rules and behavior. 
Qualitative findings revealed that participants used editing to align the AI with their knowledge and engaged with enhancements to refine thresholds and topology under their control. Together, these results suggest that editable explanations, and CoExplain in particular, foster more effective and efficient human–AI alignment than static, read-only explanations.

\section{Discussion}
We discuss Editable XAI and its implementation as CoExplain in terms of the scope, relation to other approaches, and the collaborative challenges of human–AI co-editing. Together these themes highlight how editable explanations create new opportunities for bi-directional alignment between humans and AI.

\subsection{Scope and Generalization of CoExplain}
While our implementation of \textbf{Editable XAI} as \textbf{CoExplain} targeted structured data with decision tree as the knowledge representation, we discuss how the framework of Editable XAI can be generalized.

\subsubsection{Beyond Symbolic Rules to General Human Knowledge}
CoExplain was designed specifically for human knowledge expressed as symbolic rules. However, the broader principle of Editable XAI is not limited to this representation. In many domains, experts encode knowledge through heuristic equations or domain-specific formulae, for example, a risk score calculated as \(\text{Risk} = 0.3 \times \text{Age} + 0.5 \times \text{Blood Pressure}\), which, like rules, offer a shared medium interpretable by both humans and AI. Extending editable explanations to such mathematical expressions would broaden applicability, allowing collaboration around quantitative heuristics as well as logical conditions. While our work does not address this direction directly, research on knowledge distillation~\cite{gou2021knowledge}, neurosymbolic integration~\cite{hilario2013overview}, and equation learner networks~\cite{kim2020integration} highlight promising avenues that could inspire future methods for parsing and refining equations within editable XAI frameworks.

\subsubsection{Beyond Neural Networks to Diverse AI Models}
While CoExplain is implemented for neural networks, the principle of Editable XAI as using an editable representation as a shared medium for human–AI alignment is not limited to this model class. Neural networks provide high accuracy, flexible gradient-descent training, and the ability to incorporate user-specified constraints, making them well-suited for editable explanations. Editable XAI is also valuable for glass-box models, such as decision trees or decision sets~\cite{lakkaraju2016interpretable}, where users can already inspect and directly modify model components. In high-stakes domains, combining glass-box models with editable explanations can further enhance alignment, transparency, and trust~\cite{rudin2019stop}. For other black-box models, such as recurrent or ensemble models, Editable XAI would require identifying an appropriate intermediate representation, similar to how CoExplain uses decision tree rules, that preserves model fidelity while supporting human edits and comprehension. This suggests that the concept of Editable XAI could potentially be applied more broadly across diverse model families, beyond neural networks.

\subsubsection{Beyond Structured Data to Concept-based Explanations}
\label{subsub:modality}

Our evaluation targeted binary classification tasks on structured data, though the parsing algorithm also supports multiclass classification. 
Extending to regression tasks would require new rule-to-network mappings that accommodate continuous outputs.
Extending to unstructured data can be done by first interpreting semantic concepts, then reasoning over them in a structured (logical) way.

With the importance of explaining in interpretable concepts~\cite{zhang2022towards, lim2025diagrammatization},
XAI techniques for concept-based explanations can be used to infer human interpretable concepts~\cite{ghorbani2019towards, koh2020concept, poeta2023concept}, that are then treated as tabular data to be explained and edited with CoExplain.
Recent works have integrated concept extraction and logical neural networks~\cite{ciravegna2023logic, jain2022extending}, but they are not editable.
Concept-based explanations assume that the concepts align with human prior knowledge, but if they are spurious, users may want to edit them~\cite{hu2025editable}.
Hence, future work can extend CoExplain for unstructured data by using concept-based explanations with editable concepts.

As a preliminary proof-of-concept, we demonstrate CoExplain for medical diagnosis on electrocardiogram (ECG) with ECG concept features from a time series ECG trace (see Fig.~\ref{fig:ECG}, modeling details in Appendix~\ref{Appendix:ECG-demo}).
Here, we extracted concepts (e.g., QRS Duration $t_\text{QRS}$, PR Duration $t_\text{PR}$; see Fig.~\ref{fig:ECG}a) using the NeuroKit2~\cite{Makowski2021neurokit}, and used CoExplain to train a neural network, and explain with a decision tree (Fig.~\ref{fig:ECG}b) based on those concept features.

\begin{figure}[t]
    \centering
    \includegraphics[width=8.5cm]{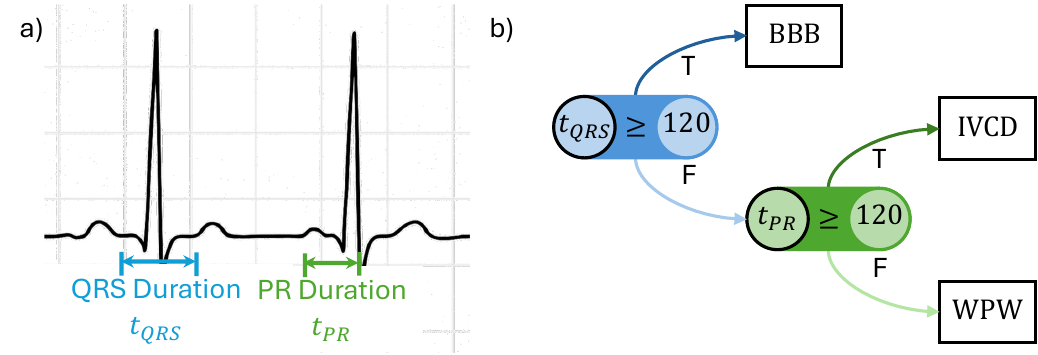}
    % \vspace{-0.7cm}
    \caption{
        Use of CoExplain on ECG signal data with feature extraction. Concept features extracted using NeuroKit2~\cite{makowski2021neurokit2}, decision tree constructed from MG Khan~\cite{khan2008rapid}. Leaf nodes denotes diagnosis, Bundle Branch Block (BBB), IntraVentricular Conduction Delay (IVCD) and Wolff-Parkinson-White syndrome (WPW).
        \Description[Demonstration of applying CoExplain on ECG signal data]{left shows an ECG wave, with annotation for QRS Duration and PR Duration, right shows a decision tree compiled using the two features.}
    }
    \label{fig:ECG}
    % \vspace{-0.5cm}
\end{figure}

\subsubsection{Beyond Monolithic Model to Modular Sub-Models}
\label{subsub:scale}

We have explored the use of CoExplain to parse and distill interpretable decision trees with feedforward neural networks.
To manage the cognitive for our participants, we had pruned the tree size to be no larger than a depth of 4 levels and 16 leaves. 
This also limited the corresponding neural networks to about 5 layers with 150 neurons in total.
Yet, the technical approach is not limited to small trees or networks, and can scale to larger models. For example, an unpruned tree of depth 10 can correspond to a NN of $20,460$ neurons.
Recent works knowledge distillation techniques (e.g., layer-wise distillation~\cite{song2021tree}, soft decision trees~\cite{frosst2017distilling}, tree embeddings~\cite{li2020tnt}) can be leveraged to further scale to the overall technical approach to accommodate modern deep models like DenseNet-169~\cite{vulli2022fine}.

In contrast, human cognition is more bounded, and this presents scalability challenges too, 
as observed for large expert systems~\cite{liao2005expert, jackson1986introduction}. 
Hence, how could users edit a truly large and complex model?
We hypothesize that abstraction~\cite{boggust2025abstraction} with modularity~\cite{zhang2022towards, lim2025diagrammatization, matsuyama2023iris} could help users to interpret the model hierarchically with sub-models.
Instead of editing a large monolithic model, the user can separately edit each sub-model and refine them along a shared module-module interface. 
This is beyond the scope of our work and should be designed and examined in future work.

\subsection{Relation to Other XAI and AI Approaches}
We situate Editable XAI within broader work on human–AI interaction, highlighting both its differences from and complementarities with interactive explanations, interactive machine learning and large language models.

\subsubsection{Editable XAI vs. Interactive Explanations}
Existing interactive explanation methods focus on enhancing user understanding by allowing queries through clarifications such as counterfactuals~\cite{metsch2024clarus}, examples~\cite{cai2019effects}, or contrastive cases~\cite{buccinca2025contrastive,teso2019explanatory}. 
While effective for inspection, these approaches typically leave the underlying reasoning fixed, limiting users’ ability to align the model with their domain knowledge. 
Editable XAI introduces a new level of interaction by making explanations themselves editable: users can directly reshape the decision logic or reasoning artifacts, which in turn updates the model’s behavior. 
CoExplain exemplifies this approach, demonstrating that editable explanations can improve bi-directional alignment, allowing users not only to understand but also to correct and align model reasoning. 
The concept of Editable XAI is general and could be applied to other explanation techniques, suggesting a broader paradigm where explanations serve as manipulable interfaces for collaborative human–AI reasoning, especially in domains where users already possess domain expertise (e.g., medicine or finance) and seek not just to interpret but also to correct or refine the model’s reasoning for alignment~\cite{boggust2025abstraction, lim2025diagrammatization}.

\begin{figure}[t]
    \centering
    \includegraphics[width=8.5cm]{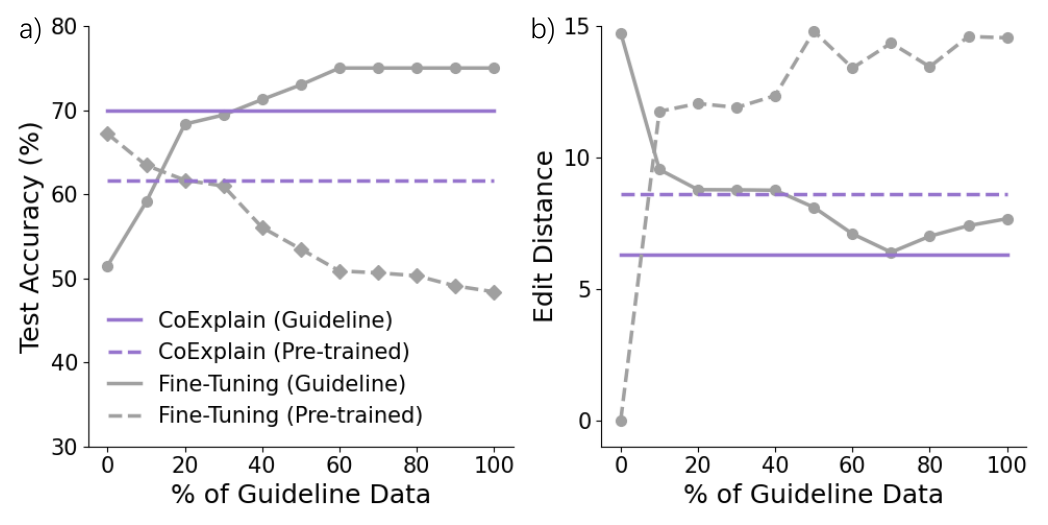}
    % \vspace{-0.8cm}
    \caption{
        Comparison of CoExplain and Direct Fine-Tuning
        showing the trade-off between effortful human labeling and alignment to distributions.
    }
    \Description[Comparison results of CoExplain and direct fine-tuning]{The x axis shows percentage of guideline line data used to fine-tune the model, subplot a) shows the comparison on test accuracy between CoExplain and fine-tuning on Guideline and Pre-trained distribution. Fine-tuning's performance on Guideline distribution exceeded that of CoExplain when more than around 30 percent of labeled data is used to fine-tune, but its performance on Pre-trained distribution dropped lower than CoExplain after 20 percent of labeled data on guideline distribution is used to fine-tune. Subplot b) shows the comparison on edit distance, fine-tuning gradually reduced the edit distance to the level of CoExplain on guideline distribution, but suffered from a much higher distance from pre-trained distribution rules than CoExplain.}
    \label{fig:fine-tune-summary}
    % \vspace{-0.8cm}
\end{figure}

\subsubsection{Editable XAI vs. Interactive Machine Learning}
% 1. labor intensive labeling
% 2. user understanding matters when aligning models
% 3. local coverage per instance vs global coverage with explanation
% 4. human label without a global view is opaque and may introduce unintended behavior resulting from the model's learning spurous correlations

Interactive machine learning (IML) approaches, such as programming by demonstration, active learning, or reinforcement learning from human feedback, improve model behavior through iterative data provision and retraining~\cite{fails2003interactive}. 
These methods are effective for expanding coverage and adapting to new tasks but typically require substantial user effort in generating labels or demonstrations~\cite{amershi2014power}. 

Editable XAI operates at a different level: rather than supplying new data, users directly modify reasoning artifacts such as rules, which can be more efficient in settings where labeling is costly, slow, or sensitive (e.g., medical diagnosis), while also providing transparency about how edits affect model logic~\cite{guo2022building}. 

Despite the labor intensity, fine-tuning with human labeling could achieve better performance than editing via XAI.
To investigate the trade-off between effort and accuracy, we compare CoExplain with direct fine-tuning using varying amounts of guideline data on the \textit{Heart Disease} task with 460 instances. 
For fine-tuning, to simulate human labeling, we used the guideline decision tree of the user study (Section~\ref{subsub:guideline}) as the label oracle.
For CoExplain, we used the average performance acquired from the user study (Section~\ref{sec:evaluation}).
Fig.~\ref{fig:fine-tune-summary} shows that the Fine-tuned model has less alignment toward the Guideline goal (low Test Accuracy, high Edit Distance) than CoExplain, but rises as more Guideline data is labeled. Only when 32.17\% of the data (148 instances) are labeled, then would its performance match CoExplain.
Assuming labeling at a rate of 6.6 seconds per instance~\cite{arora2009estimating}, this would require 16.28 minutes of human effort, which is longer than CoExplain's 3.36 minutes.

We view Editable XAI as complementary to IML and other human-in-the-loop methods: user edits can refine reasoning structures, while data-driven techniques expand coverage in areas where rules are insufficient, and interpretable glass-box models can provide foundations for further editable adjustments~\cite{garrett2023interpretable}. Such hybrid workflows promise to balance expert alignment, efficiency, and robustness by combining the strengths of explanation editing and data-driven feedback.

% \subsubsection{Editable XAI and Large Language Models}

% Large language models (LLMs) provide alternative pathways for aligning AI with human knowledge, including fine-tuning~\cite{howard2018universal}, in-context learning~\cite{dong2022survey}, and prompt engineering~\cite{sahoo2024systematic}. While these approaches can produce flexible and context-sensitive outputs, they are inherently stochastic and may not consistently follow user-intended reasoning. LLM-generated explanations are also prone to hallucination~\cite{ji2023survey}, which can compromise fidelity and trustworthiness. Editable XAI offers a complementary approach: by representing reasoning as structured, manipulable concepts, it supports predictable bi-directional alignment and allows users to directly correct or refine AI behavior. \toreview{We demonstrate in Fig.~\ref{fig:LLM-concept} about how to utilize LLM and Editable XAI's strengths and integrate them together to reduce human effort. Fig.~\ref{fig:LLM} shows a comparison between LLM and CoExplain.} Future work could explore hybrid workflows that combine the flexibility of LLMs with the structured guarantees of editable explanations, potentially leveraging LLMs to propose edits or alternative rules while maintaining user oversight and control.

% \todo{missing how you give the LLM context of the rules and dataset. Appendix~\ref{Appendix:Prompt}}

\begin{figure}[t]
    \centering
    \includegraphics[width=8.5cm]{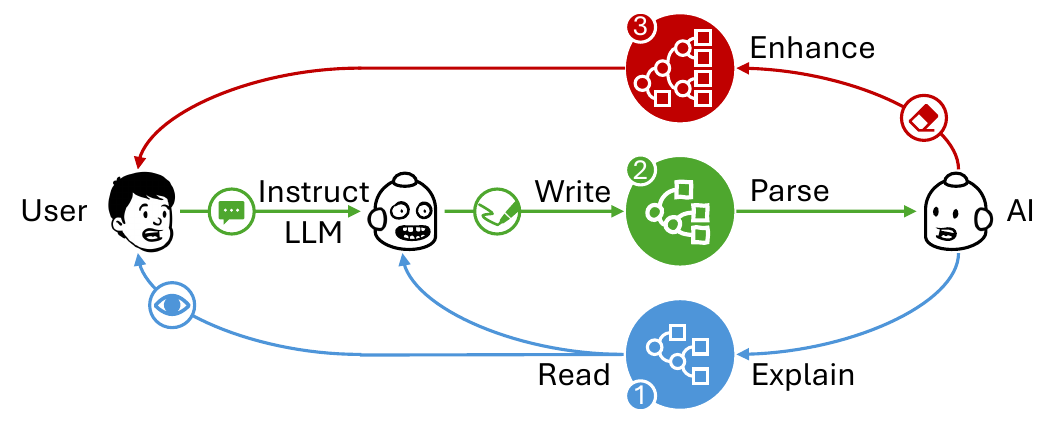}
    % \vspace{-0.8cm}
    \caption{
        Extension of conceptual overview of Fig.~\ref{fig:conceptual-overview} to include LLM usage in the workflow of CoExplain. 
        This embeds the user prompting an LLM read distilled rule explanations and suggest revisions to be parsed by Editable XAI.
    }
    \Description[Conceptual overview integrating LLM]{LLM is introduced to replace human direct editing, human on the left of the LLM agent can instruct LLM with natural language.}
    \label{fig:LLM-concept}
    % \vspace{-0.5cm}
\end{figure}

\subsubsection{Editable XAI and Large Language Models}
Large language models (LLMs) provide alternative pathways for aligning AI with human knowledge, including fine-tuning~\cite{howard2018universal}, in-context learning~\cite{dong2022survey}, and prompt engineering~\cite{sahoo2024systematic}. 
While these approaches can produce flexible and context-sensitive outputs, they are inherently stochastic and may not consistently follow user-intended reasoning. 
LLM-generated explanations are also prone to hallucination~\cite{ji2023survey}, which can compromise fidelity and trustworthiness. 

\begin{figure*}[t]
    \centering
    \includegraphics[width=15cm]{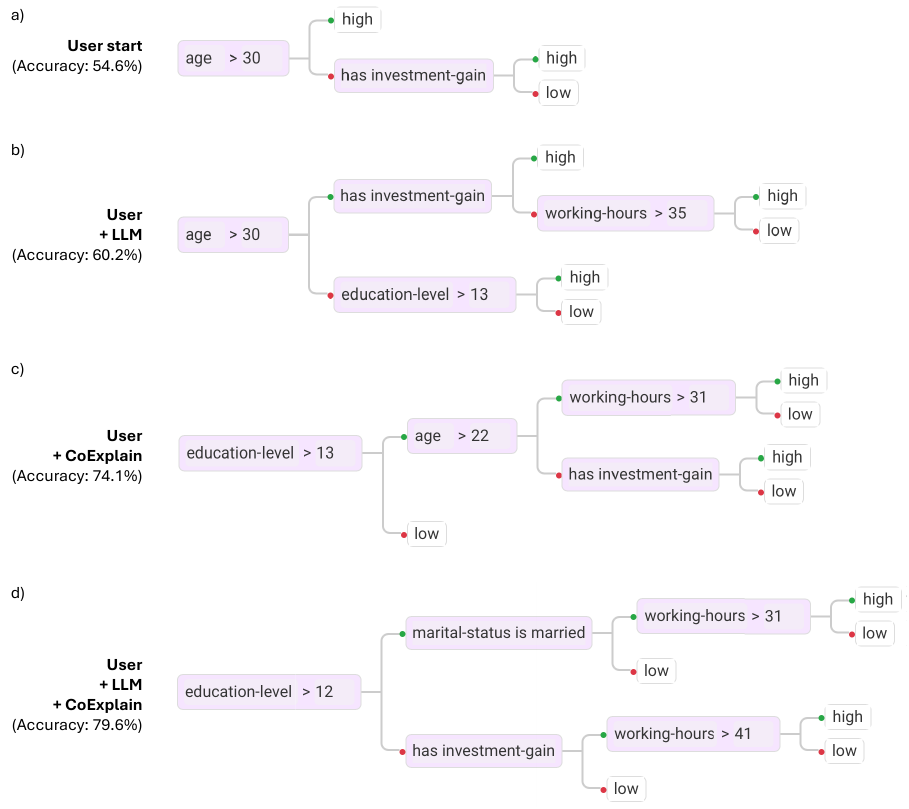}
    % \vspace{-0.3cm}
    \caption{
        Comparison of decision trees based on the Adult Income dataset as a demonstration.
        user 
        a) writes initial rules, 
        b) rewrites with LLM-prompted rule edits,
        c) receives rules after (a) enhanced by CoExplain, and
        d) receives rules after (b) enhanced by CoExplain.
        }
        \Description[Comparison of decision trees for Adult Income prediction]{Four decision trees (a–d) show rule evolution for Adult Income prediction (high/low income) with accuracy metrics:a) User start: Initial user rules (accuracy 54.6\%) start with age > 30 — predicts “high” (if true) or branches to has investment-gain (predicting “high”/“low”).b) User + LLM: User rules revised with LLM edits (accuracy 60.2\%) start with age > 30 — branches to has investment-gain (predicting “high” or working-hours > 35 for “high”/“low”) or education-level > 13 (predicting “high”/“low”).c) User + CoExplain: Rules from (a) enhanced by CoExplain (accuracy 74.1\%) start with education-level > 13 — predicts “low” (if false) or branches to age > 22 (predicting working-hours > 31 for “high”/“low” or has investment-gain for “high”/“low”).d) User + LLM + CoExplain: Rules from (b) enhanced by CoExplain (accuracy 79.6\%) start with education-level > 12 — branches to marital-status is married (predicting working-hours > 31 for “high”/“low” or “low”) or has investment-gain (predicting working-hours > 41 for “high”/“low” or “low”).
Accuracy increases across iterations (a → b → c → d), with each enhancement (LLM, CoExplain) refining rule structure.}
    \label{fig:LLM-results}
    % \vspace{-0.3cm}
\end{figure*}

Editable XAI offers a complementary approach: by representing reasoning as structured, manipulable concepts, it supports predictable bi-directional alignment and allows users to directly correct or refine AI behavior. 
LLMs can be incorporated into this workflow by providing them with contextual information about the dataset and existing rules, for example through prompt-based descriptions or in-context examples of the structured explanations (see Appendix~\ref{Appendix:Prompt}).
This enables the LLM to suggest edits, alternative rules, or explanations while staying aligned with the user's goals, explanatory schema, and data. 
Fig.~\ref{fig:LLM-concept} illustrates how combining LLMs with editable XAI can reduce human effort and support co-created editing goals.
% , and 
Fig.~\ref{fig:LLM-results} shows an exploratory results of the LLM-infused approach with enhanced tree explanations from CoExplain.

Future work could explore hybrid workflows to combine the flexibility of LLMs with the structured guarantees of Editable XAI, maintaining user control and leveraging LLM's suggestions.

\subsection{Collaboration for Bi-Directional Alignment}
Editable XAI positions explanation editing as a collaborative process. Beyond transparency, its value lies in how humans and AI negotiate shared control, adapt to each other, and co-construct reasoning over time.

\subsubsection{Bringing AI into the Human Editing Loop}
Editable XAI should be framed as a dialogic process, where explanations are shaped through iterative contributions from both human and AI~\cite{miller2019explanation}. Prior work in IML has relied heavily on human labor, such as labeling, curating, or correcting examples, while under-utilizing the AI’s capacity to optimize and propose alternatives. By enabling joint edits on a shared structure, CoExplain leverages both forms of expertise: users bring domain knowledge, while AI suggests simplifications, counterexamples, or refinements grounded in data. This transforms explanations from static artifacts into evolving negotiation spaces, supporting not just model transparency but also mutual adaptation between human reasoning and machine optimization~\cite{madumal2018towards, liao2025continual}.

% \begin{table}[h]
% \centering
% \renewcommand{\arraystretch}{1.6}
% \begin{tabular}{c|c|c}
%     & \textbf{Explanation Correct} & \textbf{Explanation Wrong} \\ \hline
% \textbf{AI Correct} &
% \begin{minipage}[c][2.2cm][c]{4.3cm}
% \centering
% (A) Both Correct\\
% \emph{Notes here...}
% \end{minipage}
% &
% \begin{minipage}[c][2.2cm][c]{4.3cm}
% \centering
% (B) Right for Wrong Reasons\\
% \emph{Notes here...}
% \end{minipage}
% \\ \hline
% \textbf{AI Wrong} &
% \begin{minipage}[c][2.2cm][c]{4.3cm}
% \centering
% (C) Model Wrong, Explanation Correct\\
% \emph{Notes here...}
% \end{minipage}
% &
% \begin{minipage}[c][2.2cm][c]{4.3cm}
% \centering
% (D) Both Wrong\\
% \emph{Notes here...}
% \end{minipage}
% \\
% \hline
% \end{tabular}
% \caption{Four-case matrix comparing AI correctness and explanation correctness.}
% \end{table}

\subsubsection{Human Mistakes and Quality of Input}
A central design challenge lies in managing the quality of human contributions. Users may provide incomplete, biased, or erroneous rules, which could propagate harmful reasoning into the system if incorporated uncritically~\cite{hilbert2012toward, pronin2007perception}. Editable XAI offers a partial safeguard by making the editable medium shared: in CoExplain, both human and AI operate on the same decision tree explanation, allowing the AI to correct mistakes through data-driven optimization, or suggest alternative edits to improve low-quality inputs.  
Still, additional challenges remain. One risk is overconfidence, where users trust their expertise even when it is limited, potentially leading to rigid or incorrect edits~\cite{chong2022human}. Another is over-reliance, where users defer excessively to the AI and accept its suggestions uncritically~\cite{buccinca2021trust}. While Editable XAI can alleviate these effects by exposing reasoning structures and making corrections visible, it cannot resolve them entirely. Addressing them will also require drawing on findings from empirical XAI research on human–AI trust~\cite{van2025selective,pareek2025s} and confidence~\cite{li2025confidence} to design safeguards that balance user agency with error prevention.  

\subsubsection{Balancing Initiative in Collaboration}
Editable XAI frames collaboration through a shared editable medium, but bi-directional alignment also depends on calibrating initiative across different stages of interaction~\cite{gebreegziabher2025supporting}. Initiative shapes not only how edits unfold, such as when humans should take the lead in editing and when the AI should propose refinements, but also how collaboration begins, such as whether rules are first authored by humans, generated by the AI, or co-constructed from partial contributions~\cite{he2025contributions}. Our study focused on the human-first setting, where users defined rules and the AI adapted through optimization. Other starting points raise new opportunities and concerns. AI-initialized rules may accelerate entry into complex domains, but risk biasing users toward machine suggestions~\cite{pataranutaporn2025synthetic}, while human-initialized rules foreground expertise but may overlook patterns that are only visible in data~\cite{gomez2024human}. Developing mechanisms to balance initiative at both the outset and during ongoing edits remains central to positioning Editable XAI as a genuine collaboration rather than a one-sided process.

\subsection{Limitations and Future Work}
Our work has several limitations.
First, CoExplain currently uses decision tree rules as the interaction medium, which effectively links human reasoning with model optimization, but suffers from loss of details like other forms of easily understandable explanations~\cite{vonder2023analysis}. 
Future work may extend Editable XAI to additional knowledge representations, such as rule sets or heuristic non-linear equations to address broader cases.

Second, we evaluated CoExplain only on structured inputs with small models, whose attributes align well with users' deliberative reasoning. The method does not directly extend to perception tasks (e.g., vision, audio) or language reasoning, which involve innate mental processes due to stimuli or low-level, practiced skills. 
Future work can 
% We plan to generalize to 
integrate other modalities through feature extraction and concept-based explanations (Section~\ref{subsub:modality}) and to larger models through modularization and abstraction (Section~\ref{subsub:scale}).

Third, although our user study included a heart disease task and several participants with medical backgrounds, we focused on usage by lay users. Future work should investigate how domain experts would use Editable XAI or what schemas they would wish to edit with. 
% expand to more domains with expert users. 
Different fields encode expertise in distinct forms, and enabling editing through domain-familiar representations may reduce cognitive load and support more trustworthy outcomes~\cite{lim2025diagrammatization}.

\section{Conclusion}
We introduced \textbf{Editable XAI} through CoExplain, a neurosymbolic framework that allows users to read, write, and enhance rule-based explanations. In a study with 43 participants, Editable XAI improved understanding and alignment compared to read-only explanations, while CoExplain further reduced effort and balanced user knowledge with near-optimal performance. These results highlight the value of writable explanations as a means of fostering bi-directional alignment and active engagement. Future work can extend Editable XAI to other explanation forms and domains, advancing more collaborative and controllable human–AI interaction.

\begin{acks}
This research is supported by the National Research Foundation, Singapore and Infocomm Media Development Authority under its Trust Tech Funding Initiative (Grant No. DTC-RGC-09), and by the Ministry of Education, Singapore (Grant No. MOE-T2EP20121-0010). Any opinions, findings and conclusions or recommendations expressed in this material are those of the author(s) and do not reflect the views of National Research Foundation, Singapore and Infocomm Media Development Authority.
\end{acks}

% \clearpage
% \balance
\bibliographystyle{ACM-Reference-Format}
\bibliography{reference}

\clearpage
\appendix
\onecolumn
\renewcommand{\thefigure}{A.\arabic{figure}}
\setcounter{figure}{0} % restart figure counter

%TC:ignore
\section{Appendix}

\subsection{Data Partitioning}
\label{Appendix:partition}
To provide participants with a unified baseline of knowledge across the three scenarios, we partitioned each dataset into two distributions: a \textbf{Guideline Distribution} and a \textbf{pre-trained Distribution}. The guideline distribution was used as the knowledge source for participants to construct their decision rules. This mimics the partial, textbook-like knowledge that cannot be directly applied to real-world practice. In contrast, the pre-trained distribution was used to train the AI model and was not revealed to participants, representing real-world data that cannot be fully anticipated from the guideline distribution.

This partitioning created a deliberate misalignment between participants’ reasoning and the AI model’s prediction behavior. Our design follows prior work on tackling Out-of-Distribution (OOD) challenges in human-AI collaboration ~\cite{liu2021towards, chiang2021you, kulinski2023towards, liu2021understanding, lai2022human}. We split by gender for \textit{Adult Income} and \textit{Heart Disease}, and by age (below/above 50) for \textit{House Price}. Distribution shifts were quantified via feature-wise Wasserstein distances and label KL divergence. \textit{Adult Income} showed moderate shifts (Wasserstein = 0.4844, KL = 0.1412). \textit{House Price} showed the largest feature shift (Wasserstein = 0.8587) but minimal label shift (KL = 0.0188), reflecting age-related housing differences without major price change. \textit{Heart Disease} showed the smallest feature shift (Wasserstein = 0.3003) but the largest label shift (KL = 0.3092), reflecting gender-based prevalence differences with stable feature distributions. Together, these datasets cover different types of distribution shifts.

\clearpage
\subsection{Demonstration of Applying CoExplain on ECG Signal Data with Feature Extraction}
\label{Appendix:ECG-demo}
We implement and generalize our CoExplain explanation method to ECG signal data as a demonstration with feature extraction.

\subsubsection{Dataset} PTB-XL~\cite{wagner2020ptb} contains 21837 multi-labeled 10-second ECG records gathered from 18885 patients. In total, PTB-XL has 71 labels consisting of diagnostic labels describing specific CVDs associated with the ECG, form labels describing ECG’s morphology, and cardiac
axis labels.

\subsubsection{Feature Extraction}
We extract interpretable ECG features using NeuroKit2~\cite{makowski2021neurokit2}, following standard clinical and signal-processing conventions. The pipeline computes time-domain intervals (e.g., RR, PR, QRS, QT), frequency-domain heart-rate–variability (HRV) metrics, and morphology-based measurements (e.g., amplitudes and segment deviations), all of which can be directly incorporated into CoExplain as editable concepts. Details of extracted features can be found in Table~\ref{tab:ECG}.

\subsubsection{Domain Rules}
\begin{figure*}[htb]
    \centering
    \includegraphics[width=0.5\linewidth]{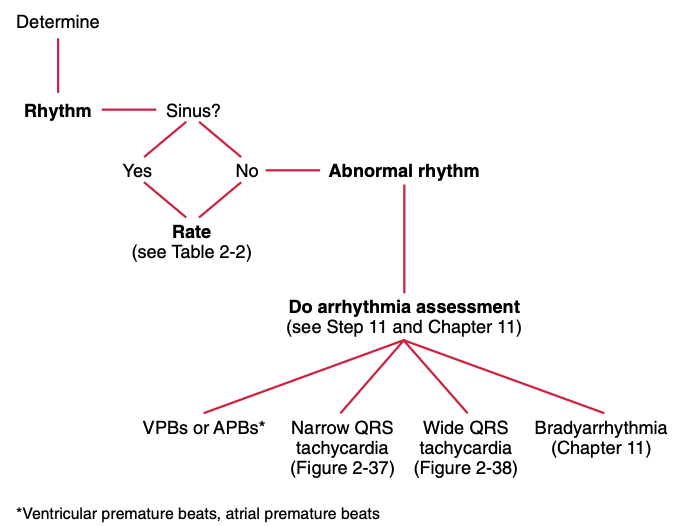}
    \caption{Example of ECG step rules used in \textit{Rapid ECG Interpretation} (MG Khan, 2008, p. 36)~\cite{khan2008rapid}. Step-by-step method for accurate ECG interpretation. Step 1: Assess rhythm and rate.}
    \label{fig:ECG-rule}
    \Description[ECG step rules for rhythm and rate assessment]{Flowchart outlining Step 1 (Assess rhythm and rate) of ECG interpretation from *Rapid ECG Interpretation* (2008). The process starts with determining rhythm: first, check if the rhythm is sinus. If “Yes”, assess the rate (referencing Table 2-2). If “No” (abnormal rhythm), proceed to an arrhythmia assessment (referencing Step 11 and Chapter 11), which branches into subcategories: VPBs or APBs (ventricular/atrial premature beats), narrow QRS tachycardia (Figure 2-37), wide QRS tachycardia (Figure 2-38), and bradyarrhythmia (Chapter 11). This step-by-step method guides accurate ECG rhythm classification.}
\end{figure*}

We constructed the decision tree rules for ECG diagnosis from \textit{Rapid ECG Interpretation} by MG Khan~\cite{khan2008rapid}. ECG diagnosis are organized into 9 interconnected steps, we converted the 9 steps flowcharts into 9 decision tree rules, each used to train a specific module of the ECG diagnosis network. Fig.~\ref{fig:ECG-rule} shows an example of the ECG diagnosis rules.

\subsubsection{Model Organization}
For each of the 9 steps, we have one neural network regularized by one decision tree rule to predict certain diagnosis with specific concept input features, steps dependent of others steps are interconnected and takes the previous steps' output as input.

\subsubsection{Training and Predicting}
We initialize the step modules with the parsing algorithm, and treat the ECG rules from~\cite{makowski2021neurokit2} as the source of human edit, and regularize the training of the neural networks with them. The prediction task performed was a multi-label task and a
total of 21 diagnoses were considered, models trained with Adam optimizer for 50 epochs. After testing on the test set, CoExplain have an accuracy of 92.02\% with an AUROC of 0.8360. 

\begin{table*}[t]
\centering
\caption{ECG Features, Abbreviations, and Explanations}
\label{tab:ECG}
\begin{tabular}{p{3.2cm} p{2.5cm} p{8cm}}
\toprule
\textbf{Feature Name} & \textbf{Abbreviation} & \textbf{Explanation} \\
\midrule
Heart Rate & HR & Heart Rate of the patient \\
Bradycardia & BRAD & Whether the patient has bradycardia (HR < 60 bpm) \\
Tachycardia & TACH & Whether the patient has tachycardia (HR > 100 bpm) \\
Sinus & SINUS & Whether the rhythm is sinus: Each P wave in lead II should be positive AND precedes a QRS complex \\
RR interval range & RR\_DIFF & max R--R interval – min R--R interval \\
PR duration & PR\_DUR & Duration of the PR segment \\
Prolonged PR & LPR & Whether the PR interval is prolonged \\
QRS duration & QRS\_DUR & Duration of the QRS complex \\
Prolonged QRS & LQRS & Whether the QRS complex is prolonged \\
Prolonged QRS for WPW & LQRS\_WPW & Whether the QRS complex is prolonged by WPW’s standards \\
Short PR & SPR & Whether the PR interval is shortened \\
ST segment amplitude & ST\_AMP\_x & Mean amplitude of ST segment in lead x \\
ST Elevation & STE\_x & Whether the ST segment is elevated in lead x \\
ST Depression & STD\_x & Whether the ST segment is depressed in lead x \\
Poor R-wave Progression & PRWP & Whether R waves are not within desired ranges for at least one lead in V1--V4 \\
Q wave duration & Q\_DUR\_x & Duration of the Q wave in lead x \\
Q wave amplitude & Q\_AMP\_x & Amplitude of Q wave in lead x \\
Pathological Q wave & PATH\_Q\_x & Whether the Q wave in lead x is pathological \\
P wave duration & P\_DUR\_x & Duration of P wave in lead x \\
P wave amplitude & P\_AMP\_x & Amplitude of P wave in lead x \\
Prolonged P wave & LP\_x & Whether the P wave is prolonged in lead x \\
Peaked P wave & PEAK\_P\_x & Whether the P wave is peaked (high amplitude) in lead x \\
Age & AGE & Age of the patient \\
Old age & AGE\_OLD & Whether the patient’s age is greater than 30 \\
Male & MALE & Whether the patient is male \\
R wave amplitude & R\_AMP\_x & Amplitude of R wave in lead x \\
S wave amplitude & S\_AMP\_x & Amplitude of S wave in lead x \\
R/S Ratio & RS\_RATIO\_x & Ratio between amplitudes of R and S waves in lead x \\
Peaked R wave & PEAK\_R\_x & Whether the R wave is peaked in lead x \\
Deep S wave & DEEP\_S\_x & Whether the S wave is deep (low amplitude) in lead x \\
Dominant R wave & DOM\_R\_x & Whether R wave amplitude is greater than that of the S wave \\
Dominant S wave & DOM\_S\_x & Whether S wave amplitude is greater than that of the R wave \\
T wave amplitude & T\_AMP\_x & Amplitude of T wave in lead x \\
Inverted T wave & INVT\_x & Whether the T wave is inverted in lead x \\
Sum of QRS & QRS\_SUM & The QRS area above the baseline minus the QRS area below \\
Positive QRS & POS\_QRS\_x & The QRS is positive in lead x \\
Normal cardiac axis & NORM\_AXIS & Whether the patient has a normal cardiac axis \\
Left axis deviation & LAD & Whether the patient’s cardiac axis deviates toward the left \\
Right axis deviation & RAD & Whether the patient’s cardiac axis deviates toward the right \\
\bottomrule
\end{tabular}
\end{table*}

\clearpage
\subsection{Prompt from Comparing LLM with CoExplain}
\label{Appendix:Prompt}
\paragraph{Prompt}
\noindent
Do not use any previous chat’s memory; only provide an output in the input form of JSONLogic.  
Suggest an enhancement on this rule to improve accuracy on the attached dataset.  
While staying faithful and making changes minimal, you are not required to preserve the rule exactly.  
You may refine thresholds, modify structure, prune branches, or extend the rule with attributes from the dataset.

\vspace{0.75em}
\paragraph{Input}
\begin{verbatim}
{"if":[
  {">":["marital-status",0.5]},
  {"if":[
    {">":["education-level",13]},
    "high",
    {"if":[{">":["working-hours",36]},"high","low"]}
  ]},
  {"if":[
    {">":["investment-gain",0.5]},
    {"if":[
      {">":["education-level",13]},
      {"if":[{">":["age",30]},"high","low"]},
      "low"
    ]},
    {"if":[{">":["working-hours",36]},"high","low"]}
  ]}
]}
\end{verbatim}

\vspace{0.75em}
\paragraph{Output}
\begin{verbatim}
{"if":[
  {">":["marital-status",0.5]},
  {"if":[
    {">":["education-level",12]},
    "high",
    {"if":[
      {">":["working-hours",35]},
      {"if":[{">":["age",28]},"high","low"]},
      "low"
    ]}
  ]},
  {"if":[
    {">":["investment-gain",0]},
    {"if":[
      {">":["education-level",12]},
      {"if":[
        {">":["working-hours",40]},
        "high",
        {"if":[{">":["age",32]},"high","low"]}
      ]},
      "low"
    ]},
    {"if":[{">":["hours-per-week",38]},"high","low"]}
  ]}
]}
\end{verbatim}

\clearpage
\subsection{Edit History Snapshots}

\begin{figure*}[h]
    \centering
    \includegraphics[width=0.6\linewidth]{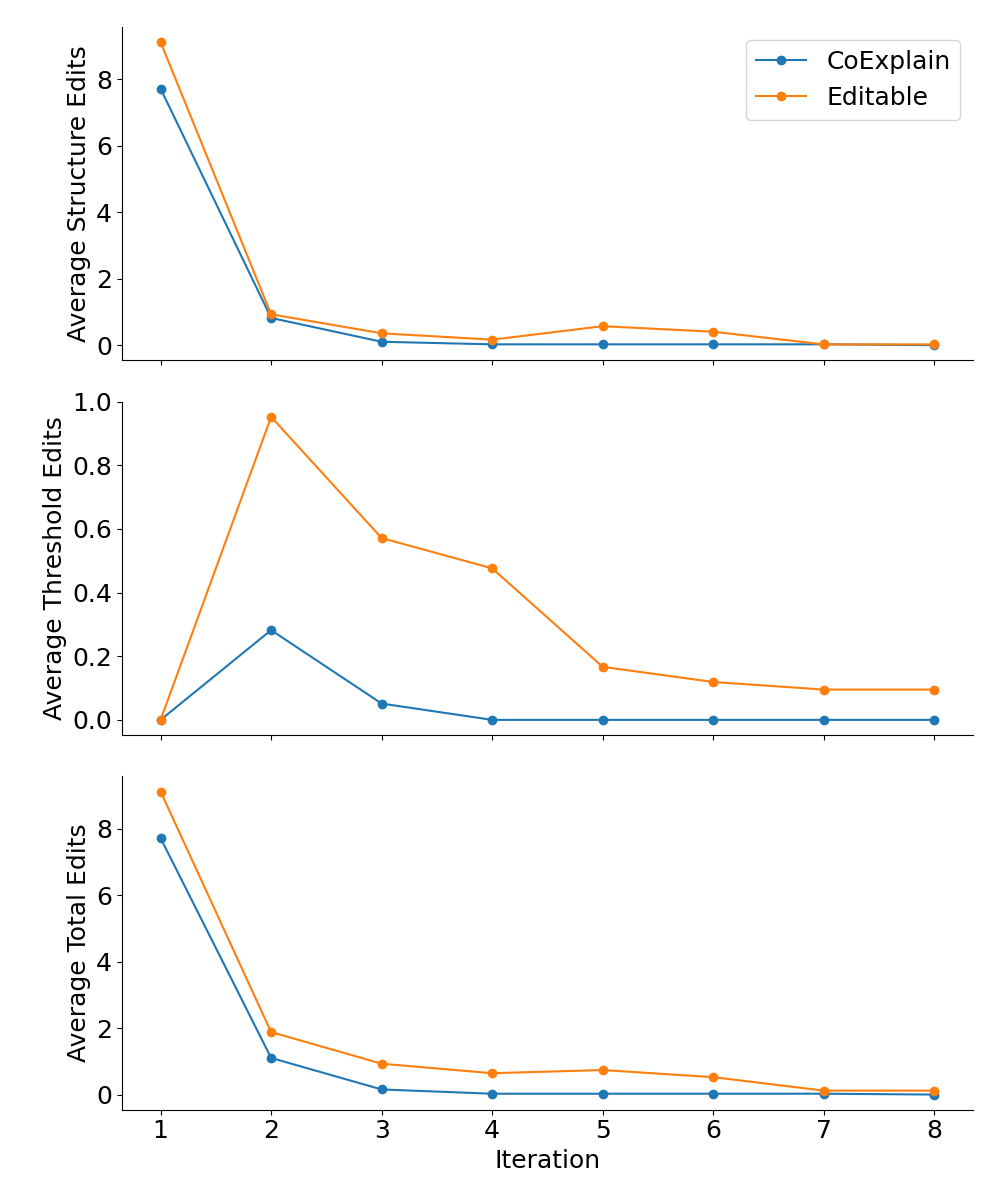}
    \caption{Human editing effort from Editable and CoExplain}
    \label{fig:history}
    \Description[Human editing effort for Editable and CoExplain]{Three plots (stacked vertically) show human editing effort across 8 iterations for “CoExplain” (blue line) and “Editable” (orange line): 1. Average Structure Edits: Both start with high edits (CoExplain ~8, Editable ~9), then drop sharply by Iteration 2 and stabilize near 0 for subsequent iterations.  2. Average Threshold Edits: Editable starts at 0, peaks near 1.0 at Iteration 2, then declines gradually; CoExplain starts at 0, rises slightly at Iteration 2, then drops to 0 and stays stable.  3. Average Total Edits: Mirrors the structure edits trend: high initial edits (CoExplain ~8, Editable ~9), a sharp drop by Iteration 2, and stabilization near 0 for later iterations. The plots illustrate that editing effort (structure, threshold, total) is concentrated in early iterations, with minimal effort required after Iteration 2 for both interfaces.}
\end{figure*}

Fig.~\ref{fig:history} shows the difference on human editing effort between Editable and CoExplain, we report the averaged structure editing operations, threshold editing operations and total editing operations per edit iteration. Results suggest that most editing operations are on structure, while threshold edits comes in later iterations as minor refinements, Editable users adjust their threshold more frequently than CoExplain users. Editable users takes more iterations than CoExplain users to finalize their model, and continue to make minor changes on thresholds during later iterations.

\begin{figure*}[h]
    \centering
    \includegraphics[width=0.8\linewidth]{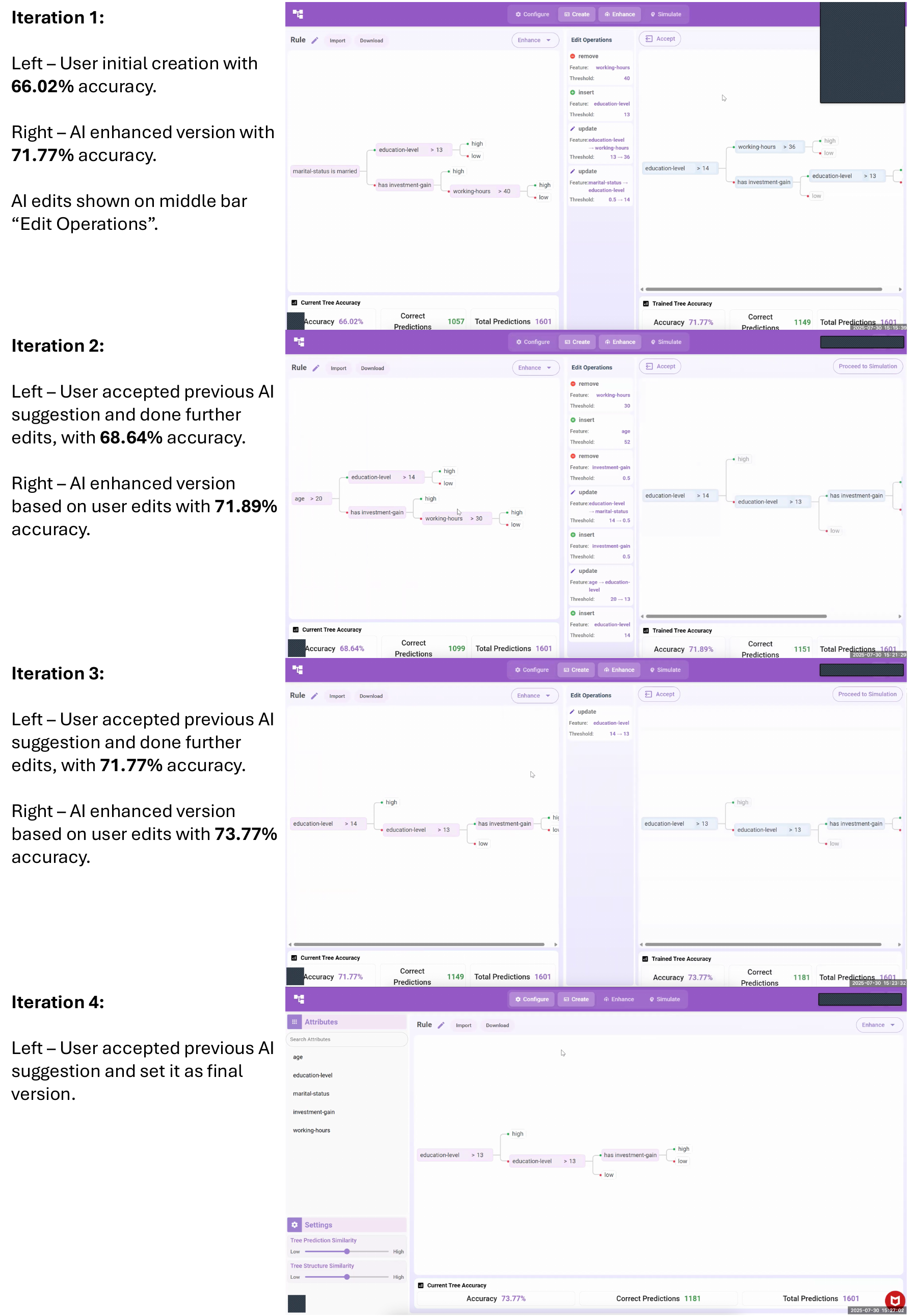}
    \caption{Edit history snapshot from C15 on adult income task.}
    \label{fig:snapshot}
    \Description[Edit history snapshot for Adult Income task (C15)]{A step-by-step edit history (4 iterations) for the Adult Income task, showing user and AI rule versions with accuracy metrics:  
Iteration 1: Left = User’s initial rule (66.02\% accuracy); Right = AI-enhanced rule (71.77\% accuracy). AI edits are listed in a “Edit Operations” panel.  
Iteration 2: Left = User’s rule (accepted prior AI suggestions + further edits, 68.64\% accuracy); Right = AI-enhanced rule (based on user edits, 71.89\% accuracy).  
Iteration 3: Left = User’s rule (accepted prior AI suggestions + further edits, 71.77\% accuracy); Right = AI-enhanced rule (based on user edits, 73.77\% accuracy).  
Iteration 4: User accepts the prior AI suggestion as the final AI version (73.77\% accuracy).  

Each iteration pairs the user’s current rule (left) with the AI’s enhanced version (right), tracking accuracy improvements and edit workflows via interface screenshots (including rule visualizations, accuracy stats, and edit logs).}
\end{figure*}

Fig.~\ref{fig:snapshot} shows an editing history snapshot from the user study with C15, who spent 4 iterations with CoExplain to reach a final decision tree rule. From the edit history, user started with a self-created decision tree rule of relatively low accuracy of 66.02\%, and used the CoExplain enhancement to refine the rules, AI edits are shown on the middle bar of the screenshot. For the first iteration, AI made 4 edits on C15's creation, and increased the accuracy to 71.77\%, without drastic change on user's original creation. C15 accepted CoExplain's editing and choose to further refine on the AI's creation, together with AI enhancement, the accuracy gradually increased to 73.77\%, and C15 finalized the model after the third iteration.

\clearpage
\subsection{Read-only Explanations}
\label{Appendix:readonly}

\begin{figure}[htb]
    \centering
    \includegraphics[width=0.7\linewidth]{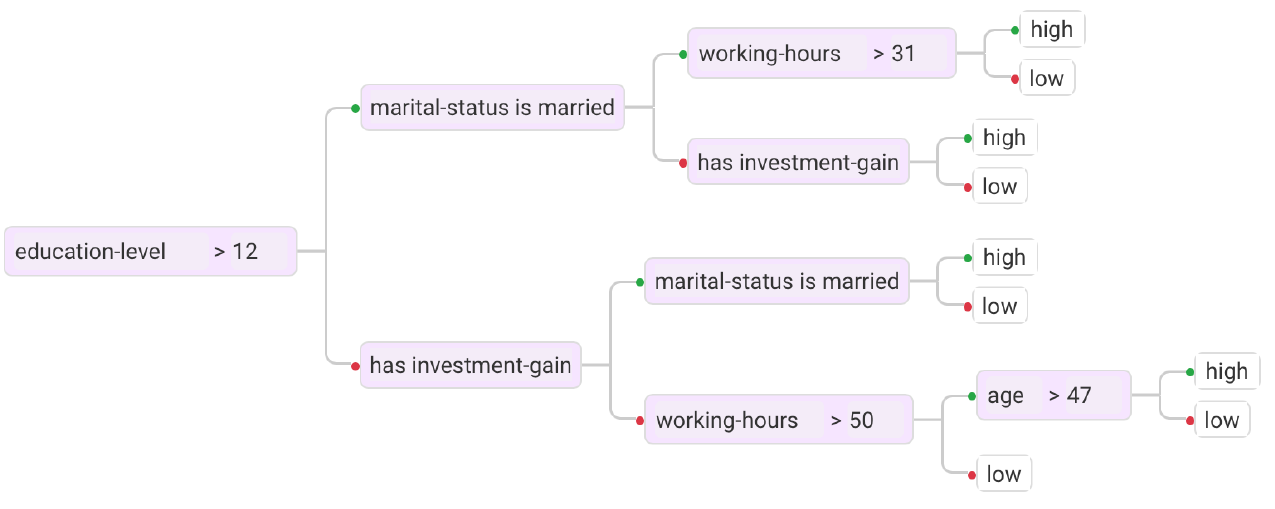}
    \caption{Read-only explanation for Adult Income Prediction Task.}
    \label{fig:readonly-income}
    \Description[Read-only explanation for Adult Income Prediction Task]{Fully described in the caption.}
\end{figure}

\begin{figure}[htb]
    \centering
    \includegraphics[width=0.7\linewidth]{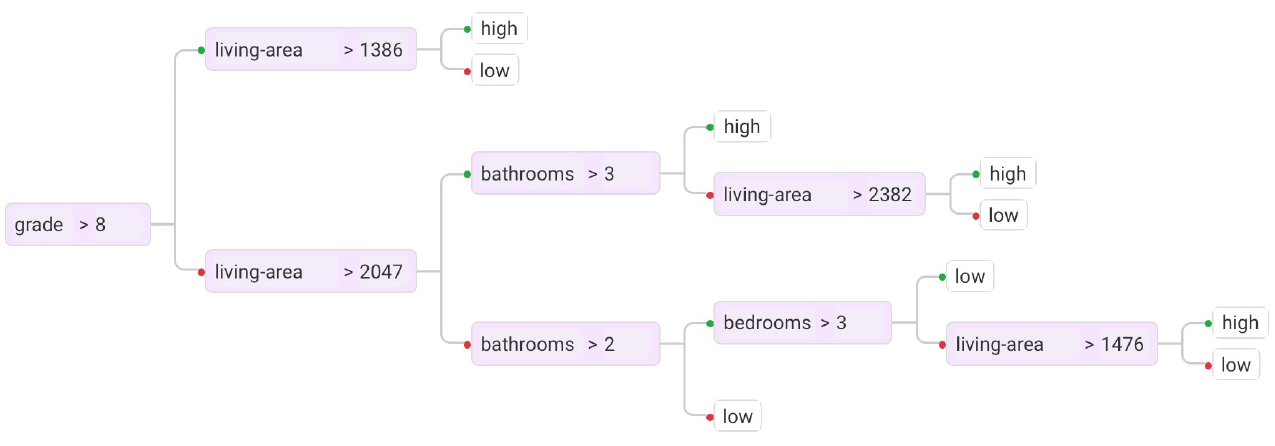}
    \caption{Read-only explanation for House Price Prediction Task.}
    \label{fig:readonly-house}
    \Description[Read-only explanation for House Price Prediction Task]{Fully described in the caption.}
\end{figure}

\begin{figure}[htb]
    \centering
    \includegraphics[width=0.7\linewidth]{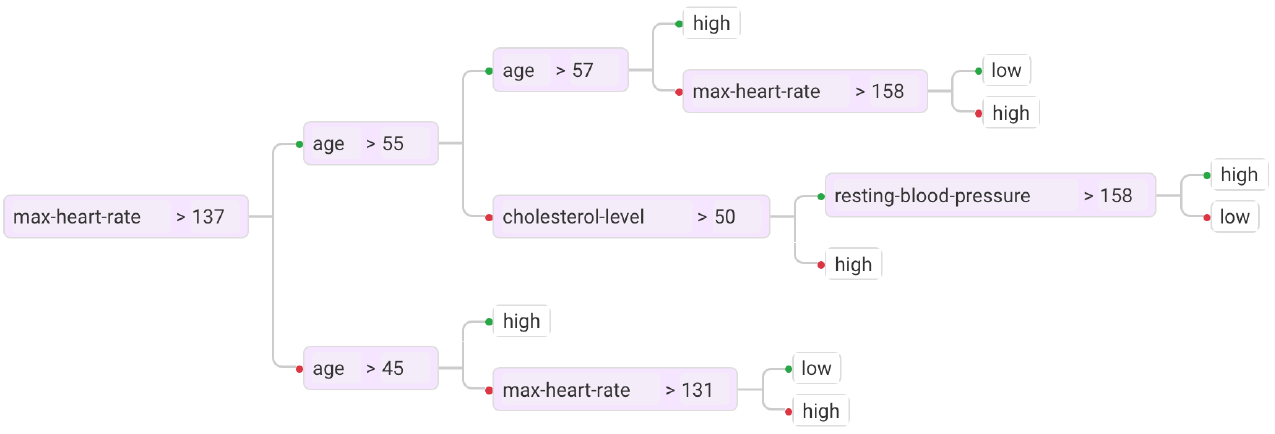}
    \caption{Read-only explanation for Heart Disease Prediction Task.}
    \label{fig:readonly-heart}
    \Description[Read-only explanation for Heart Disease Prediction Task]{Fully described in the caption.}
\end{figure}

\clearpage
\subsection{Examples of User Created Decision Tree Rules}

\begin{figure}[htb]
    \centering
    \includegraphics[width=0.8\linewidth]{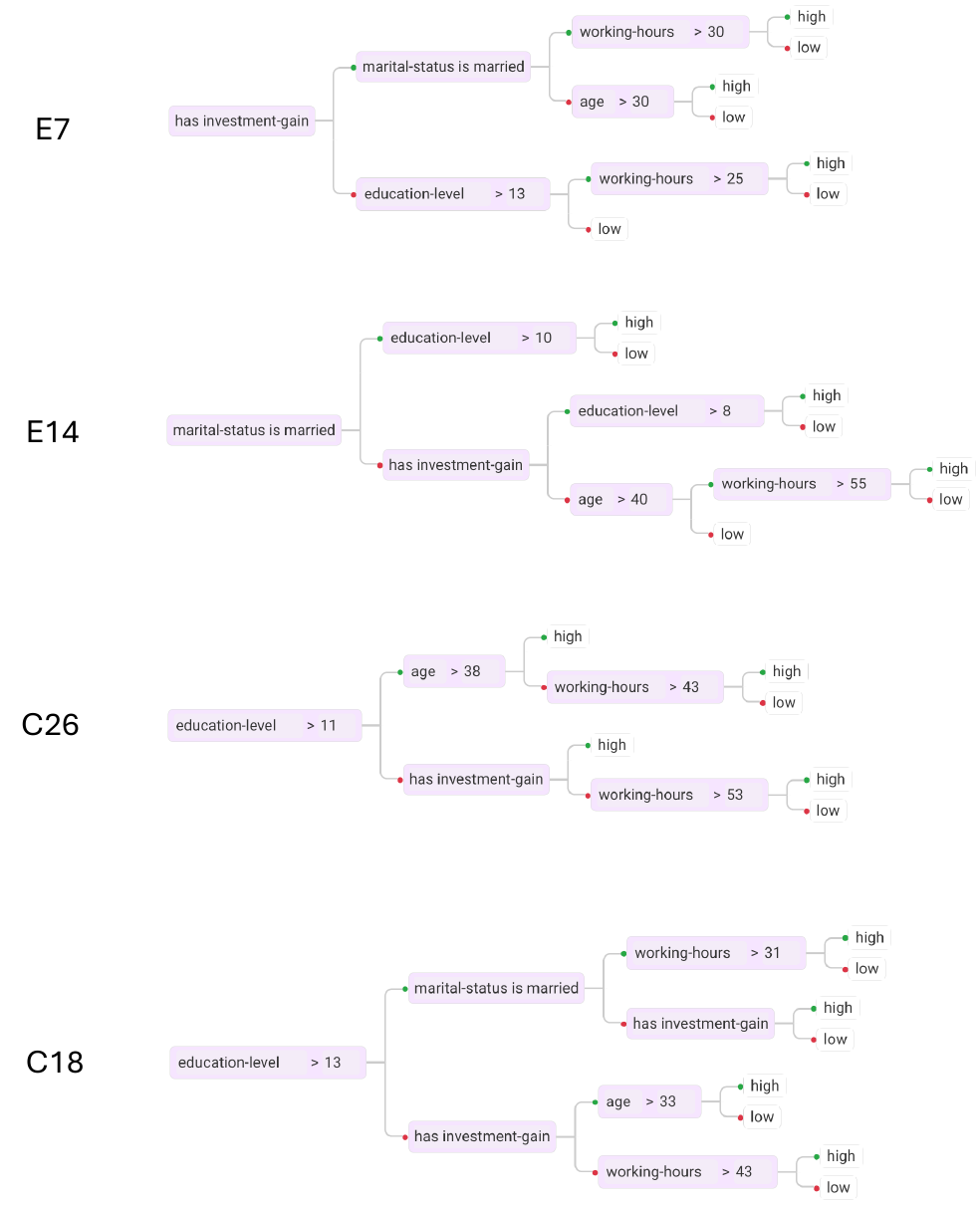}
    \caption{Examples of user created decision tree rules on the Adult Income Prediction Task}
    \label{fig:user-income}
    \Description[User created rules for Adult Income Prediction Task]{Two examples from Editable and two from CoExplain.}
\end{figure}

\begin{figure}[htb]
    \centering
    \includegraphics[width=0.85\linewidth]{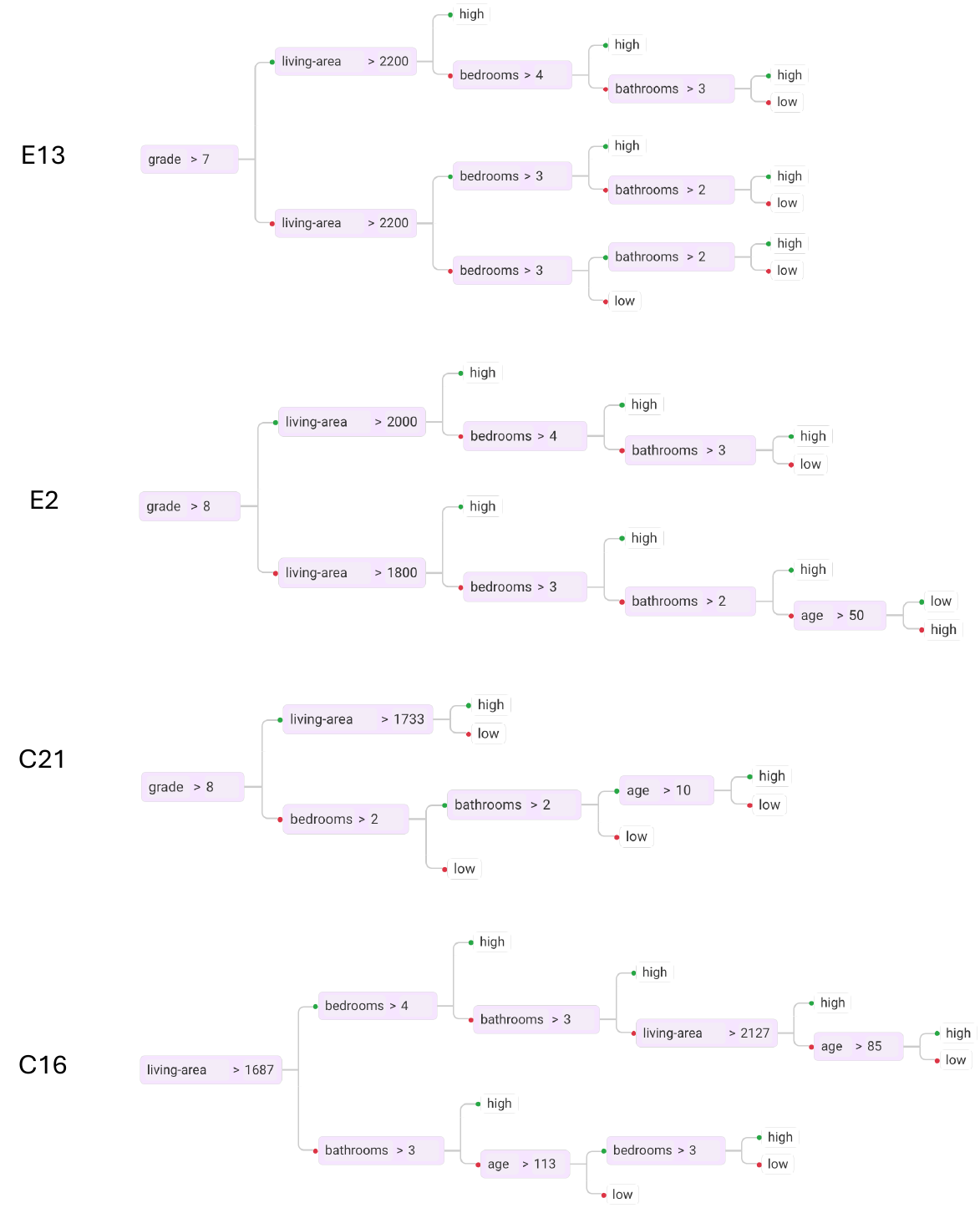}
    \caption{Examples of user created decision tree rules on the House Price Prediction Task}
    \label{fig:user-house}
    \Description[User created rules for House Price Prediction Task]{Two examples from Editable and two from CoExplain.}
\end{figure}

\begin{figure}[htb]
    \centering
    \includegraphics[width=0.95\linewidth]{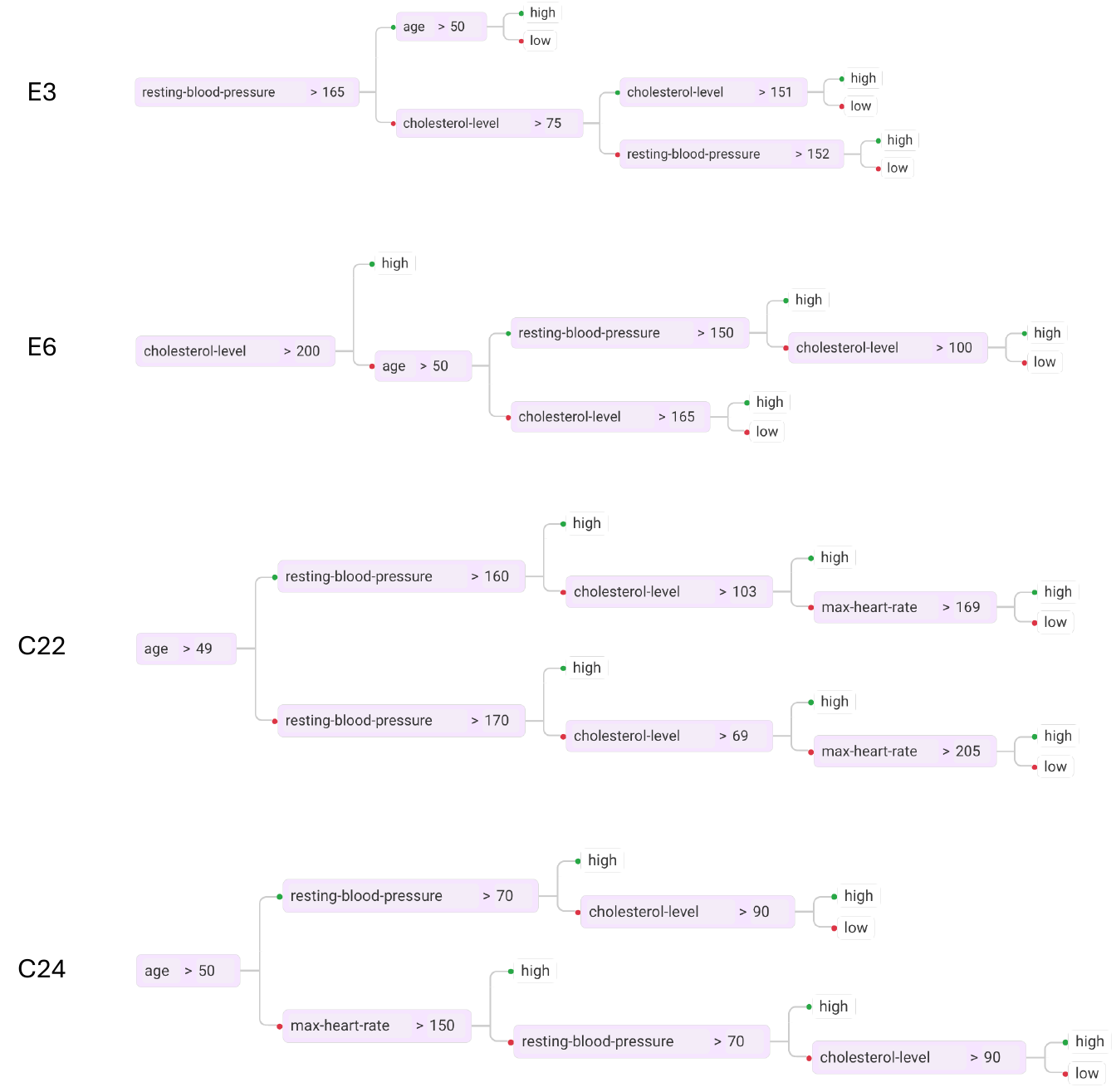}
    \caption{Examples of user created decision tree rules on the Heart Disease Prediction Task}
    \label{fig:user-heart}
    \Description[User created rules for Heart Disease Prediction Task]{Two examples from Editable and two from CoExplain.}
\end{figure}

\clearpage
\subsection{Survey for the User Study}
\label{Appendix:survey}

\begin{figure}[h]
    \centering
    \includegraphics[width=0.8\linewidth]{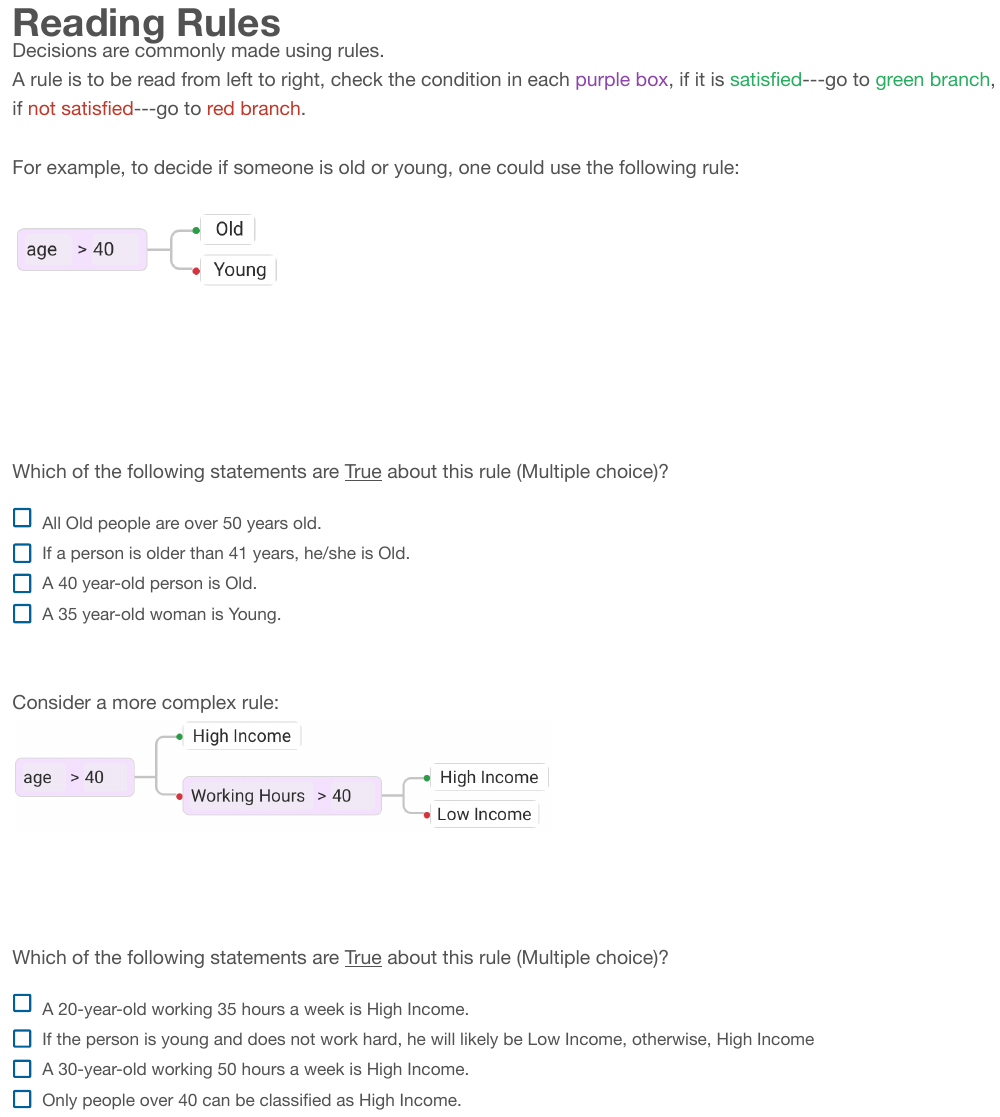}
    \caption{Tutorial on reading the rules and screening questions to check the users' interpretation.}
    \label{fig:tutorial-rule}
    \Description[Tutorial on reading decision rules with screening questions]{Tutorial titled “Reading Rules” explaining how to interpret decision rules: rules are read left to right, with conditions in purple boxes—satisfied conditions follow the green branch, unsatisfied follow the red branch. A simple example uses “age greater than 40” to classify “Old” (green branch) or “Young” (red branch), followed by multiple-choice questions about this rule (e.g., “If a person is older than 41 years, he/she is Old”). A more complex example uses “age greater than 40” and “Working Hours greater than 40” to classify “High Income” or “Low Income”, with another set of multiple-choice questions (e.g., “A 30-year-old working 50 hours a week is High Income”). The figure includes decision tree diagrams for both examples and screening questions to check user interpretation.}
\end{figure}

\aptLtoX{\begin{figure}
    \centering
    \includegraphics[width=0.8\linewidth]{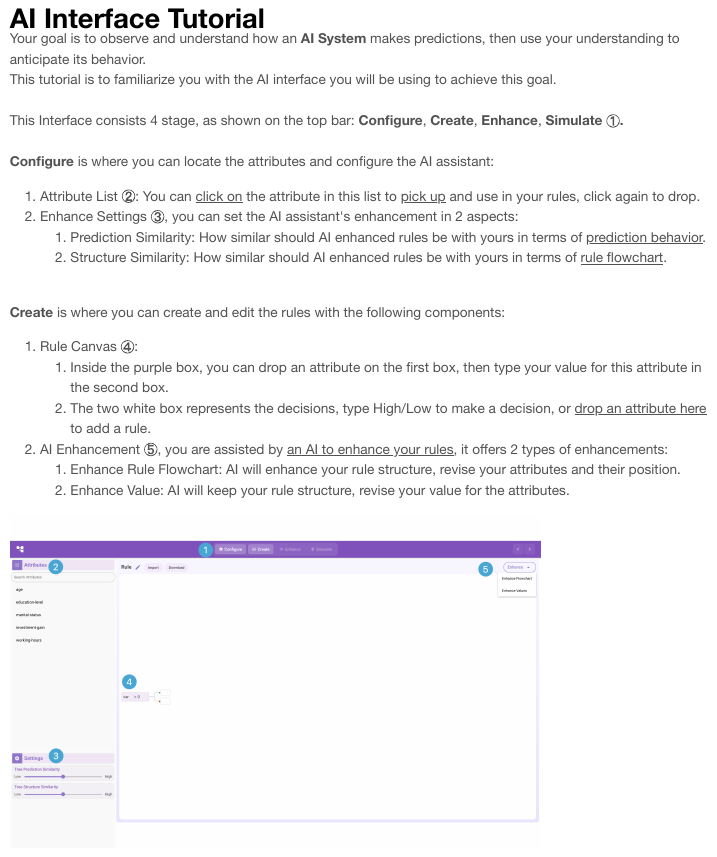}
    \includegraphics[width=0.8\linewidth]{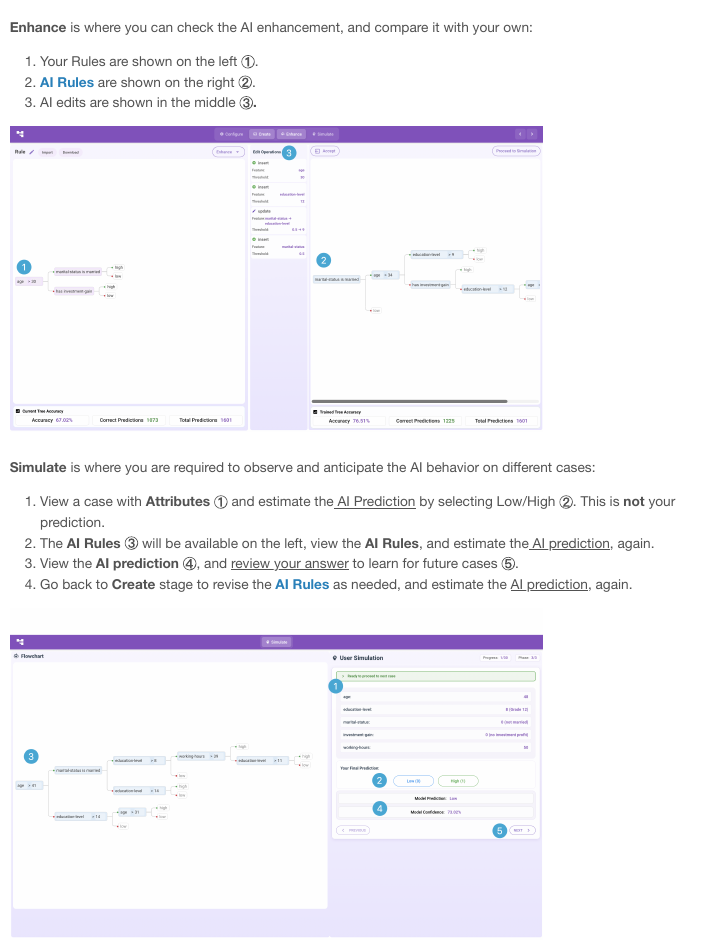}
    \caption{Tutorial on the AI interface.}
    \label{fig:tutorial-interface-a}
    \Description[AI Interface Tutorial]{The tutorial introduces an AI interface with four stages in a top bar: Configure, Create, Enhance, Simulate. In the Configure stage, there is an Attribute List where attributes can be clicked to pick up for use in rules (and clicked again to drop), and Enhance Settings to adjust “Prediction Similarity” (how similar AI-enhanced rules’ prediction behavior is to the user’s) and “Structure Similarity” (how similar AI-enhanced rules’ flowchart structure is to the user’s). The Create stage includes a Rule Canvas where attributes are dropped into a purple box (with a value typed in a second box), and white boxes accept “High/Low” decisions or additional attributes to extend rules; it also offers AI Enhancement with two options: “Enhance Rule Flowchart” (AI revises rule structure, attributes, and their positions) and “Enhance Value” (AI keeps structure but revises attribute values). A screenshot below the text labels interface elements (numbered 1–5) matching these components: the top bar stages, attribute list, enhance settings, rule canvas, and AI enhancement menu. The "Enhance" stage allows users to check AI enhancements and compare them with their own rules: user rules are shown on the left, AI rules on the right, and AI edits in the middle. The "Simulate" stage requires users to observe and anticipate AI behavior for different cases: view a case’s attributes and estimate the AI prediction by selecting Low/High (not the user’s own prediction); AI rules are available on the left to re-estimate the prediction; then view the actual AI prediction and review the answer to learn for future cases. Users can also return to the "Create" stage to revise AI rules and re-estimate predictions. Screenshots depict these stages: the "Enhance" interface shows side-by-side visualizations of user and AI rules with performance metrics, while the "Simulate" interface includes case attribute displays, prediction selection buttons, AI rule visualizations, and prediction feedback elements.}
\end{figure}}{\begin{figure}
    \centering
    \includegraphics[width=0.8\linewidth]{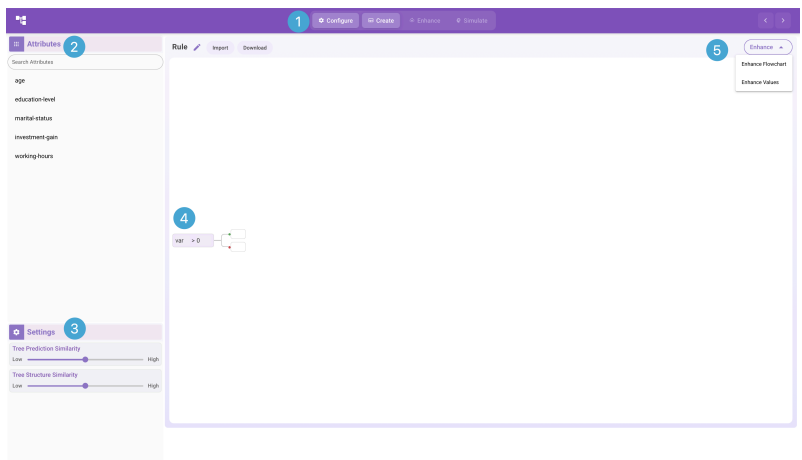}
    \caption{Tutorial on the AI interface.}
    \label{fig:tutorial-interface-a}
    \Description[AI Interface Tutorial]{The tutorial introduces an AI interface with four stages in a top bar: Configure, Create, Enhance, Simulate. In the Configure stage, there is an Attribute List where attributes can be clicked to pick up for use in rules (and clicked again to drop), and Enhance Settings to adjust “Prediction Similarity” (how similar AI-enhanced rules’ prediction behavior is to the user’s) and “Structure Similarity” (how similar AI-enhanced rules’ flowchart structure is to the user’s). The Create stage includes a Rule Canvas where attributes are dropped into a purple box (with a value typed in a second box), and white boxes accept “High/Low” decisions or additional attributes to extend rules; it also offers AI Enhancement with two options: “Enhance Rule Flowchart” (AI revises rule structure, attributes, and their positions) and “Enhance Value” (AI keeps structure but revises attribute values). A screenshot below the text labels interface elements (numbered 1–5) matching these components: the top bar stages, attribute list, enhance settings, rule canvas, and AI enhancement menu.}
\end{figure}
\begin{figure}\ContinuedFloat
    \centering
    \includegraphics[width=0.8\linewidth]{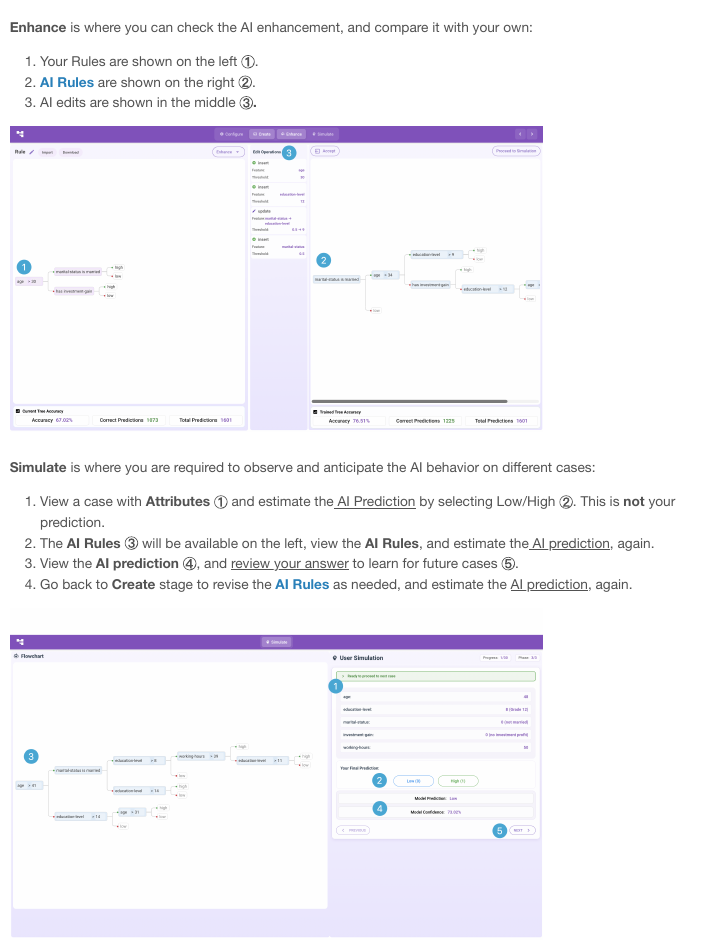}
    \caption{(continued)}
    \label{fig:tutorial-interface-b}
    \Description[Continued AI Interface tutorial]{The "Enhance" stage allows users to check AI enhancements and compare them with their own rules: user rules are shown on the left, AI rules on the right, and AI edits in the middle. The "Simulate" stage requires users to observe and anticipate AI behavior for different cases: view a case’s attributes and estimate the AI prediction by selecting Low/High (not the user’s own prediction); AI rules are available on the left to re-estimate the prediction; then view the actual AI prediction and review the answer to learn for future cases. Users can also return to the "Create" stage to revise AI rules and re-estimate predictions. Screenshots depict these stages: the "Enhance" interface shows side-by-side visualizations of user and AI rules with performance metrics, while the "Simulate" interface includes case attribute displays, prediction selection buttons, AI rule visualizations, and prediction feedback elements.}
\end{figure}}

\begin{figure}
    \centering
    \includegraphics[width=0.8\linewidth]{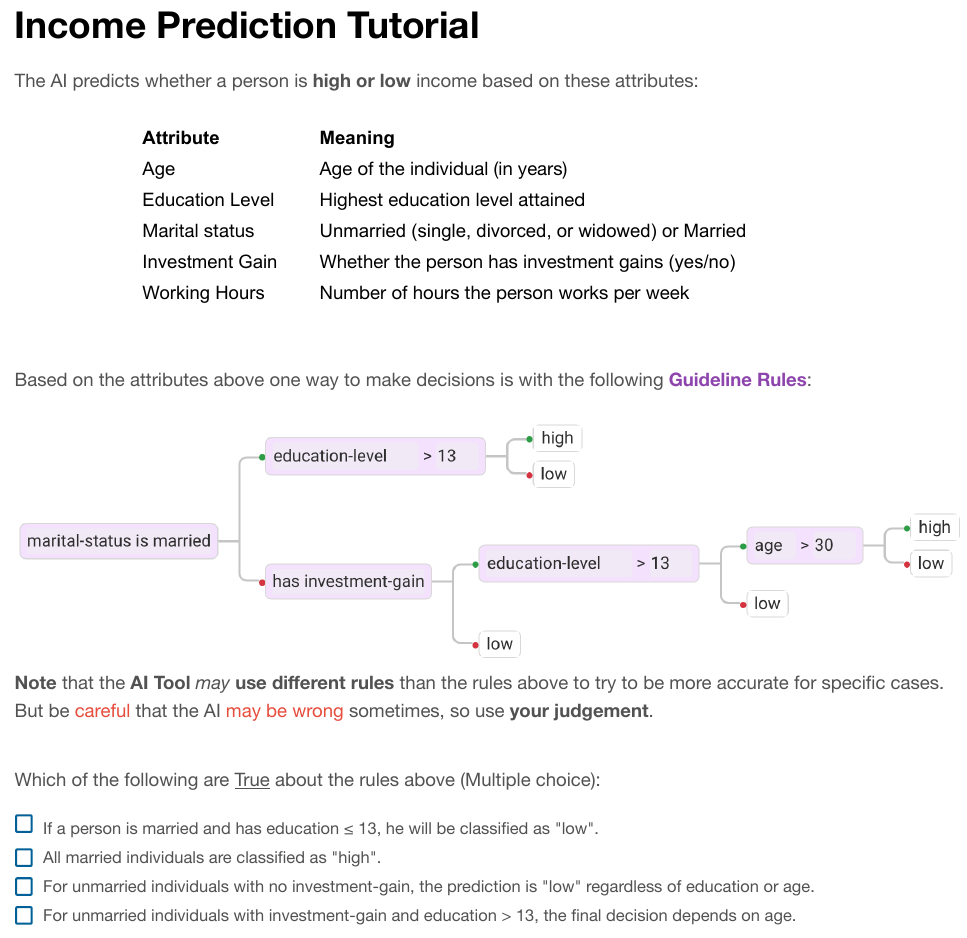}
    \caption{Tutorial on the Adult Income Prediction Task and screening questions to check the users' interpretation.}
    \label{fig:income}
    \Description[Adult Income Prediction Tutorial]{The tutorial explains an AI system that predicts if a person has high or low income using attributes: Age (in years), Education Level (highest level attained), Marital status (Unmarried or Married), Investment Gain (yes/no), and Working Hours (per week). It presents “Guideline Rules” via a decision tree: starting with marital-status is married. If true, check education-level greater than 13 — predicting “high” (if true) or “low” (if false). If marital-status is married is false, check has investment-gain: if false, predicts “low”; if true, check education-level greater than 13 — if false, predicts “low”; if true, check age greater than 30 — predicting “high” (if true) or “low” (if false). A note warns the AI tool may use different rules for accuracy (and can be incorrect), so users should use their judgment. Multiple-choice questions check understanding (e.g., “If a person is married and has education less than or equal to 13, he will be classified as 'low'”; “For unmarried individuals with investment-gain and education greater than 13, the final decision depends on age”).}
\end{figure}

\begin{figure}
    \centering
    \includegraphics[width=0.8\linewidth]{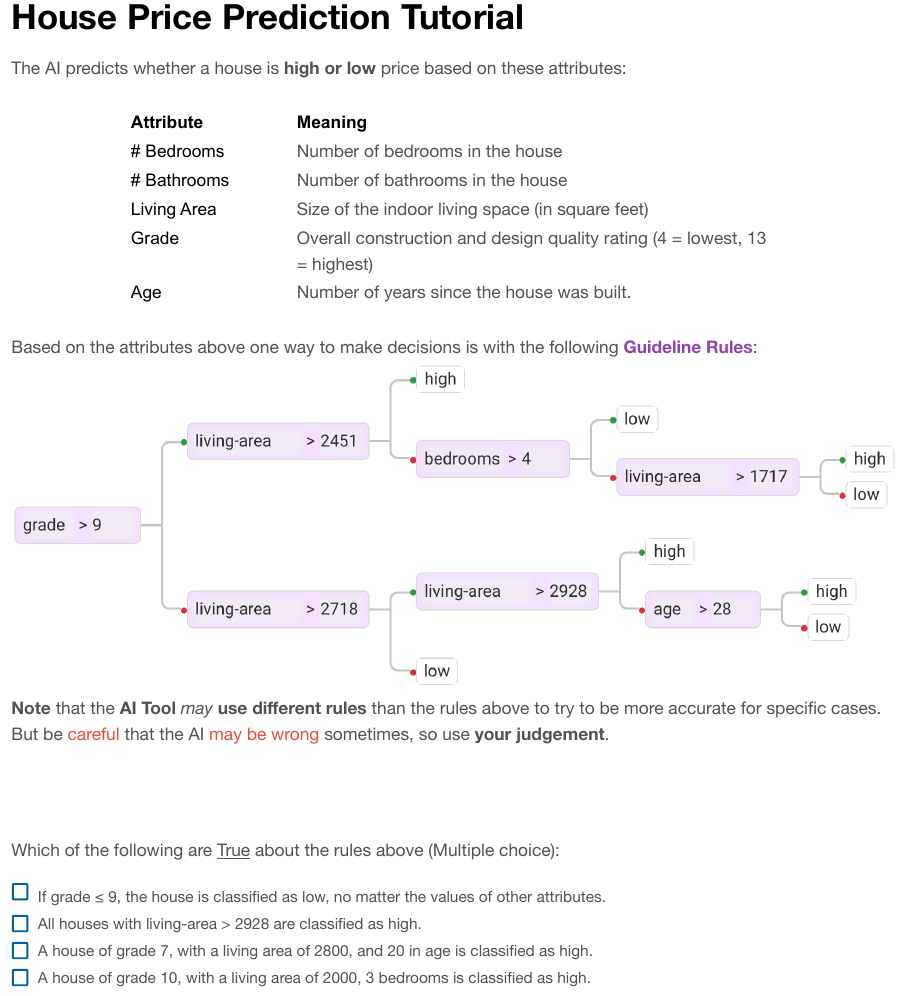}
    \caption{Tutorial on the House Price Prediction Task and screening questions to check the users' interpretation.}
    \label{fig:house}
    \Description[House Price Prediction Tutorial]{The tutorial explains an AI system that predicts whether a house has a high or low price using attributes: \# Bedrooms (number of bedrooms in the house), \# Bathrooms (number of bathrooms in the house), Living Area (size of indoor living space in square feet), Grade (overall construction and design quality rating where 4 is lowest and 13 is highest), and Age (number of years since the house was built). It presents “Guideline Rules” via a decision tree: starting with grade greater than 9. If true, check living-area greater than 2451 — predicting “high” (if true) or branching to bedrooms greater than 4 (which predicts “low” if true, or living-area greater than 1717 to predict “high” or “low”). If grade greater than 9 is false, check living-area greater than 2718 — if true, check living-area greater than 2928 (predicting “high” if true, or age greater than 28 to predict “high” or “low”); if living-area greater than 2718 is false, predicts “low”. A note warns the AI tool may use different rules for accuracy (and can be wrong), so users should use judgment. Multiple-choice questions verify understanding (e.g., “If grade less than or equal to 9, the house is classified as low, no matter the values of other attributes”; “A house of grade 10, with a living area of 2000, 3 bedrooms is classified as high”).}
\end{figure}

\begin{figure}
    \centering
    \includegraphics[width=0.8\linewidth]{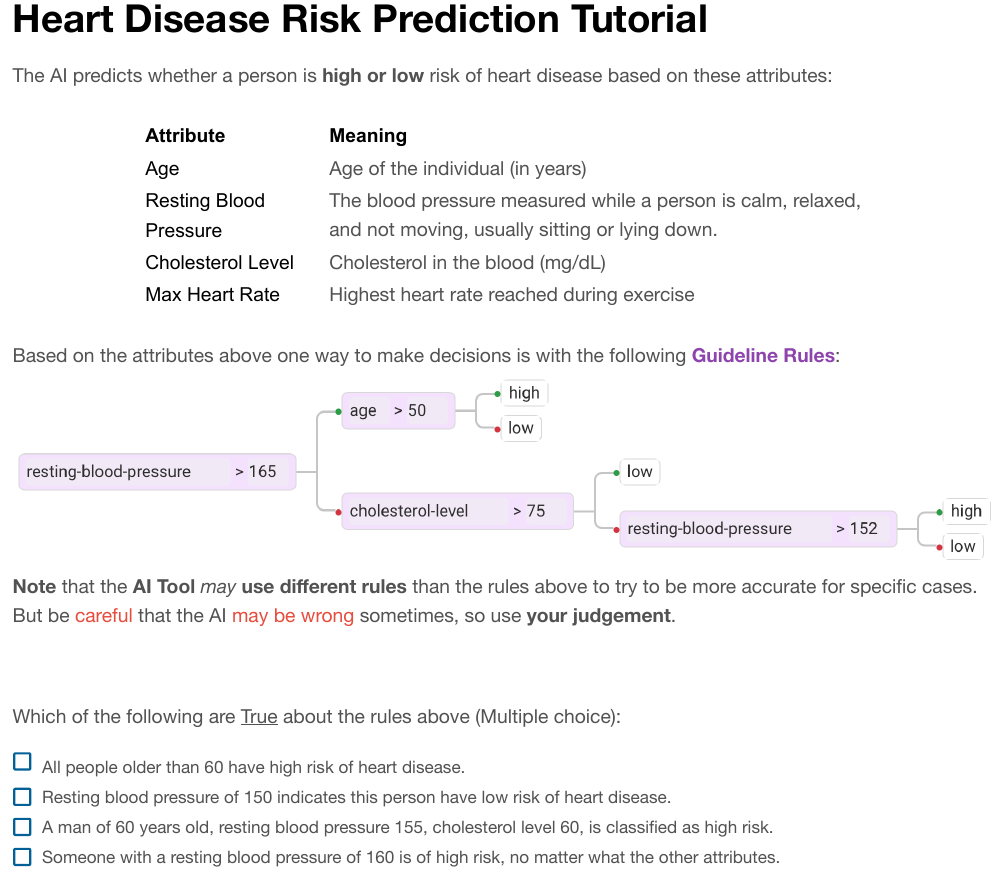}
    \caption{Tutorial on the Heart Disease Prediction Task and screening questions to check the users' interpretation.}
    \label{fig:heart}
    \Description[Heart Disease Risk Prediction Tutorial]{The tutorial explains an AI system that predicts whether a person has high or low risk of heart disease using attributes: Age (in years), Resting Blood Pressure (measured while calm/relaxed, usually sitting/lying down), Cholesterol Level (in mg/dL), and Max Heart Rate (highest during exercise). It presents “Guideline Rules” via a decision tree: starting with resting-blood-pressure greater than 165. If true, check age greater than 50 — predicting “high” (if true) or “low” (if false). If resting-blood-pressure greater than 165 is false, check cholesterol-level greater than 75 — if false, predicts “low”; if true, check resting-blood-pressure greater than 152 — predicting “high” (if true) or “low” (if false). A note warns the AI tool may use different rules for accuracy (and can be wrong), so users should use judgment. Multiple-choice questions verify understanding (e.g., “Resting blood pressure of 150 indicates this person have low risk of heart disease”; “A man of 60 years old, resting blood pressure 155, cholesterol level 60, is classified as high risk”).}
\end{figure}

\clearpage
\begin{figure}
    \centering
    \includegraphics[width=0.8\linewidth]{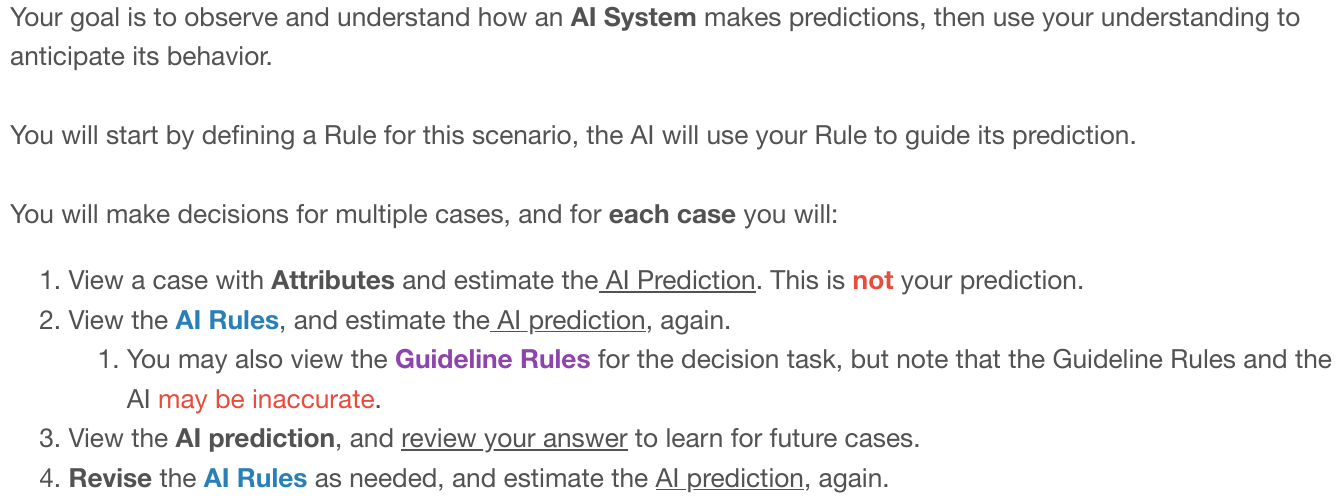}
    \caption{Task description for Editable and CoExplain explanation.}
    \label{fig:task-editable}
    \Description[Task description for Editable and CoExplain explanation]{Text outlining the task for “Editable” and “CoExplain” interfaces: The goal is to observe and understand how an AI system makes predictions, then use that understanding to anticipate its behavior. Users begin by defining a rule, which the AI uses to guide its predictions. For each case, users perform four steps: 1) View a case’s attributes and estimate the AI’s prediction (noting this is not their own prediction). 2) View the AI’s rules (and optionally view guideline rules for the task, with a note that guideline rules and the AI may be inaccurate), then re-estimate the AI’s prediction. 3) View the AI’s actual prediction and review their answer to learn for future cases. 4) Revise the AI’s rules as needed, then re-estimate the AI’s prediction.}
\end{figure}

\begin{figure}
    \centering
    \includegraphics[width=0.8\linewidth]{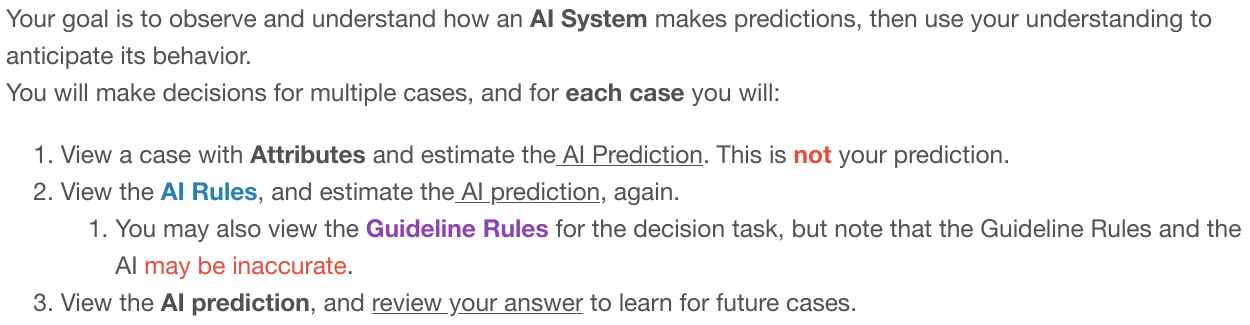}
    \caption{Task description for Read-only explanation.}
    \label{fig:task-readonly}
    \Description[Task description for Read-only explanation]{Text outlining the task for the “Read-only” interface: The goal is to observe and understand how an AI system makes predictions, then use that understanding to anticipate its behavior. For each case, users perform three steps: 1) View a case with attributes and estimate the AI prediction (noting this is not their own prediction). 2) View the AI rules (and optionally view guideline rules for the decision task, with a note that guideline rules and the AI may be inaccurate), then re-estimate the AI prediction. 3) View the AI prediction and review their answer to learn for future cases.}
\end{figure}

\begin{figure}
    \centering
    \includegraphics[width=0.8\linewidth]{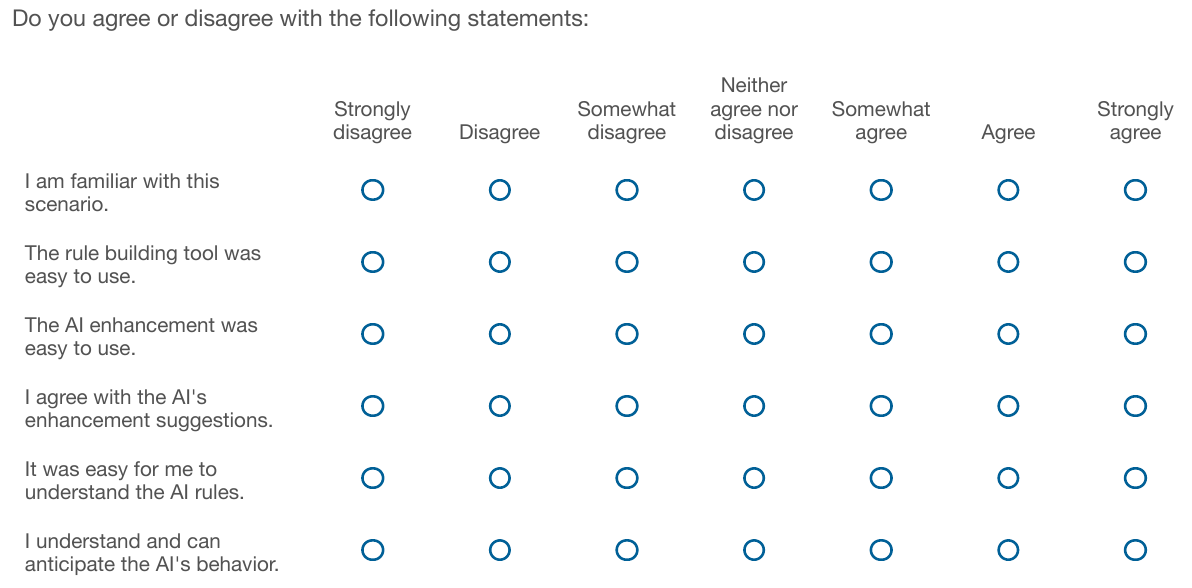}
    \caption{Likert Scale for perceived rating.}
    \label{fig:likert}
    \Description[Likert scale for perceived rating]{A Likert scale used to gather agreement ratings for statements related to user perception. The scale spans from “Strongly disagree” to “Strongly agree”, with intermediate options: “Disagree”, “Somewhat disagree”, “Neither agree nor disagree”, “Somewhat agree”, and “Agree”. The statements presented are: “I am familiar with this scenario”, “The rule building tool was easy to use”, “The AI enhancement was easy to use”, “I agree with the AI's enhancement suggestions”, “It was easy for me to understand the AI rules”, and “I understand and can anticipate the AI's behavior”. Each statement is paired with radio buttons for selecting a response on the scale.}
\end{figure}

%TC:endignore

\end{document}